\documentclass[12pt,preprint]{aastex}
\newcommand\kms{km s$^{-1}$}
\newcommand\msun{\ifmmode{M_{\odot}}\else $M_{\odot}$\fi}
\newcommand{\teff}{$T_{\rm eff}$} 

\begin{document}

\title{Elemental Abundance Ratios in Stars of the Outer 
Galactic Disk.\ I.\ Open Clusters\footnote{This
paper makes use of observations obtained at the National Optical Astronomy
Observatory, which is operated by AURA, Inc., under contract from
the National Science Foundation. We also employ data products
from the Two Micron All Sky Survey, which is a joint project
of the University of Massachusetts and the Infrared Processing
and Analysis Center/California Institute of Technology, funded by the
National Aeronautics and Space Administration and the National
Science Foundation.}}
\author{David Yong}
\affil{Department of Physics \& Astronomy, University of North
Carolina, Chapel Hill, NC 27599-3255; email: yong@physics.unc.edu}

\author{Bruce W.\ Carney}
\affil{Department of Physics \& Astronomy, University of North
Carolina, Chapel Hill, NC 27599-3255; email: bruce@physics.unc.edu}

\author{Maria Lu\'{i}sa Teixera de Almeida}
\affil{Department of Physics \& Astronomy, University of North
Carolina, Chapel Hill, NC 27599-3255; email: luisa@oal.ul.pt}

\begin{abstract}
We summarize radial velocity studies of selected
stars in the old, distant clusters Berkeley~20,
Berkeley~21, NGC~2141, Berkeley~29, and Berkeley~31. 
Cluster members are identified using optical and infrared
color-magnitude diagrams as well as radial velocities
derived from
high-resolution echelle spectra. Three members of M67
were observed similarly, and those velocities compare extremely
well with prior measures. Mean cluster radial velocities are determined.
We also employ the highest quality spectra to analyze the
chemical compositions of all six clusters for [Fe/H]
as well as abundances of ``$\alpha$'' elements, iron-peak
elements, and those
synthesized in either the $s$-process or the $r$-process.
In Be 21, our observed star is found to be rotating rapidly and 
overabundant in lithium, the second Li-rich star found in this sparse cluster.

We confirm the lack of correlation between abundance and age.
For the outer disk, the abundance gradient for [Fe/H] deviates
from the trend defined near the solar neighborhood. Rather
than declining with increasing Galactocentric distance, [Fe/H] 
appears to reach a ``basement'' at [Fe/H]$\approx-$0.5 
beyond $R_{\rm GC}\approx$ 10 to 12kpc. 
Our radial abundance distribution for [Fe/H] is not inconsistent with the
radial abundance discontinuity exhibited by Cepheids. 
We find enhanced [O/Fe], [$\alpha$/Fe], and [Eu/Fe] in the outer disk 
revealing a rapid star formation history. The outer disk also exhibits
enhancements for $s$-process elements. We compare the open
cluster compositions with the thin disk, thick disk, halo, bulge,
and dwarf spheroidals. None of these stellar populations perfectly
matches the abundance ratios of the outer disk open clusters.
Several key points arise from these comparisons. (1) [O/Fe] and 
[$\alpha$/Fe] resemble the thick disk. (2) [Na/Fe] and [Al/Fe]
are enhanced relative to the thin disk. (3) [Ni/Fe] and [Mn/Fe] are in accord
with the thin disk while [Co/Fe] may be slightly enhanced. 
(4) The neutron capture elements indicate different
ratios of $s$-process to $r$-process material with no cluster showing 
a pure $r$-process distribution. (5) An unusual pattern exists among
the $\alpha$ elements with [$<$Mg+Ti$>$/Fe] 
enhanced while [$<$Si+Ca$>$/Fe] is normal. 
Similar abundance ratios have been reported for Galactic bulge giants 
and indicate a common but not necessarily shared 
nucleosynthetic history between the bulge and the outer disk. 
Enhanced ratios of [Al/Fe] and [Co/Fe] offer another possible similarity 
between the bulge and the outer disk. 

An intriguing, but tentative, conclusion is that the outer disk
open cluster abundance ratios are consistent with 
the outer disk being formed via a merger event. The 
basement in [Fe/H] and enhanced [$\alpha$/Fe] 
suggest that the outer disk formed from a reservoir of gas with a star 
formation history distinct from the solar neighborhood. That the
open clusters may be associated with an accreted dwarf galaxy is
appealing since the clusters are young and have [$\alpha$/Fe] ratios indicating 
a rapid star formation history. However, the high [$\alpha$/Fe] ratios are unlike 
those seen in any current dwarf galaxies at the same [Fe/H]. Therefore, the 
open clusters may have formed as a result of star formation triggered by a
merger event in the outer disk. The ages of the outer disk open clusters would 
then be a measure of when the merger occurred. However, Be 29 is a candidate
merger member while Be 31 is not. One problem with the merger scenario is that 
open clusters with presumably very different origins have similar and unusual 
compositions. 

\end{abstract}


\section{INTRODUCTION}

How do galactic disks arise and evolve? Disks are both common yet
relatively easily destroyed in inter-galactic tidal encounters
or mergers with comparable mass systems, so it would appear that
most disks have undergone relatively undramatic evolution. Putting
aside the origin of the angular momentum, which presumably arises from
significant tidal encounters as galaxies were forming early in
the Universe's history, what are the major drivers of galactic
chemical and dynamical evolution and what observational data
do we need to test our ideas? Models of galactic evolution have identified
a number of key processes, including the star formation rate, the initial
mass function, infall of material, and the degree and extent of
recycling of the products of stellar nucleosynthesis. In the solar
neighborhood the data to test the models are extensive, including
ages of clusters and some individual stars. One of the earlier
puzzles, the lack of a relationship in the local thin disk population 
between age and mean metallicity (cf.\ \citealt{bdp93,nordstrom04})
may have been resolved \citep{bdp03} 
once the thin disk and thick disk stars are considered separately.
Nonetheless, other puzzles remain. 
For example, the remaining scatter in metallicity at a given age (or,
conversely, the scatter in age at a fixed metallicity), is a major
puzzle. It also appears that despite the existence of a relationship
between age and metallicity, and the presence of a radial 
metallicity gradient in the disk,
the detailed abundance patterns [X/Fe] at a given [Fe/H] are essentially 
identical for all thin disk stars \citep{bdp03}. Open clusters
do not appear to display signs of an age-metallicity relation 
\citep{friel95,friel02,chen03,salaris04}, but it would be interesting to
try to distinguish them in terms of thin disk and thick disk membership,
and see if the solution of \citeauthor{bdp03} to the age-metallicity
relationship applies to the clusters. 

Detailed chemical abundance patterns offer valuable
additional clues in our stellar ``excavations" and interpretation
of our Galactic history. For example, when star formation begins
in an ensemble of gas, it is enriched first in the
products ejected by the evolution of the shortest-lived stars.
Massive stars are thought to eject higher fractions of $r$-process
nucleosynthesis, as well as the ``$\alpha$" elements (oxygen,
magnesium, silicon, calcium, sulfur, and, perhaps, titanium). 
A little later ($\approx 10^{8}$ years) the 
intermediate mass stars begin to contribute
via ejecta from asymptotic giant branch stars, and the $s$-process
elements begin to appear in significant quantities. Only later
($\approx 10^{9}$ years)
do the iron peak elements arise in greater abundances as the Type~Ia
supernovae begin to enrich the interstellar medium. The mass
function alters some of the relative ratios of the these various elements,
and the star formation rate helps determine the timing of their
appearance, and, hence, in a ``closed box'' model, 
the metallicity at which they are first seen.
For example, at the lowest metallicities, when star formation has
just begun, we expect to see high [$\alpha$/Fe] and [Eu/Fe] ratios
(Eu being almost entirely produced by the $r$-process). In systems
in which the star formation is intense, and the metallicity rises
rapidly, we would expect to see these high ratios continue up
to high metallicities. A much slower star formation rate may
result in contributions from Type~Ia supernovae appearing at much
lower metallicities.
Infall and mixing both contaminate the interstellar medium abundance
mixture, but [X/Fe] ratios,
and the scatter in them at fixed [Fe/H] or age, can help to
sort out all these effects.

While the chemical abundances data are extensive in the solar neighborhood, our
knowledge of other parts of the Galactic disk is still relatively
primitive. The study of the mean metallicity gradient, 
$\Delta$[Fe/H]/$\Delta R_{GC}$
(where $R_{GC}$ is the distance from the Galactic center), provides some
information, but its use is limited in disentangling all the contributing
factors. (Recent work and references of prior work are summarized
nicely by \citealt{friel02,andrievsky02b,chen03}.)
Additional uncertainty arises because the samples of different
classes of objects (H~II regions, B~stars, planetary nebulae, open clusters)
are analyzed using very different methods, enabling subtle systematic
effects to magnify or diminish the true metallicity gradients. Further,
we would very much like to compare the results as a function of age,
if at all possible, so that we can at least measure some sort of time
derivative in the metallicity gradient and, preferably, in [X/Fe] ratios
as well. The possibility of an on-going merger event in the outer disk
\citep{ibata03,yanny03,frinchaboy04,martin04} underscores
the importance of pushing our spectroscopic studies of element-to-iron
ratios for samples of stars with known ages out to large Galactocentric
distances.

For these reasons we began a program in 1997 to identify stars with
fairly large Galactocentric distances, and whose temperatures
were similar enough so that we could analyze their spectra using
identical tools. For young stars, we chose to study Galactic Cepheids,
which are readily identifiable to large distances and, it appears,
amenable to traditional spectroscopic analyses 
\citep{fry97,andrievsky02a,andrievsky02b,andrievsky02c,luck03}.
For old stars, we must generally
rely on open clusters and, due to extinction and distance, their brightest
members, luminous K giants. We report our results for five such clusters
in this paper. In a forthcoming paper, we will report results for three
field K giants that appear to lie in the direction of the southern Galactic 
warp \citep{carney93}, and, in a third paper, on abundances for roughly
two dozen distant Cepheids.

\section{SELECTION OF CLUSTERS}

\citet*{phelps94} undertook a systematic search
of open clusters, employing the morphology of available
color-magnitude diagrams to identify the oldest clusters.
They relied on the observed index, $\delta V$, defined to
be the $V$ magnitude difference between two reference
points in the color-magnitude diagram. The brighter
point is the mean level
of the red horizontal branch or the red clump so commonly
seen in older and/or metal-rich clusters. The fainter
point could have been the main sequence turn-off, but this
is not always easy to estimate when clusters show a
``blue hook" in that regime. The presence
of binary stars may additionally complicate the
measurement of the main sequence turn-off magnitude.
\citet{phelps94} therefore chose to rely on the
inflection point seen in the subgiant branch, lying
between the region of the main sequence turn-off and the
base of the red giant branch. \citet{phelps94} 
have provided an empirical, and necessarily approximate,
relationship between $\delta V$ and clusters' ages, defining
a ``morphological age index" (MAI). We chose to study
clusters with $\delta V$ values of 1.6 or greater, 
which lead to MAI values (and approximate cluster ages)
of 3 Gyr or greater. The ages we adopt are taken from
a re-calibration of the MAI by \citet{salaris04},
as were the initial estimates of the clusters' metallicities,
[Fe/H]\footnote{The adopted reddenings and [Fe/H] values agree generally
with those from \citet{friel02}, except for Be~21, where they found
[Fe/H]~=~$-0.62$ due to a difference in the adopted reddening to the
cluster.}. Table~\ref{tab:clusters} summarizes the basic
data for the clusters and the sources of the photometry. 
We note that improved abundance estimates and good 
color-magnitude diagrams (optical and infrared) should 
enable more accurate age determinations. 

Since we wish to probe the outer disk, we also relied
on the cluster positions and distance estimates from
\citet{phelps94}, so that our program clusters
would have Galactocentric distances of 12~kpc or greater.
Given practical limits of observing time, our final list
includes five old open clusters, Be~20, Be~21, NGC~2141,
Be~29, and Be~31. 

We have also observed three bright stars in the local
old open cluster M67. Since our goal was to determine
cluster membership using radial velocities before we
obtained longer exposure, higher signal-to-noise (S/N) spectroscopic
observations, we decided to derive velocities for
very well-studied stars in the cluster. Furthermore,
high-S/N spectroscopy of M67 stars provides us
with a direct comparison of the abundance pattern
of an old open cluster with a Galactocentric distance
comparable to the Sun (and comparable in age) with
those of the outer disk clusters, using the same
spectroscopic facilities, line lists, and analysis
procedures.

We selected program stars within each cluster on the
basis of available optical color-magnitude diagrams. Subsequent
to the observing for this program, the all-sky release of the 2MASS survey
became available, and since infrared photometry provides
a useful check on optical photometry for clusters suffering
relatively high extinction, we show in 
Figures~\ref{fig:be20vi}-\ref{fig:be31jk} the available
optical and infrared data. Our program stars are identified
by the open squares that surround their plotted positions
in the Figures.
The available photometric data for our program stars
are given in Table~\ref{tab:photometry}.

\section{SPECTROSCOPIC OBSERVATIONS}

We employed the echelle spectrographs and the 4-meter
telescopes at the Kitt Peak National Observatory (KPNO) and
the Cerro Tololo Inter-American Observatory (CTIO) during four
different observing runs in December 1997, January 1998,
and January 1999. In all cases we employed the
long red cameras and the 31.6 lines~mm$^{-1}$
echelle gratings. GG495 filters were used to block
second-order blue light. The wavelength coverage
was different at CTIO (5200-7940 \AA) than at KPNO
(4825-8000 \AA), due to the use of differing
cross-dispersers, G181 at CTIO (316 lines~mm$^{-1}$)
and G226 at KPNO (226 lines~mm$^{-1}$). The
slit was opened to 150 microns, providing a width of
1.0\arcsec\ on the sky, and yielding a spectral
resolving power of 28,000 and a dispersion of 0.07 pixel
at 5800\AA, providing two pixels per resolution element.

The observing routine included 20 quartz lamp exposures
to provide data for flat-fielding, and 15 zero-second
exposures (to provide ``bias" frames). Th-Ar hollow
cathode lamp spectra were taken before and after each
stellar exposure, and at least one radial velocity
standard star was observed every night. Table~\ref{tab:vradstandards}
lists the four stars upon which we relied to provide
radial velocity standards. All are K giants, similar at
least approximately in temperature and gravity to our
program stars. Details of the observations of the
program stars is given in Table~\ref{tab:spectra}.

The initial observations of cluster members involved
relatively short exposure times, not sufficient to
undertake a comprehensive abundance analysis, but
always sufficient to enable us to measure radial
velocities with uncertainties likely to be slightly smaller
than the anticipated cluster internal velocity dispersions.
Following analyses of these observations, longer
exposures were obtained for selected stars. The fainter
stars, $V > 14.7$ mag, had S/N levels of as low as 60
per pixel, or about 85 per resolution element at
7100 \AA. But we generally were able to achieve S/N
levels of about 100 per pixel for our program clusters'
abundance analysis targets, and 200 for the M67 stars.

The spectroscopic data were reduced using the
IRAF\footnote{IRAF (Image Reduction and Analysis
Facility) is distributed by the National Optical Astronomy
Observatory, which is operated by the Association of Universities
for Research in Astronomy, Inc., under contract with the National
Science Foundation.} packages IMRED, CCDRED, and ECHELLE to correct
for the bias level, trim the overscan region, divide by the normalized
flat field, remove scattered light, extract individual
orders, fit the continuum, apply a wavelength solution using
the Th-Ar spectra (and determine a systematic correction using
the observed radial velocity standard). 

\section{RADIAL VELOCITIES AND CLUSTER MEMBERSHIP}

\subsection{Measurement of Radial Velocities}

Radial velocities were measured with the task FXCOR in the
IRAF package RV. FXCOR employs Fourier transform cross-correlations
between one-dimensional spectra with wavelength solutions for
a program star and a template, preferably one with high S/N
and a similar spectrum as the program star. Our wavelength
dispersions were rebinned to a common log-linear dispersion.
We set the 200 pixels on the short and long wavelength edge of
each order to zero intensity but the edges were apodized using
a cosine bell curve. The influence of noise was suppressed
using a ramp filter for the Fourier transforms. The cross-correlation
process produces a series of peaks in the output power spectrum,
and the dominant peak was chosen to be that relevant to the
radial velocity shift. That peak was fit using a Gaussian function.
We obtained velocities for many orders for each exposure, and these
were averaged to produce our final results, which are given
in Table~\ref{tab:spectra}. 

\subsection{Comparisons with Prior Results}

We first compare our results to those obtained by other workers
using high-resolution spectra, specifically for the clusters
M67 and Be~21.

The stars we have studied in M67 are cluster members, by any
criterion. They occupy plausible locations in both the optical
and infrared color-magnitude diagrams (Figures~\ref{fig:m67bv}
and \ref{fig:m67jk}). Stars 105, 108, and 141 (using the
original \citealt{fagerholm06} nomenclature) are numbered 1016, 978,
and 1010 by \citet{sanders77}. His study of proper motions of
stars in the cluster assigned membership probabilities of
93\%, 95\%, and 96\% for the three stars, respectively. The extensive
study of radial velocities in the cluster by \citet{mathieu86}
likewise confirms cluster membership for all three stars, and also
shows no signs of binary companions. Their 33, 327, and 25
radial velocities of stars 105, 108, and 141 yield mean radial
velocities of +34.3, +34.7, and +33.6 \kms. Our single measures
of the radial velocities are in excellent agreement, with
a mean difference, in the sense of our results minus those
of \citet{mathieu86}, of only $-0.9 \pm 0.2$ \kms\ 
($\sigma~=~0.3$ \kms). We conclude that our radial velocity
measurement procedures are reliable.

Unfortunately, we do not obtain such good agreement for Be~21 when we
compare our results with those of \citet{hill99}. They
observed four stars, whose radial velocities were very
consistent with each other, implying that all are members
of the cluster. The positions of the stars in a $VI_{C}$ color-magnitude
are likewise consistent with membership. But the average
velocity for the four stars is $+12.4 \pm 0.3$ \kms\ 
($\sigma~=~0.6$ \kms). Our four stars likewise appear to
be cluster members, based on the $VI_{C}$ and $JK$ color-magnitude
diagrams, and a small scatter in the measured radial velocities.
But our mean velocity is $+1.1 \pm 1.0$ \kms\ ($\sigma~=~2.0$ \kms).
[Note: while the photometry cited in Table~1 of \citet{hill99}
is consistent with cluster membership, the actual identifications
of the stars appear to be erroneous. The photometric values
from \citet{tosi98} for the stars listed are not consistent
with any of the values cited. Possibly a different set of
identifications were used? The star \citet{hill99}
listed as T9, for example, is almost certainly T67.] 
The 11~\kms\ offset
between the results of \citet{hill99} and our results
is disturbing, but, given our results for M67, we
are confident of our results. Further, we believe the small 
scatter in the velocities of all the stars from both
studies gives us some
assurance that all four stars are cluster members. 

We have two stars in common with the radial velocity study
of Be~20 by \citet{friel02}. They obtained
$V_{\rm rad}$~=~$+84 \pm 10$ \kms\ for star 22 (compared
to our $+78.9 \pm 1.2$ \kms), and $+67 \pm 10$ \kms\ for
star 28 (compared to our $+80.6 \pm 1.1$ \kms). The agreement
is reasonable given the measurement uncertainties.

\citet{minniti95} measured radial velocities for 15 stars in
the field of NGC~2141, with typical measurement uncertainties
of $\pm 5$ \kms\ for each star. We have three stars in common
with his work, numbers 514, 1286, and 1821. The mean
radial velocity for these three stars from \citet{minniti95}
is $+12.9 \pm 1.9$ \kms\ ($\sigma~=~3.3$ \kms), while the
mean velocity from Table~\ref{tab:spectra}, excluding star 1997,
is $+23.7 \pm 0.6$ \kms\ ($\sigma~=~1.0$ \kms). While our
radial velocities were lower than those of \citet{hill99},
they are higher than those of \citet{minniti95}. Again, however,
the differences are not large in the mean.
\citet{friel02} also measured the radial velocity
for star 1997 in NGC~2141, obtaining $+58 \pm 7$ \kms.
Our value, $-7.9 \pm 2.4$ \kms, differs very significantly
from theirs. We note that this star lies very close to
star 1333, and disentangling which star is which, and
to which star the photometry belongs, makes this situation
difficult to judge.

Two program stars in Be 31 (886 and 666) were also studied by 
\citet{friel02} and the radial velocity measurements are in
good agreement. They obtained $+61 \pm 10$ \kms\ for star 886 (compared
to our $+56.6 \pm 0.3$ \kms) and $+67 \pm 10$ \kms\ for star 666 (compared to our
$+74.0 \pm 0.4$ \kms). They concluded that star 666 was a non-member due to 
its high metallicity.  

\subsection{Membership Determination and Other Issues}

Proper motions are also useful cluster membership indicators,
but there are few data available for relatively distant
clusters. \citet*{dias02} have employed the
TYCHO2 catalog to measure proper motions for individual stars
and for clusters whose distances from the Sun exceed 1~kpc.
While measurements are available for some stars in NGC~2141 and
Be~31, none of those measured are common to our program stars.

In Table~\ref{tab:spectra} we identify the stars we believe to be cluster
members with asterisks. We believe all four stars 
observed in Be~20 are cluster members, based on the
small dispersion in radial velocity, and the plausible positions
of the stars in the color-magnitude diagrams in Figures~\ref{fig:be20vi}
and \ref{fig:be20jk}. (The spread in $J-K$ colors seen for stars
22 and 28 are consistent with the photometric uncertainties.)
The same conclusions are drawn for the four Be~21 program stars
(see Figures~\ref{fig:be21bv} and \ref{fig:be21jk}, 
and Table~\ref{tab:spectra}).

For NGC~2141, neglecting the two blended stars, 1333 and 1997,
the remaining six stars all appear to be cluster members, again
based on the color-magnitude diagrams (Figures~\ref{fig:ngc2141vi}
and \ref{fig:ngc2141jk})
and the small range
in radial velocities. Stars 1333 and 1997 may also prove to
be members. Indeed, it is interesting to note that 
while the measured radial velocities for star 1997 differ
dramatically (as noted above), the
average of our measured radial velocity for star 1997
and that measured by \citet{friel02} is consistent
with the cluster mean velocity. Could it be a spectroscopic
binary? This would imply a large orbital velocity amplitude
and, hence, a short period and a small orbital separation
compared to the extended radius of the red giant.

All seven program stars in Be~29 also appear to be members,
again based on the color-magnitude diagrams (Figures~\ref{fig:be29bv}
and \ref{fig:be29jk}) and the small range in radial velocities.

Our luck with the color-magnitude diagrams does not extend to
the final cluster, Be~31. Both the optical 
and infrared color-magnitude diagrams
are complex, and the distribution in radial velocities further
confuses cluster membership determinations. We suggest that
stars 886 and 728 are cluster members, with 
$<V_{\rm rad}>$~=~$+55.7 \pm 0.7$ \kms. It is interesting that
the mean radial velocity for all five stars is consistent
with this result, $+60.0 \pm 4.5$ \kms, although the
scatter is large ($\sigma~=~10.1$ \kms). We may have been
unfortunate enough to have included an uncommonly large fraction
of binary systems in our radial velocity measures. In the case
of star 666, which appears to lie near the tip of the cluster's
red giant branch, it is interesting that the 2MASS survey found
the image of the star to be elongated. Interestingly, 
\citet{friel02} also found
a large dispersion in velocity for this cluster and a large
number of field star contaminants.

We summarize the mean radial velocities for the five distant
open clusters in Table~\ref{tab:clustervelocities}. We do not
include a mean result for M67 because it would add nothing
to the comprehensive work of \citet{mathieu86}.

\section{CLUSTERS' REDDENING AND DISTANCE ESTIMATION}

The determination of interstellar reddening and extinction
is especially challenging for stars and clusters that
are concentrated to the Galactic plane, especially when we
seek objects at large distances from the Sun. 
Since A$_{K}$~=~0.11~A$_{V}$, we can study
highly extinguished clusters in a a much more uniform manner if
we rely on infrared photometry. 

Recently, \citet*{carney04} derived
a relation between the mean $J-K$ color index of red clump
stars and a cluster's metallicity:
\begin{equation}
\label{eq:jk0}
<J-K>_{0}~=~(0.170 \pm 0.026) [Fe/H] + (0.596\pm 0.016).
\end{equation}
Estimates of $<J-K>$ for clusters' red clump stars, combined
with an estimate of their metallicity, may therefore be 
combined to derive E($J-K$) [=~0.52~E($B-V$); \citealt{rieke85}].

\begin{figure}
\plotone{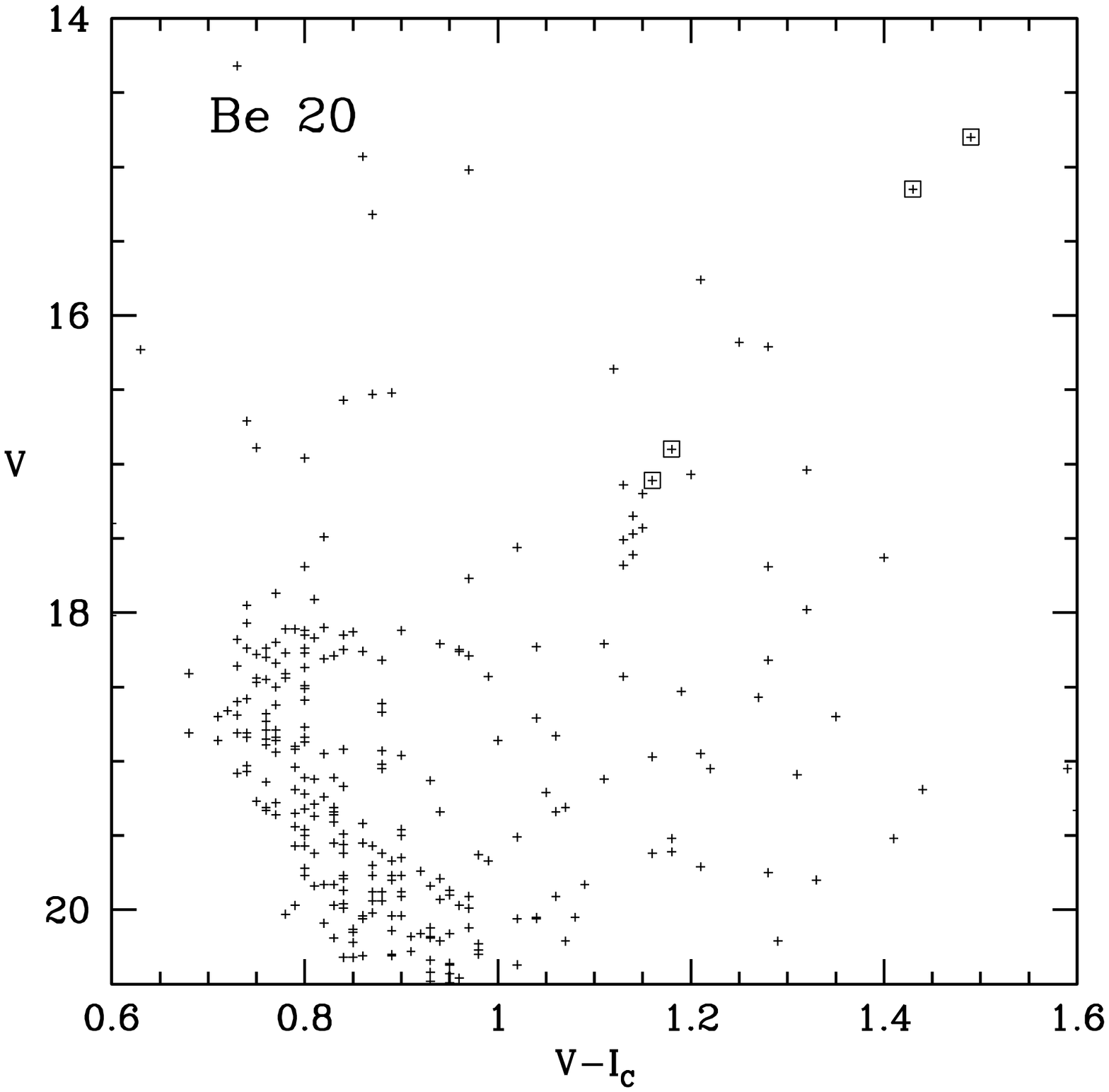}
\caption{We distinguish by enclosed squares the stars observed, using the
$VI_{C}$ data from \citet{macminn94}. \label{fig:be20vi}}
\end{figure}

\begin{figure}
\plotone{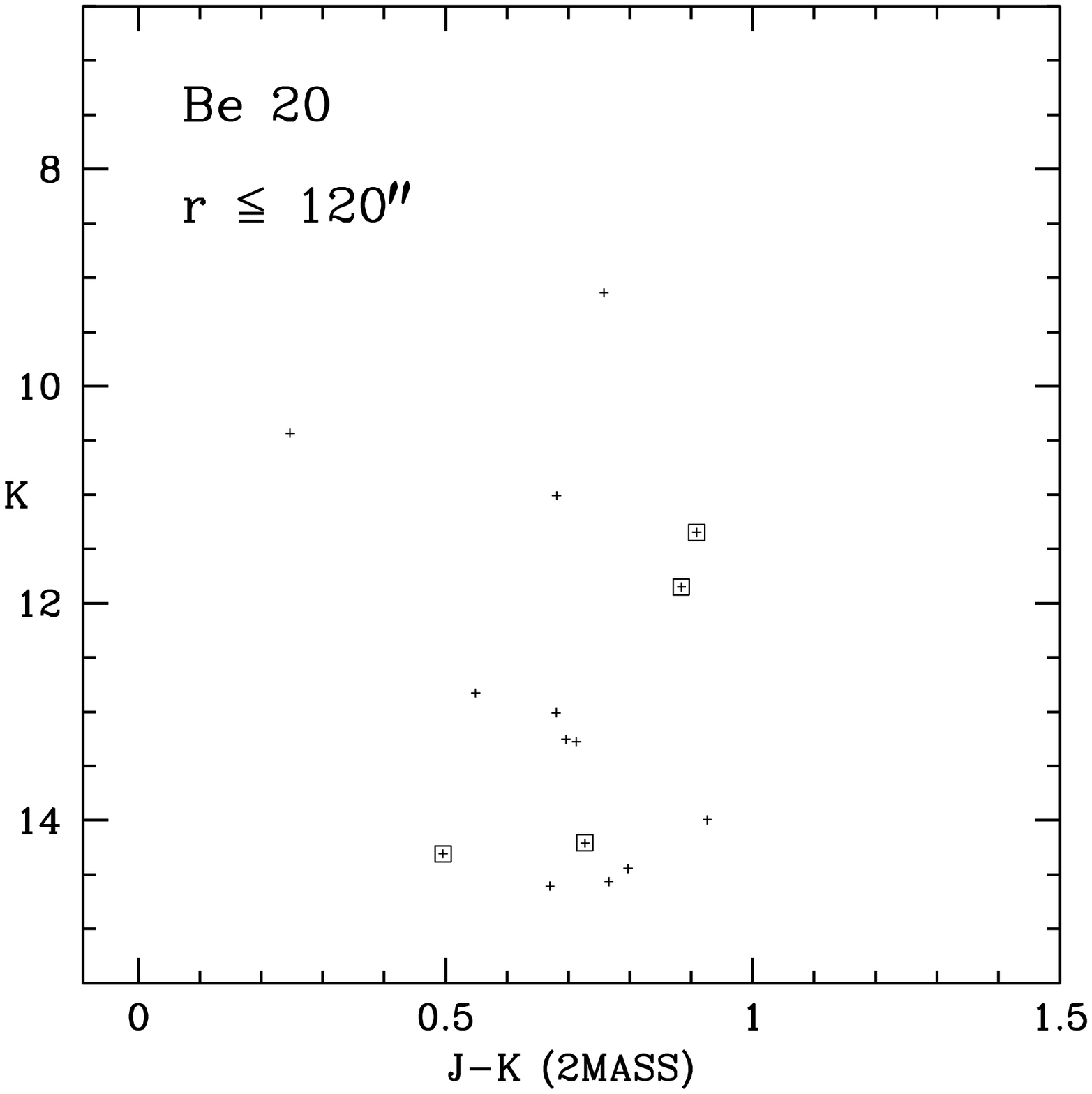}
\caption{We distinguish by enclosed squares the stars observed, using the
$JK$ data from the 2MASS survey. \label{fig:be20jk}}
\end{figure}

\begin{figure}
\plotone{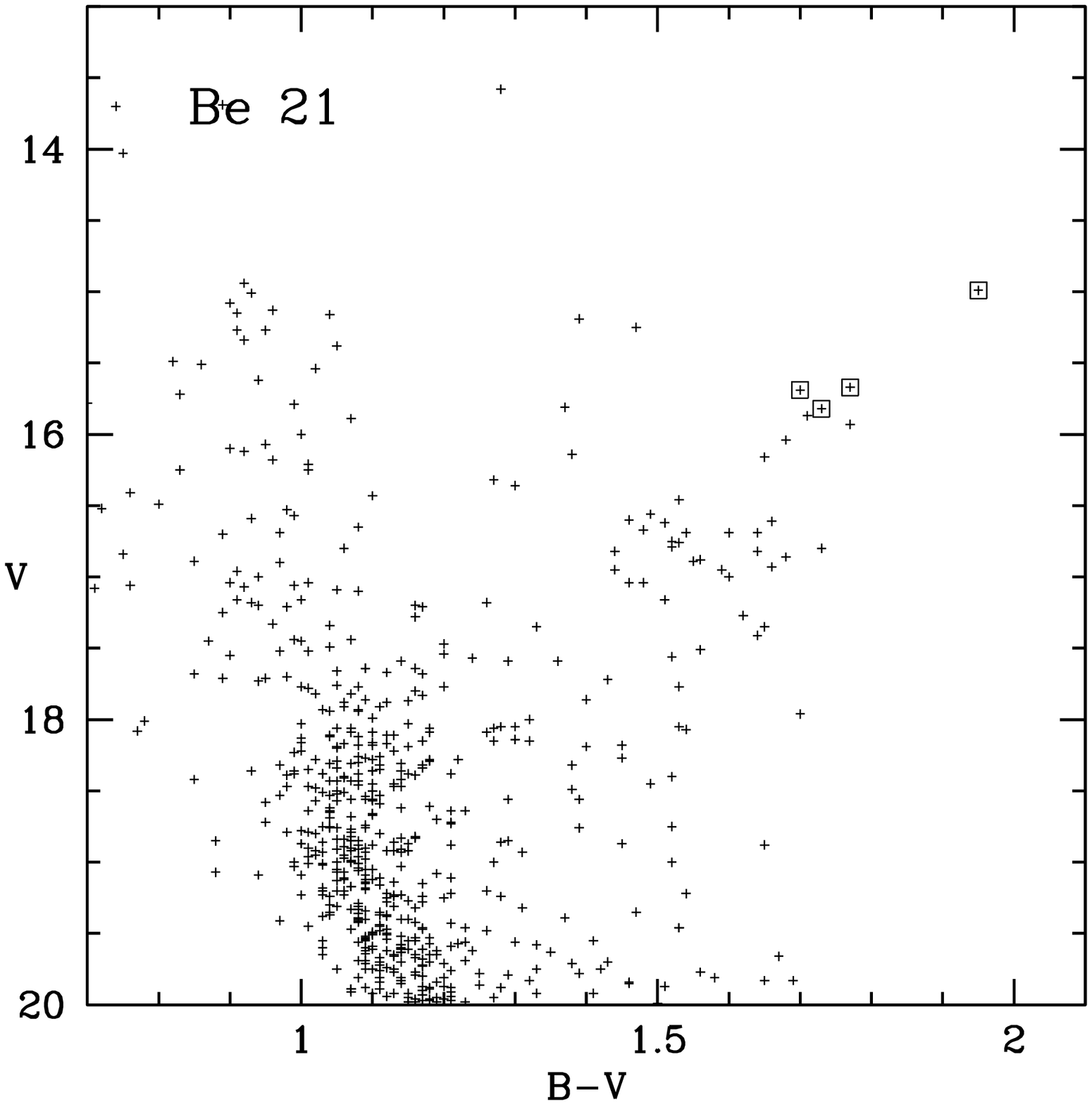}
\caption{We distinguish by enclosed squares the stars observed, using the
$BV$ data from \citet{tosi98}. \label{fig:be21bv}}
\end{figure}

\begin{figure}
\plotone{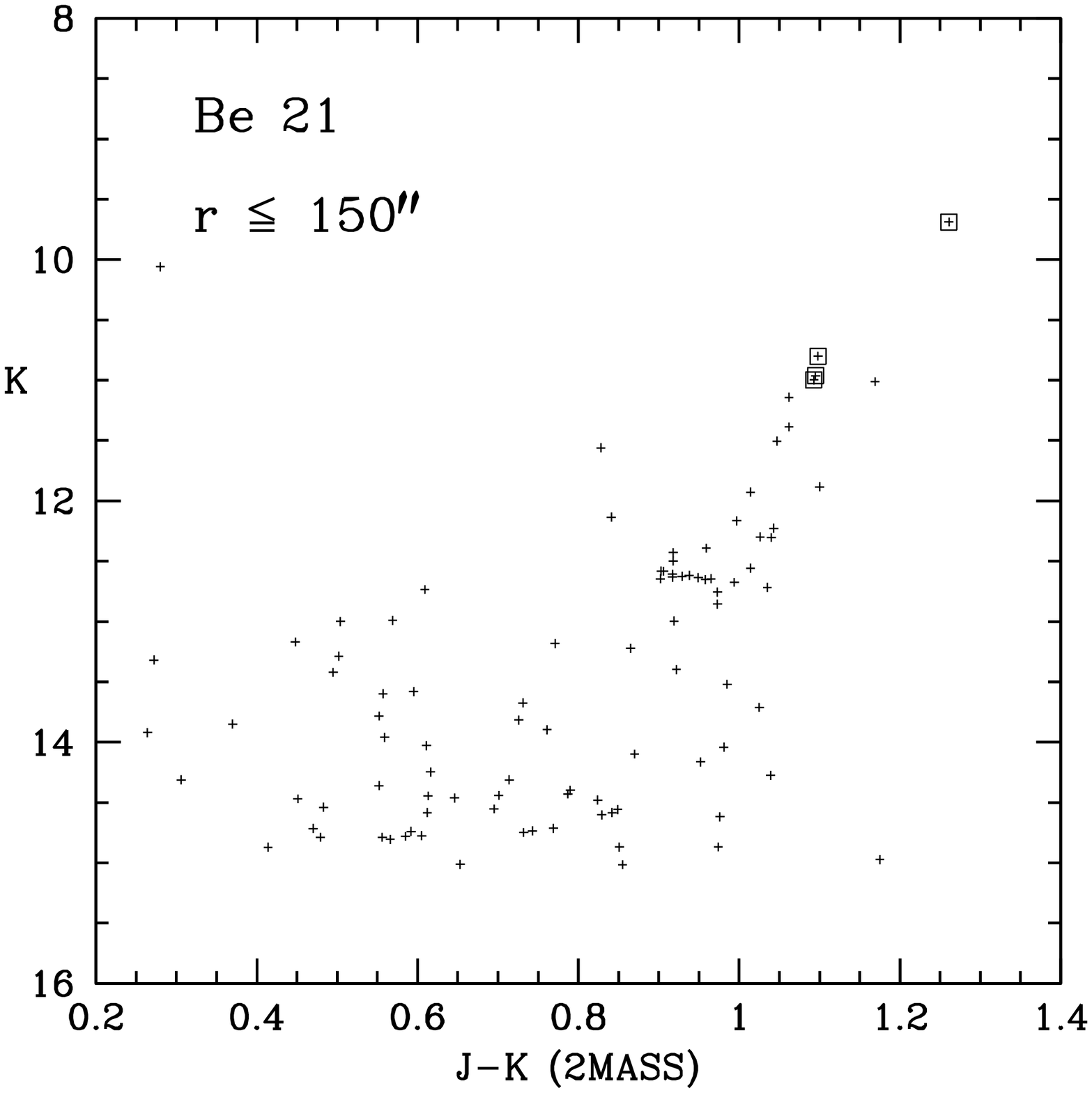}
\caption{We distinguish by enclosed squares the stars observed, using the
$JK$ data from the 2MASS survey. \label{fig:be21jk}}
\end{figure}

\begin{figure}
\plotone{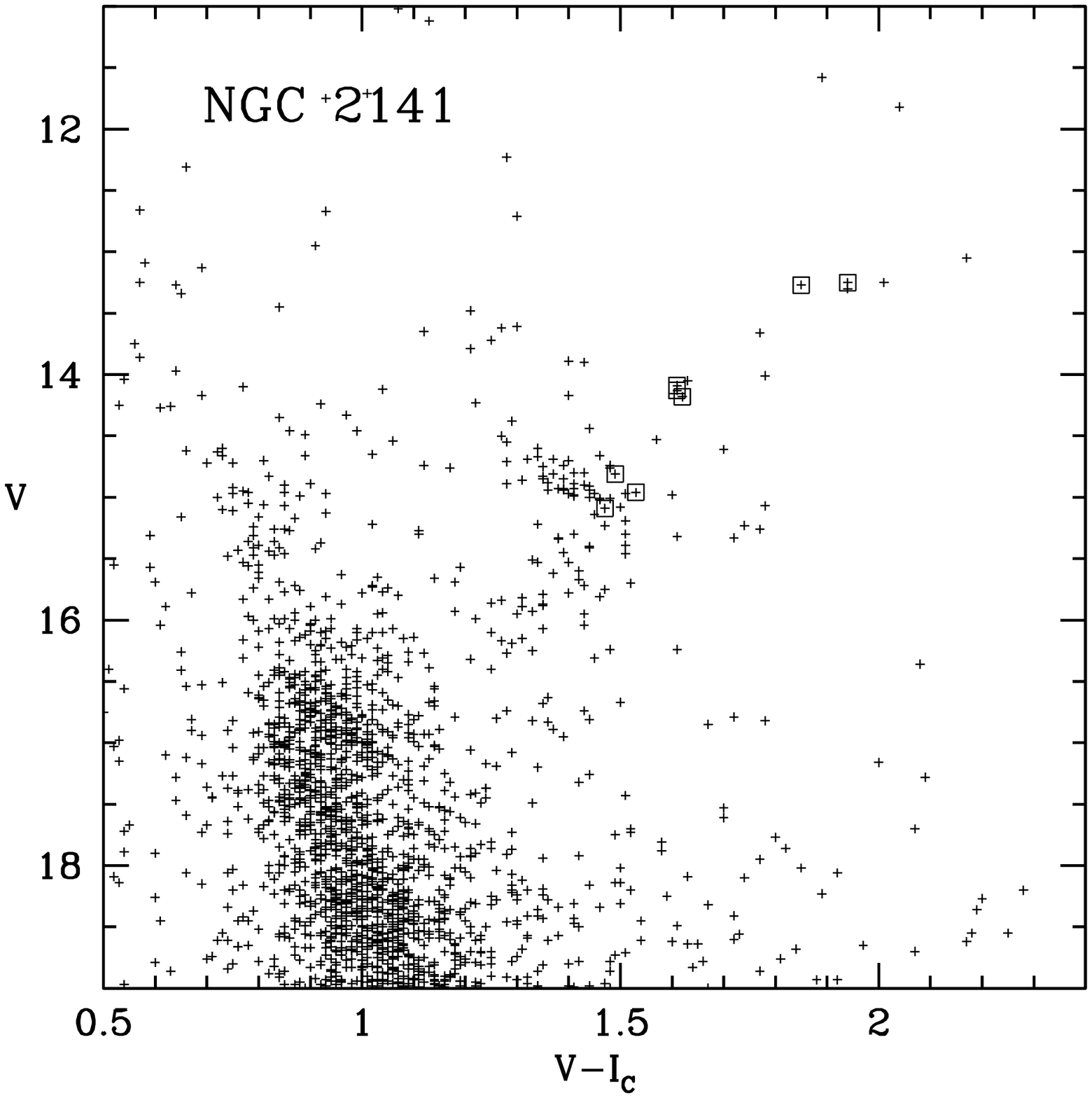}
\caption{We distinguish by enclosed squares the stars observed, using the
$VI_{C}$ data from \citet{rosvick95}. \label{fig:ngc2141vi}}
\end{figure}

\begin{figure}
\plotone{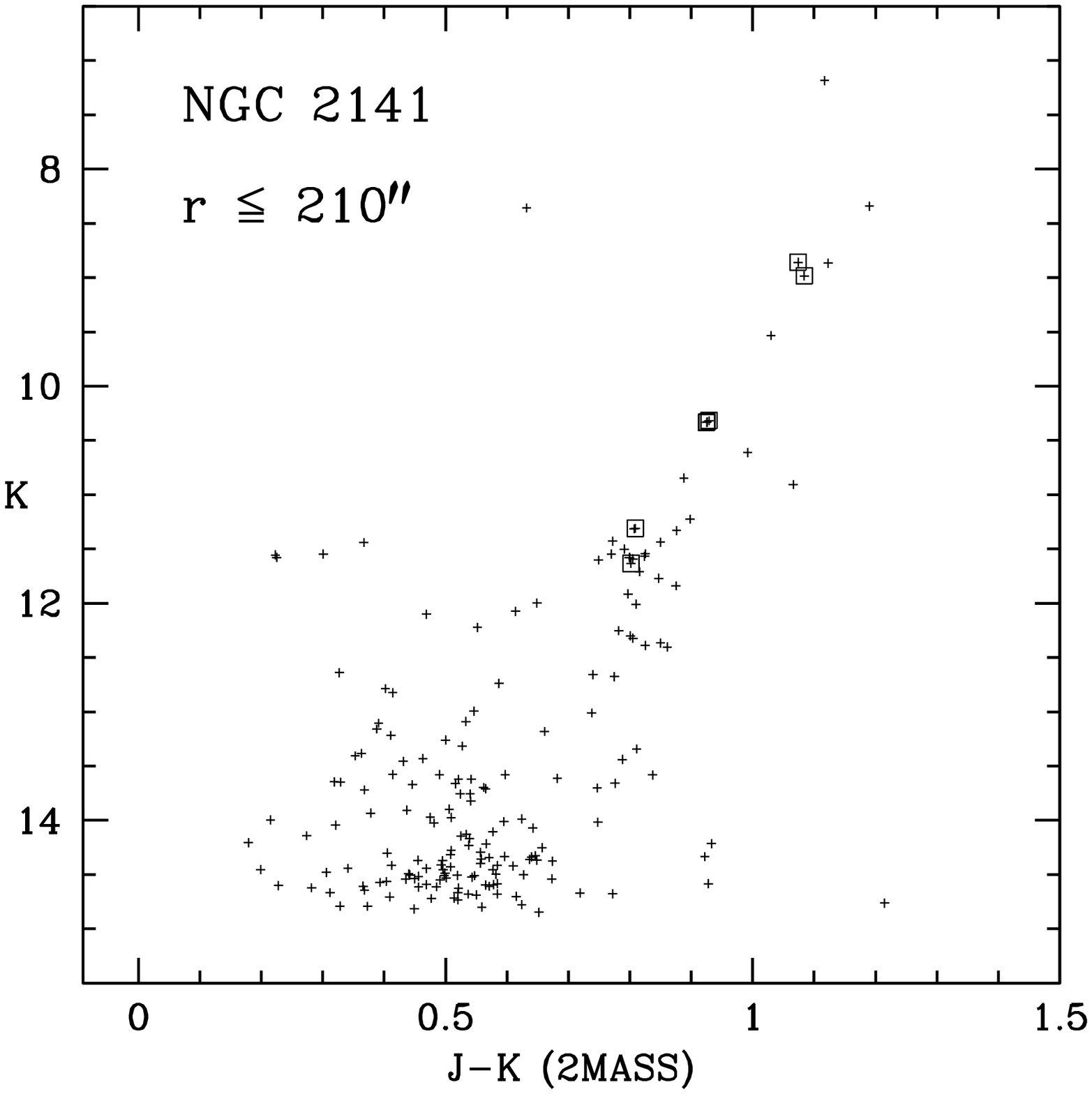}
\caption{We distinguish by enclosed squares the stars observed, using the
$JK$ data from the 2MASS survey. \label{fig:ngc2141jk}}
\end{figure}

\begin{figure}
\plotone{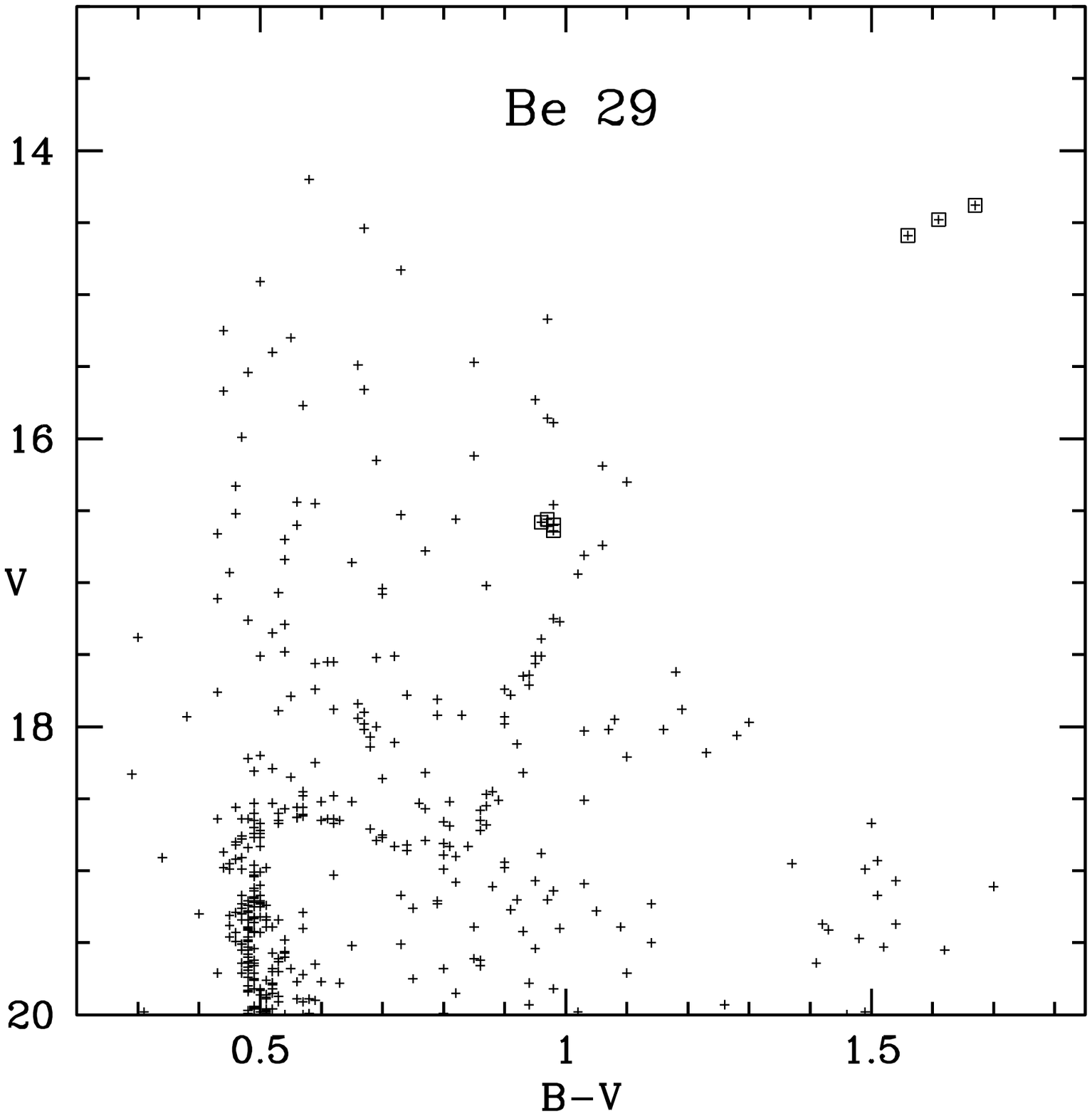}
\caption{We distinguish by enclosed squares the stars observed, using the
$BV$ data from \citet{kaluzny94}. \label{fig:be29bv}}
\end{figure}

\begin{figure}
\plotone{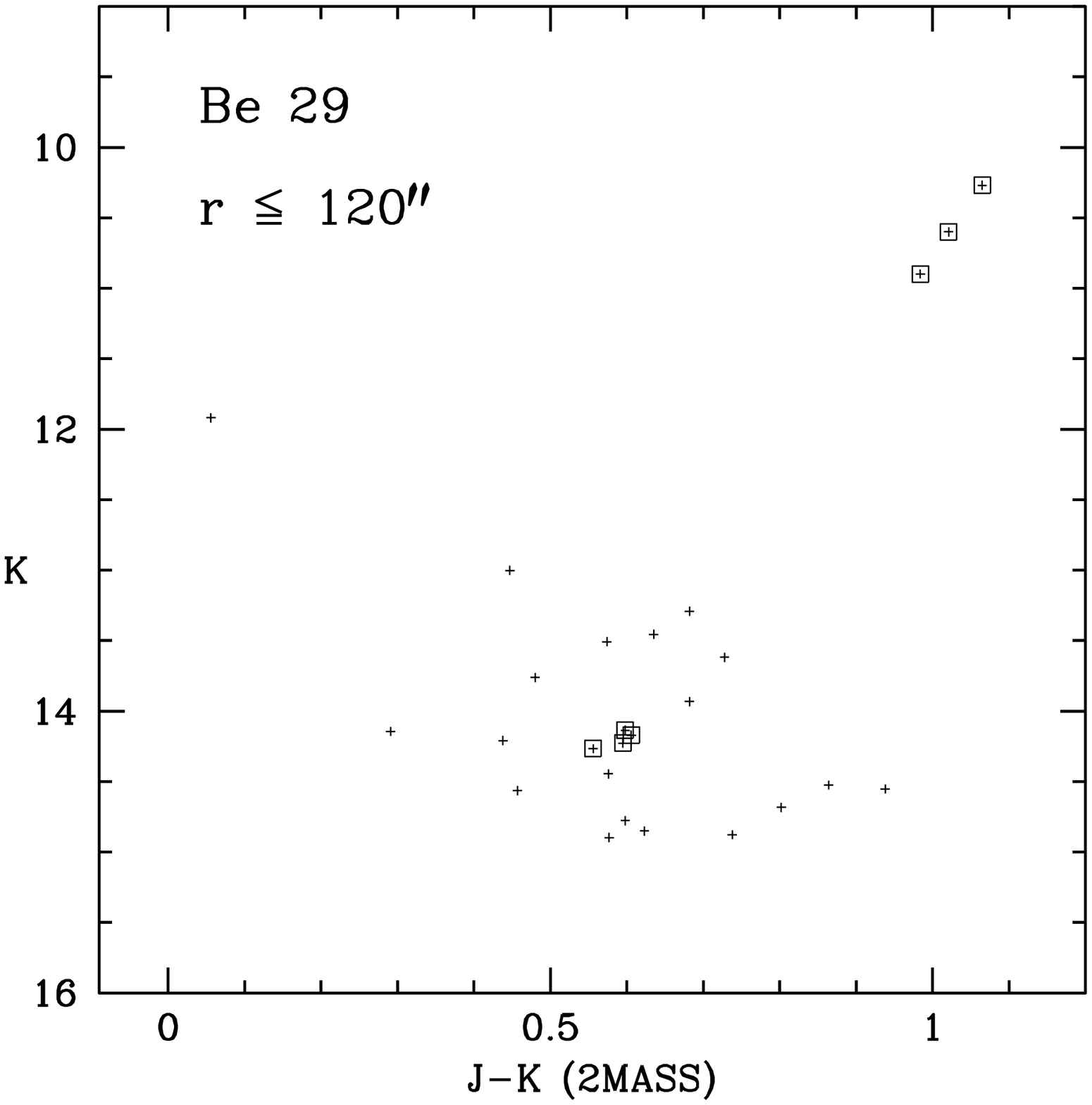}
\caption{We distinguish by enclosed squares the stars observed, using the
$JK$ data from the 2MASS survey. \label{fig:be29jk}}
\end{figure}

\begin{figure}
\plotone{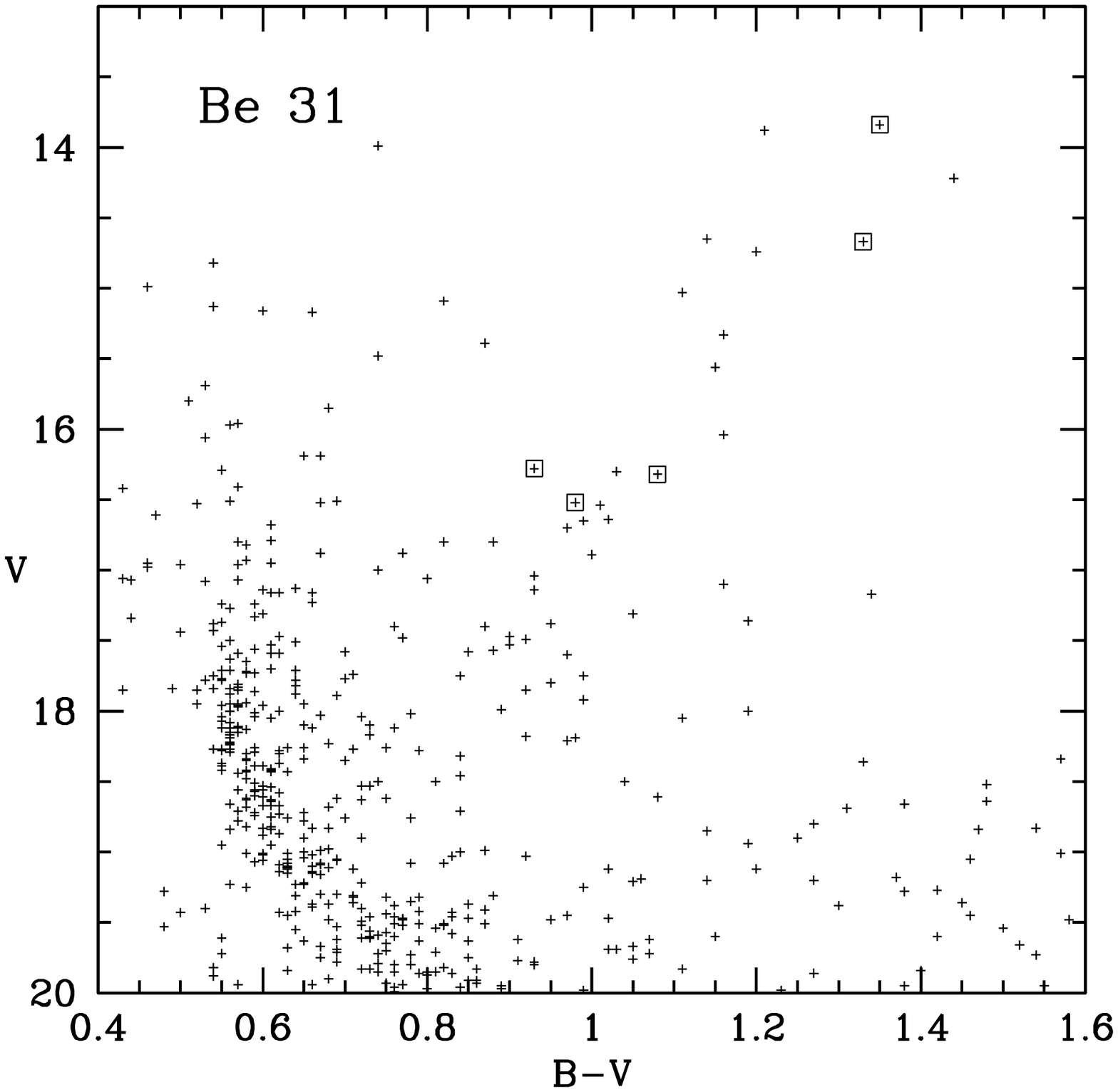}
\caption{We distinguish by enclosed squares the stars observed, using the
$BV$ data from \citet{guetter93}. \label{fig:be31bv}}
\end{figure}

\begin{figure}
\plotone{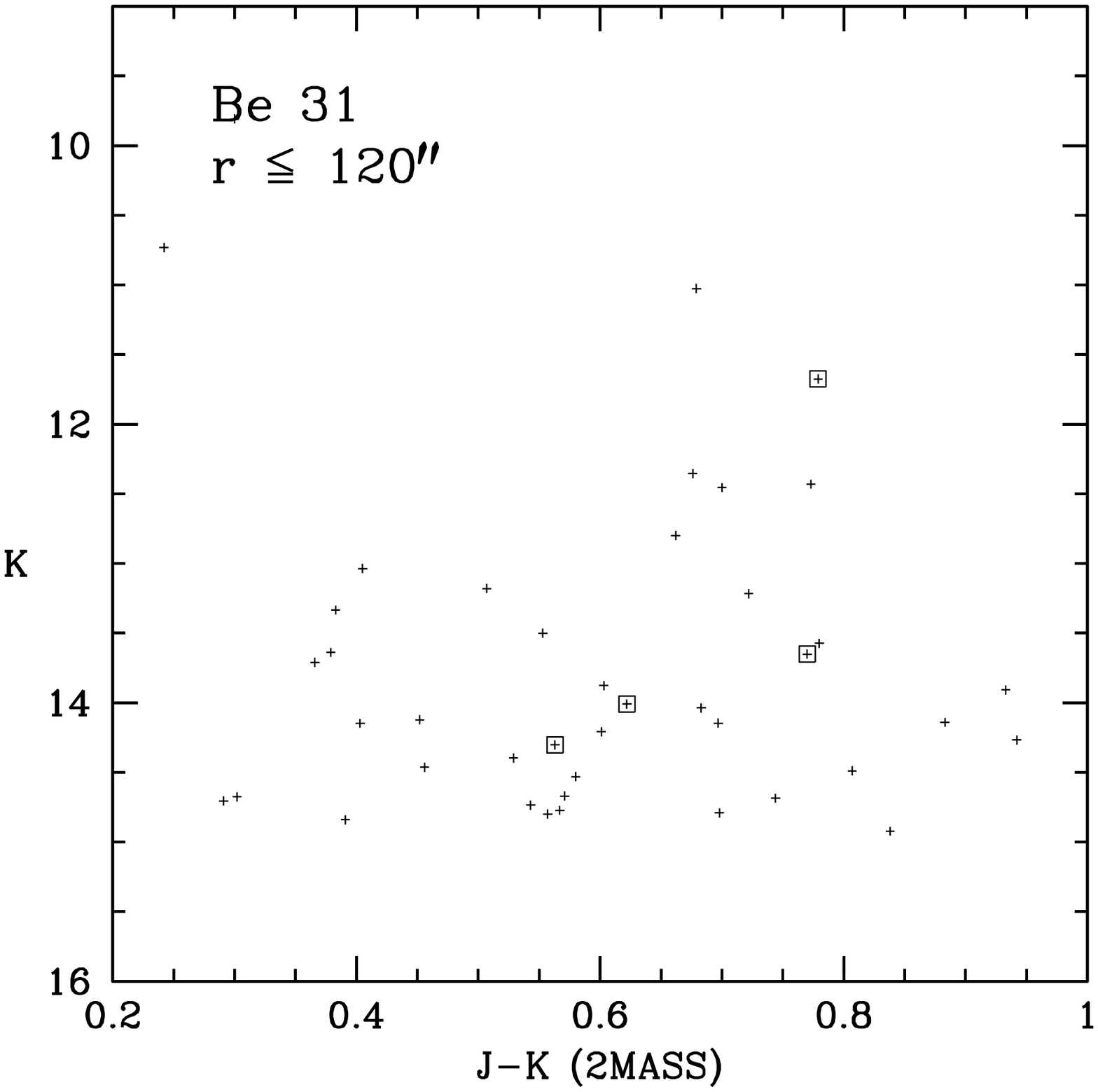}
\caption{We distinguish by enclosed squares the stars observed, using the
$JK$ data from the 2MASS survey. \label{fig:be31jk}}
\end{figure}

\begin{figure}
\plotone{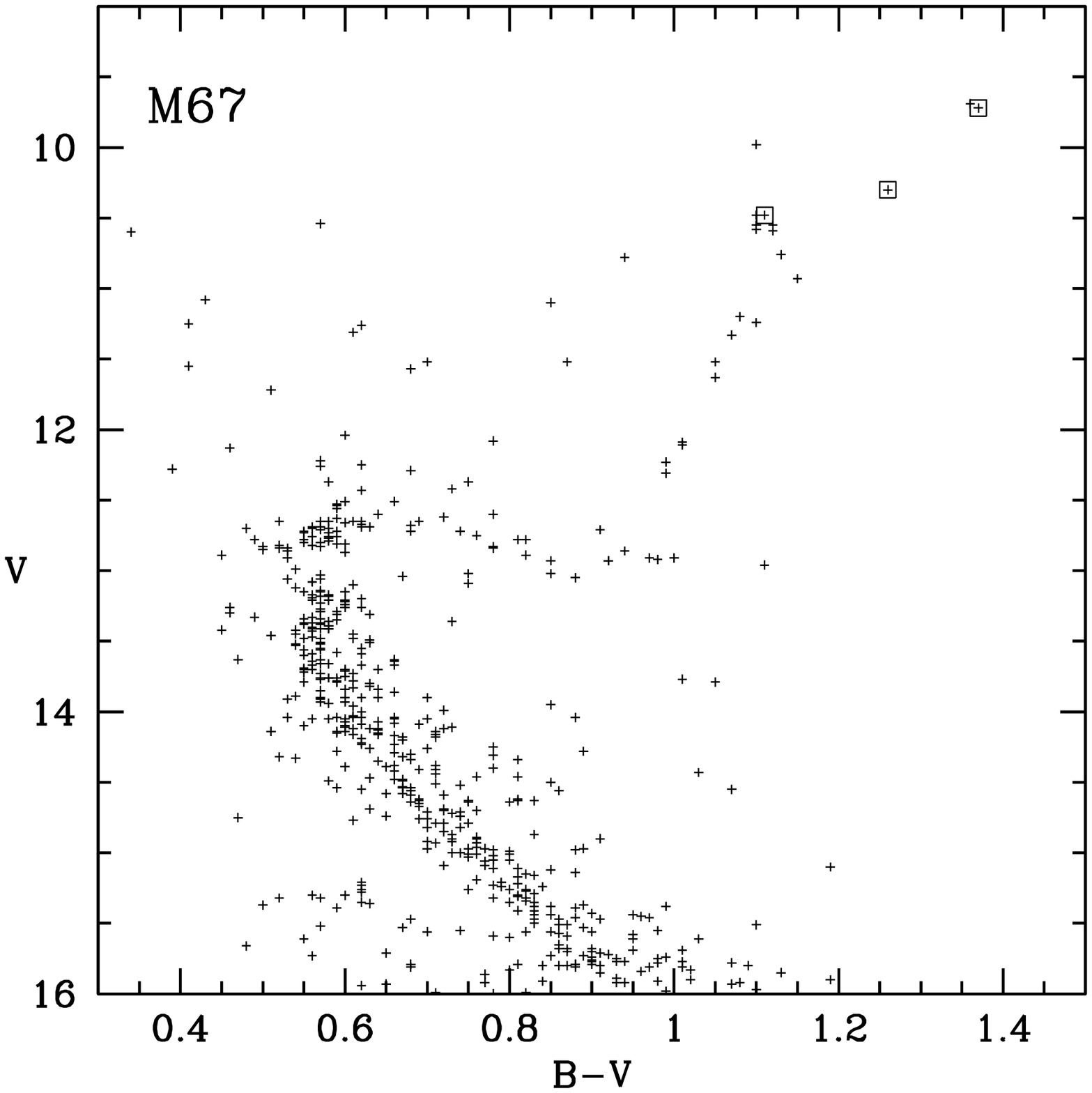}
\caption{We distinguish by enclosed squares the stars observed, using the
$BV$ data from \citet*{montgomery93}. \label{fig:m67bv}}
\end{figure}

\begin{figure}
\plotone{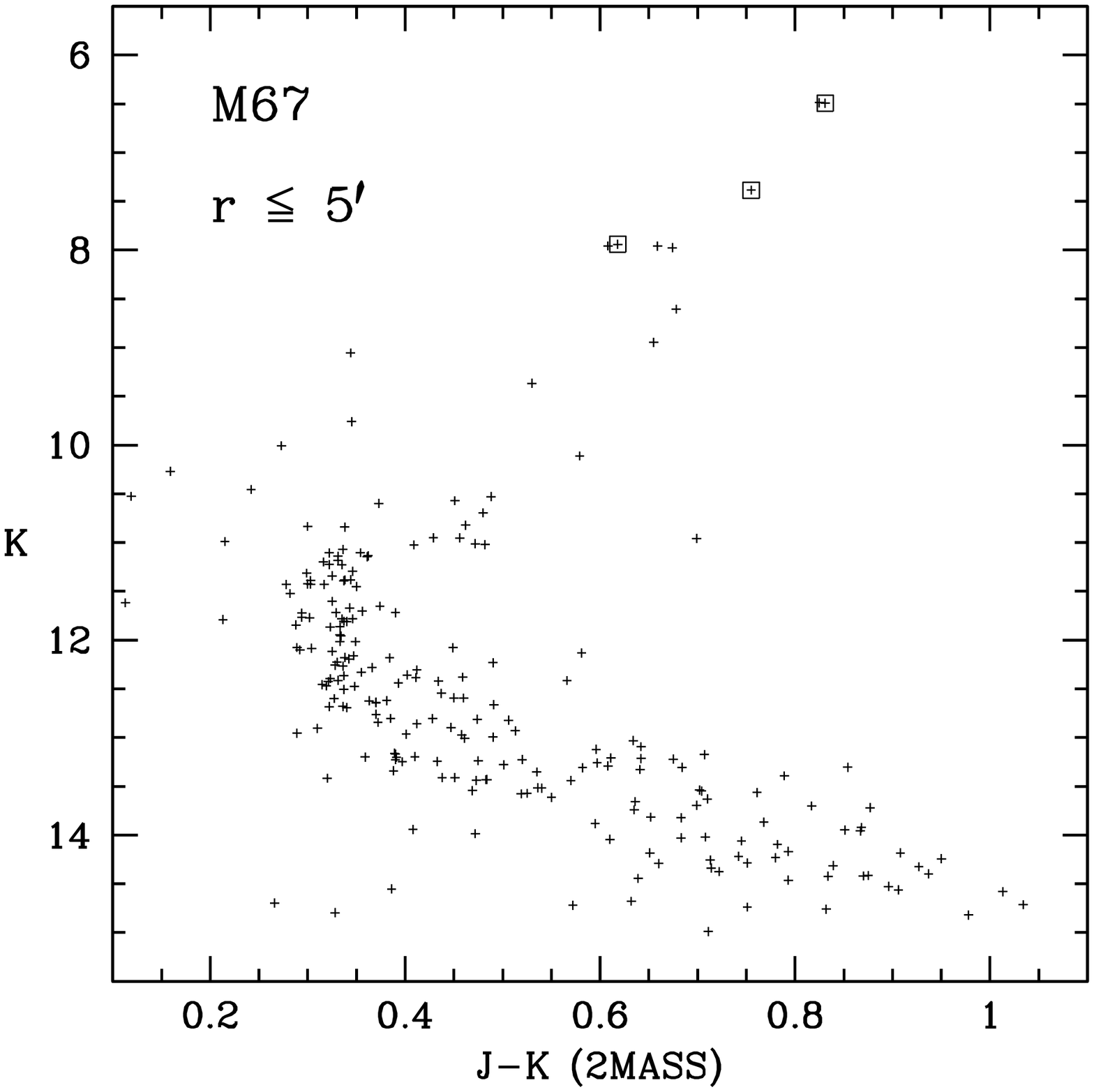}
\caption{We distinguish by enclosed squares the stars observed, using the
$JK$ data from the 2MASS survey. \label{fig:m67jk}}
\end{figure}

Based on Hipparcos parallax data, \citet{alves00} 
argued that the mean luminosity
of the red clump stars, $M_{K}(RC)$, is not sensitive to metallicity,
at least for [Fe/H] $> -1.0$. Specifically, $<M_{K}(RC)> \approx -1.61$ mag.
This is an especially powerful tool given the apparent lack of
sensitivity of $<M_{K}(RC)>$ to both metallicity and
extinction.
Using the data used to construct Figures~\ref{fig:be20vi}-\ref{fig:be31jk},
we apply these tools to our program clusters, as well as M67
(used to calibrate Equation~\ref{eq:jk0}). A few comments
are in order, however.

Be~20 (Figure~\ref{fig:be20vi}) does not show a clear red clump, but
Figure~\ref{fig:be20jk} suggests that three stars with $K \approx 13.2$
and $J-K \approx 0.7$ may define such a locus. We provisionally
identify those three stars as constituting the red clump
for Be~20, but we flag the derived reddening and distance
moduli as uncertain via colons in Table~\ref{tab:redclump}.

\begin{figure}
\plotone{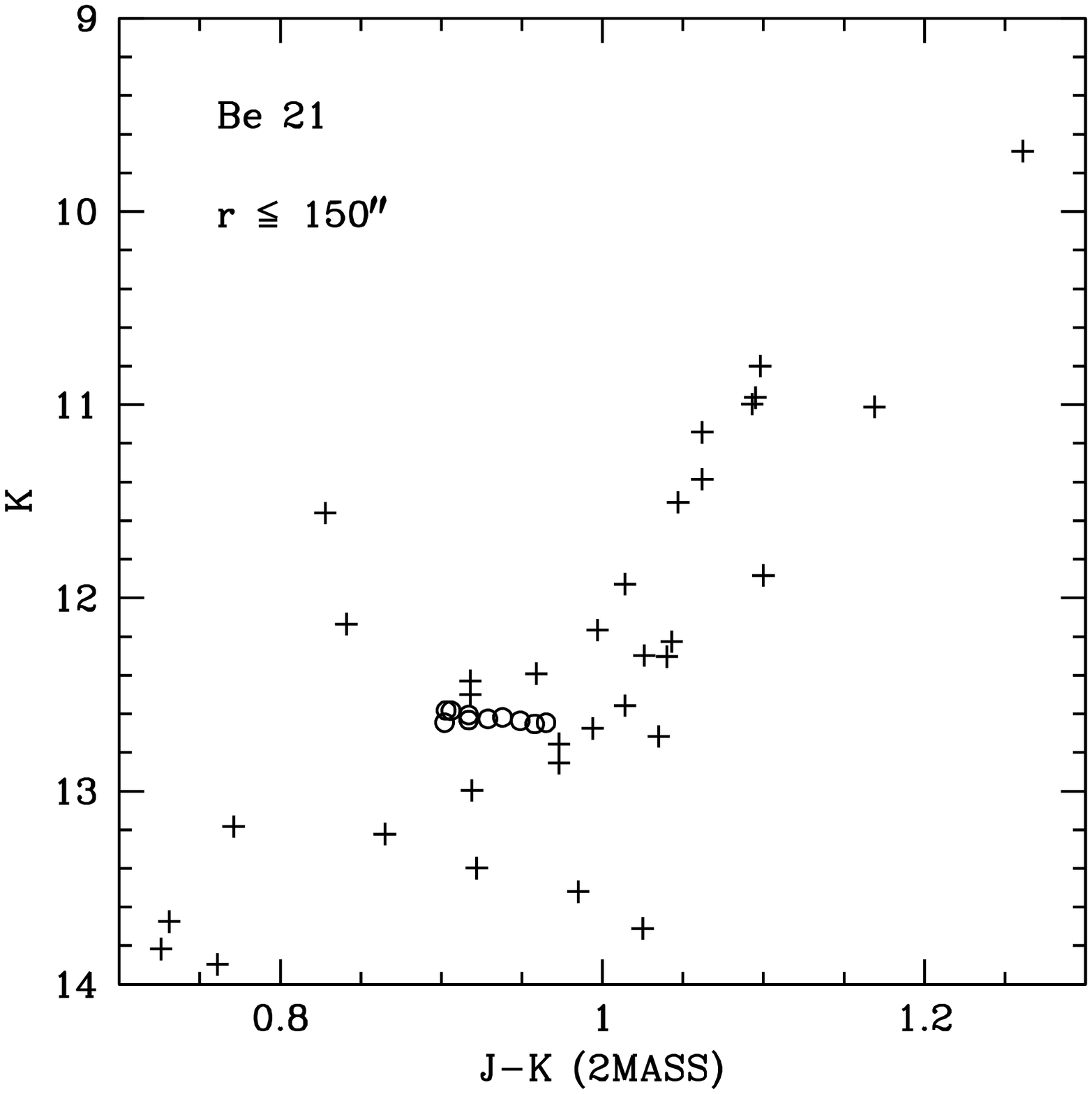}
\caption{This is an expanded view of Figure~\ref{fig:be21jk},
with stars we believe to represent the red clump of Be~21
identified as open circles. \label{fig:be21rhb}}
\end{figure}

Be~21 suffers from large and differential reddening, making
the infrared approach particularly valuable. We show in
Figure~\ref{fig:be21rhb} an expanded view of Figure~\ref{fig:be21jk},
with the red clump stars identified as open circles. It is
interesting how well the red clump is identified in the infrared
compared to the optical regime (Figure~\ref{fig:be21bv}), which
we believe illustrates the relative effects of differential
reddening in the optical compared to the infrared.

\begin{figure}
\plotone{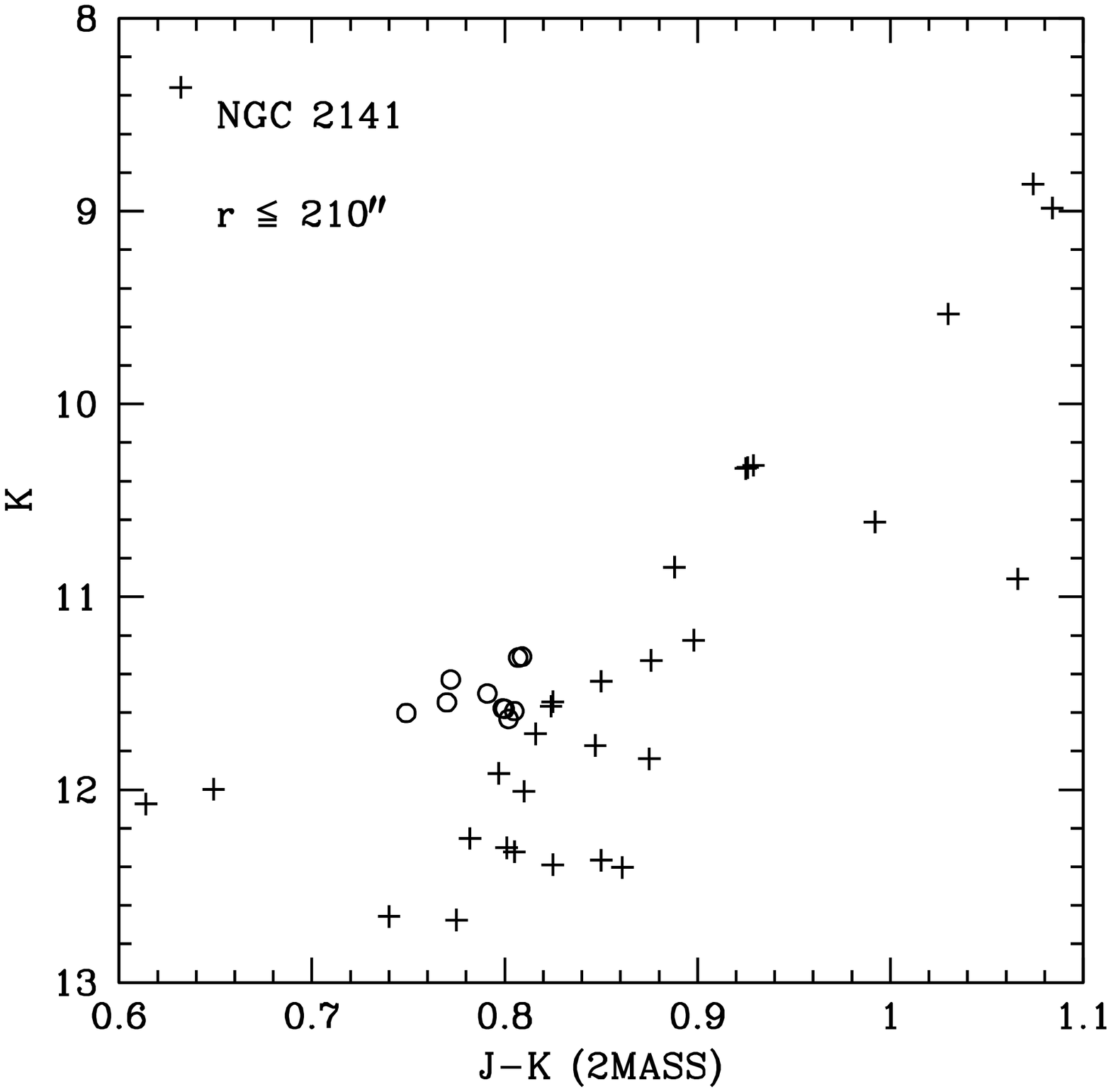}
\caption{This is an expanded view of Figure~\ref{fig:ngc2141jk},
with stars we believe to represent the red clump of NGC~2141
identified as open circles. \label{fig:ngc2141rhb}}
\end{figure}

While not as heavily extinguished as Be~21, NGC~2141 nonetheless
suffers the same effect. Again, a comparison of Figures~\ref{fig:ngc2141vi}
and \ref{fig:ngc2141jk} shows the tighter grouping in magnitude
of the red clump stars, which are identified more clearly in
Figure~\ref{fig:ngc2141rhb}. 

\begin{figure}
\plotone{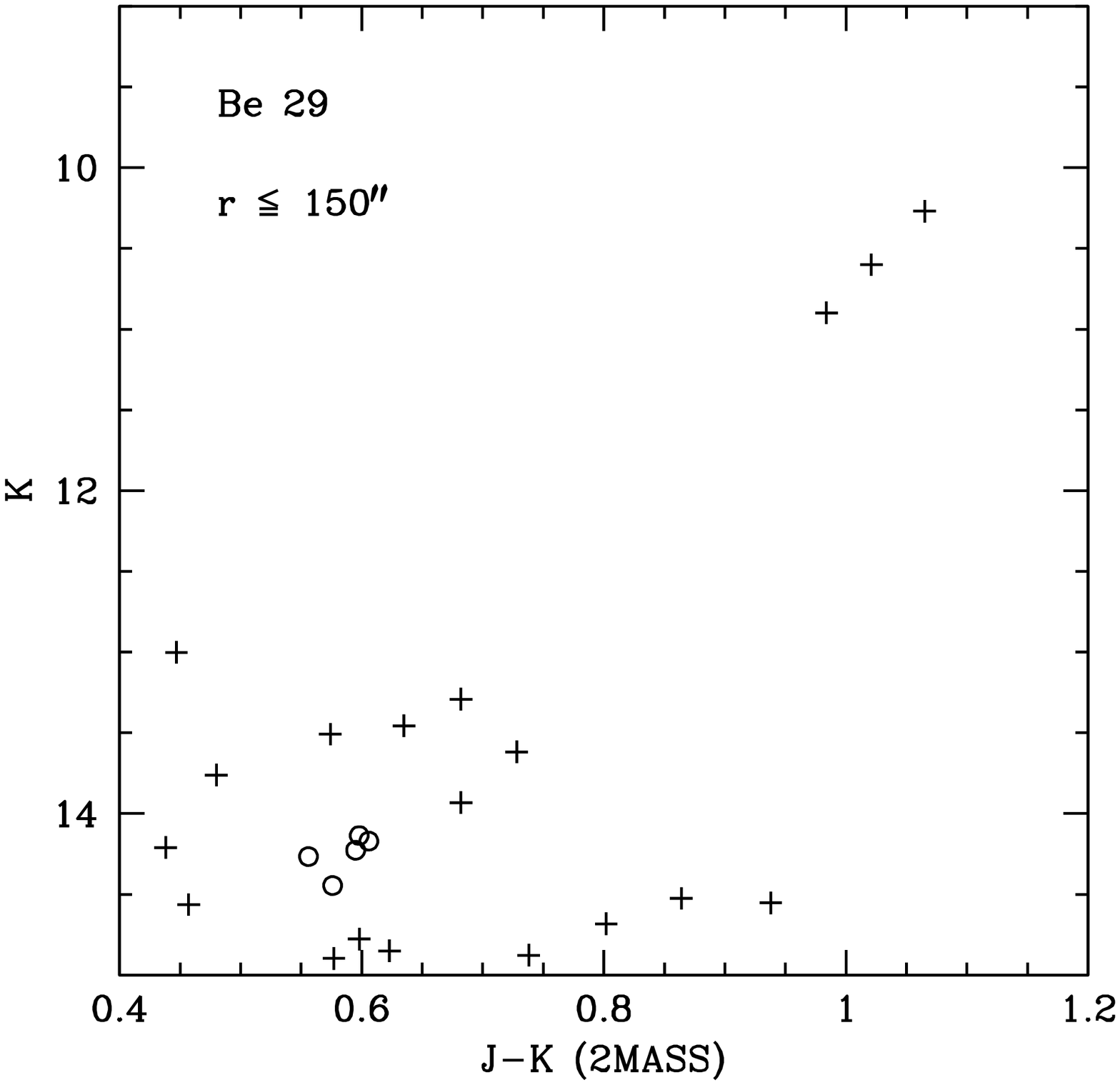}
\caption{This is an expanded view of Figure~\ref{fig:be29jk},
with stars we believe to represent the red clump of Be~29
identified as open circles. \label{fig:be29rhb}}
\end{figure}

For Be~29, we identify five red clump stars, and, happily, in this
case we have obtained radial velocities for four of them
(see Figure~\ref{fig:be29rhb}). All four stars for which
we have measured radial velocities are cluster members, as expected
from their locations in the color-magnitude diagrams
(see Table~\ref{tab:spectra}). 
\citet{tosi04} recently studied Be~29,
the most distant open cluster, and derived a Galactocentric
distance of 21.4 to 22.6 kpc, in good agreement with our
distance estimate. 

Unfortunately, we are unable to identify red clump stars unambiguously
in Figures~\ref{fig:be31bv} and \ref{fig:be31jk}. 
We provide distance and reddening estimates based on the work of 
\citet{guetter93}. 

As a check, it is of interest to derive the reddening and distance
to M67. This is not a test, however, because M67 is one of the
calibrating clusters for Equation~\ref{eq:jk0}. \citet{janes94}
quote E($B-V$)~=~0.05 mag and d~=~0.77 kpc. Agreement appears
to be very good indeed.

Table~\ref{tab:redclump} (columns 3 and 4) 
summarizes our {\em preliminary} values
of the reddenings and distances for our program clusters. These
are not {\em final} values because it is our primary purpose
here to derive cluster metallicities using high-resolution, high-S/N
echelle spectroscopy, and any changes to [Fe/H] will alter the reddening
estimates. Distance estimates will not be strongly affected, for reasons
discussed above.

\section{ELEMENTAL ABUNDANCES}

\subsection{Photometric Estimation of Atmospheric Parameters}

With estimates for reddening and distance, we are now able
to make preliminary determinations of the temperatures and gravities
for the stars for which we have obtained high-S/N spectra.

We adopt the color-temperature relations for giants from
\citet*{alonso99}, and rely exclusively
on $V-K$. We transform the 2MASS photometry from Table~\ref{tab:photometry}
to the ``TCS" system using transformation equations from
\citet{carpenter01} and \citet*{alonso94},
although, to first order, ($V-K_{\rm 2MASS}$)~=~($V-K_{\rm TCS}$).
\citet{alonso99} also include relations between colors and
bolometric corrections, and, again, we rely on the $V-K$ photometry.
With the distance, temperature, and bolometric corrections, we are
able to calculate log $g$ values if we adopt a mass for the red giants.
Since the clusters have ages relatively like those of the Sun
and are more metal-poor, we adopt 1.0~\msun\ for all the clusters.
Because the stellar mass enters only as the logarithm, our derived
log $g$ values are not especially sensitive to this choice.

We present these photometric estimates in columns 2 and 3 of
Table~\ref{tab:parameters}. They are useful starting points
for the more refined spectroscopic estimates. In the case
of Be~31, we have adopted the reddening and distance estimates
of \citet{guetter93}.

\subsection{Analyses}

The next step in our analysis is to check the photometric estimates for
the effective temperature (\teff) and surface gravity (log $g$) and possibly
derive new values based on the spectroscopy. 
Equivalent widths (EW) were measured for a selection of Fe I and Fe II lines using
routines in IRAF. The primary source of $gf$ values for the Fe I lines  
are the accurate laboratory measurements performed by the Oxford group 
(e.g., \citealt{blackwell79feb,blackwell79fea,blackwell80fea,
blackwell86fea,blackwell95fea} and references therein). 
To supplement our list of Fe I lines, a subset were drawn from 
\citet{thoren00} and \citet*{paulson03}. 
For Fe II, we used the $gf$ values from \citet{biemont91}.  
Model atmospheres were computed with the ATLAS9 program \citep{atlas9}. Using
the current version of the local thermodynamic equilibrium (LTE) stellar
line analysis program MOOG \citep{moog}, we computed abundances for each
line based on the measured EW. We set \teff~by forcing the abundances from Fe I
lines to be independent of the lower excitation potential, i.e., excitation
equilibrium. Next the microturbulence ($\xi_{\rm t}$) 
was determined from the condition that Fe I lines show no trend versus EW.
Finally, we adjusted log $g$ until the abundances from Fe I and Fe II lines
agreed, i.e., ionization equilibrium. 
Taking the photometric estimates for \teff~and log $g$ as our initial model,
we iterated 
until a self consistent set of parameters was obtained. Our spectroscopic 
stellar parameters are presented in Table \ref{tab:parameters} and are 
close to the photometric estimates. In the subsequent abundance analysis,
we adopt the spectroscopic stellar parameters. 

Next we considered lines of O, Na, Mg, Al, Si, Ca, Ti, Mn, Co, 
Ni, Rb, Zr, Ba, La, and Eu measuring EWs when possible. 
We present our line lists in Table \ref{tab:line}.
The $gf$ values for O were taken from \citet{asplund04}. For Ca
and Ti, the $gf$ values were taken from the Oxford group 
\citep{smith81ca,blackwell82tia,blackwell83tia,blackwell86tia}. 
An inverted solar analysis was used to determine the Si $gf$ values
assuming log~$\epsilon$(Si)~=~7.55.  
For Na, Mg, Al, Ni, and Zr, we relied upon the $gf$ values from \citet{ramirez02}. 
For Mn, the $gf$ values were taken from \citet{mn} and include the effects of
hyperfine splitting. The Co and Ba $gf$ values were taken from 
\citet{prochaska00} and include hyperfine splitting. The $gf$ values for Rb
were taken from \citet{tomkin99} with hyperfine splitting and isotopic
shifts taken from \citet{lambert76} where we assumed a solar isotope ratio. 
For La and Eu, the $gf$ values were taken from \citet*{la} and \citet{eu} 
respectively and we account for hyperfine and isotopic splitting assuming
a solar isotopic mix for Eu. 
Hyperfine structure must be taken into account
for certain odd-Z elements since magnetic interactions between the electrons and
the nucleus split the absorption lines into multiple components. This splitting
inhibits saturation and neglect of hyperfine structure can lead to 
abundance overestimates. 

According to \citet{burris00}, Zr, Ba, and La are synthesized primarily through 
the $s$-process where 81\%, 85\%, and 75\% of their solar abundance is 
attributable to the $s$-process with the remaining 19\%, 15\%, and 25\% being
produced by the $r$-process respectively. The opposite case holds for
Eu where 97\% of the solar abundance arises from the $r$-process with
only 3\% attributable to the $s$-process. For Rb, the solar 
abundance is due to 50\% $s$-process and 50\% $r$-process, though as we
discuss later, Rb offers insight into the neutron density
at the site of the $s$-process (and therefore mass of the AGB star that
synthesized the $s$-process elements). 

\begin{figure}
\epsscale{0.8}
\plotone{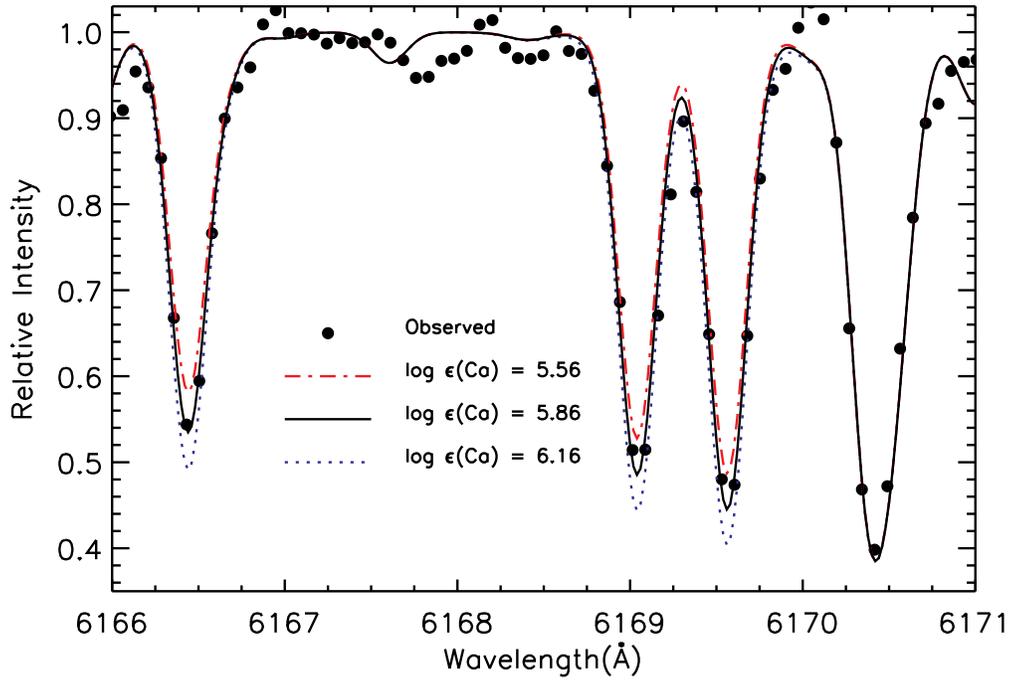}
\caption{Synthetic spectra with different Ca abundances for 
Be 29 988. \label{fig:ca}}
\end{figure}

\begin{figure}
\epsscale{0.8}
\plotone{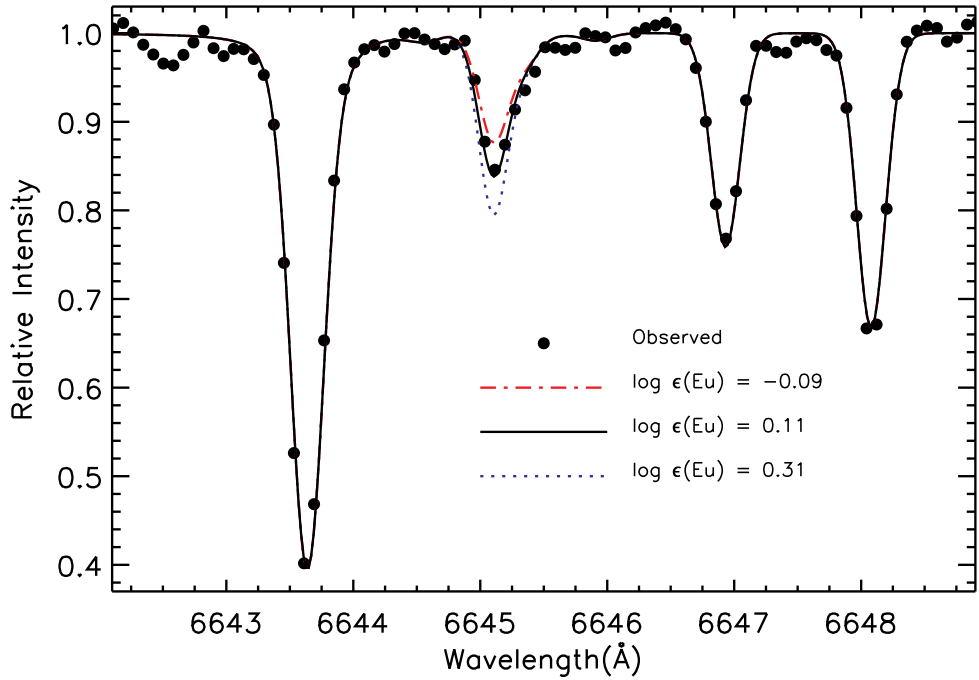}
\caption{Synthetic spectra with different Eu abundances for 
Be 29 673. \label{fig:eu}}
\end{figure}

Unlike Fe, the species 
studied in this work (O-Eu) present only a handful of lines in our 
spectra. To ensure that the derived abundances were not affected by blends, we
generated synthetic spectra using MOOG to determine the abundance for every line
in every star. An 8\AA~window centered on the feature of interest was synthesized 
(contributing lines were taken from \citealt{kurucz95}) and the abundances were
adjusted until the observed spectrum was best fit. Guided by the syntheses,
we occasionally needed to readjust the continuum by 1-2\% to provide an optimal
fit between observed and synthetic spectra. While spectrum
synthesis is considerably more time consuming than EW analysis, the 
benefit is that this technique allows us to derive abundances 
for blended lines from which a traditional EW analysis is not feasible,
e.g., Rb abundances can only be attained from spectrum synthesis due to blending
from a Si line. Since the cool giants studied in this paper have 
crowded spectra, spectrum synthesis increased the
number of lines we could analyze as well as our confidence in the derived 
abundances. In Figures \ref{fig:ca} and \ref{fig:eu}, we present examples of 
our synthetic spectra fits to measure elemental abundances in two stars. In 
general, for the lines from which we derived abundances from EWs, the abundances
from synthetic spectra were in good agreement. 
We also synthesized a handful of Fe lines and found no major 
differences between abundances derived from EWs and from spectrum synthesis.

To check our line lists and analysis techniques, elemental abundances were 
derived for the Sun using our adopted line lists and EWs measured in the 
\citet{kurucz84} Solar atlas. 
A model atmosphere was generated using ATLAS9 with the
following parameters 
\teff~=~5770 (K) and log $g$~=~4.44 (cgs) where we adopted $\xi_{\rm t}$~=~0.85 (\kms)
to remove trends between Fe I abundances and EW. 
In Table \ref{tab:solar} we present our solar abundances. There is 
a good agreement 
between our derived abundances and the \citet{grevesse98} values. 
{\it For the program stars, the elemental abundances are referenced to the 
``solar'' values found in our study.} 

As an additional check of our line lists and analysis techniques, we 
analyzed Arcturus since it is a very well studied 
red giant at a similar evolutionary stage
as the open cluster giants. We measured EWs from the \citet{hinkle00} 
Arcturus atlas and computed a model atmosphere with \teff~=~4300 (K) and 
log $g$~=~1.50 (cgs) using ATLAS9 adopting a microturbulence 
$\xi_{\rm t}$~=~1.56 (\kms) to remove trends between Fe I abundances and EW. In
Table \ref{tab:arc} we compare our abundances with those derived by
\citet*{peterson93} and \citet{carraro04}.  We find no 
major differences between the abundances derived from the different
studies. We also compared our derived abundances for Arcturus, 
log~$\epsilon$(X), with \citet{smith00} and again find a good agreement.

Having shown that our analysis techniques reproduce the abundance 
distributions in two well studied stars, the Sun and Arcturus, we present
in Table \ref{tab:abund} our measured elemental abundance ratios for
the open cluster giants. Typical
internal errors for our spectroscopic model parameters are \teff$\pm$100 K, 
log $g\pm$0.3 dex, and $\xi_{\rm t}\pm$0.2 \kms. In Table \ref{tab:err} 
we present the abundance dependences on the model parameters.
The radial velocities of the stars were sufficiently large that the night
sky emission did not affect the stellar O absorption lines near 6300\AA. Due 
to the formation of the CO molecule, the measured O abundances depend on the 
assumed C abundance. Since we do not 
know the carbon abundances, we assumed a solar ratio [C/Fe]~=~0.0. If we assumed
[C/Fe]~=~+0.3, our derived O abundances [O/Fe] increase by +0.06 dex.

\citet{tautvaisiene00} have conducted a thorough abundance analysis of
six clump stars and three red giants in the nearby open cluster M67. Their
analysis of 25 elements included those observed in this study. In Table
\ref{tab:m67}, we compare the mean abundances and find a good agreement
between the different works further reinforcing our confidence in the adopted
line lists and analysis procedures. 

\citet{carraro04} measured abundances for two stars in Be 29. Their mean 
iron abundance [Fe/H]~=~$-$0.44 is in fair agreement with our value 
[Fe/H]~=~$-$0.54. For the alpha elements, their mean abundance [$\alpha$/Fe]~=~0.17
is also in good agreement with our value [$\alpha$/Fe]~=~0.22. 

\subsection{Be 21 T406, a rapidly rotating Li-rich star}

\begin{figure}
\epsscale{0.8}
\plotone{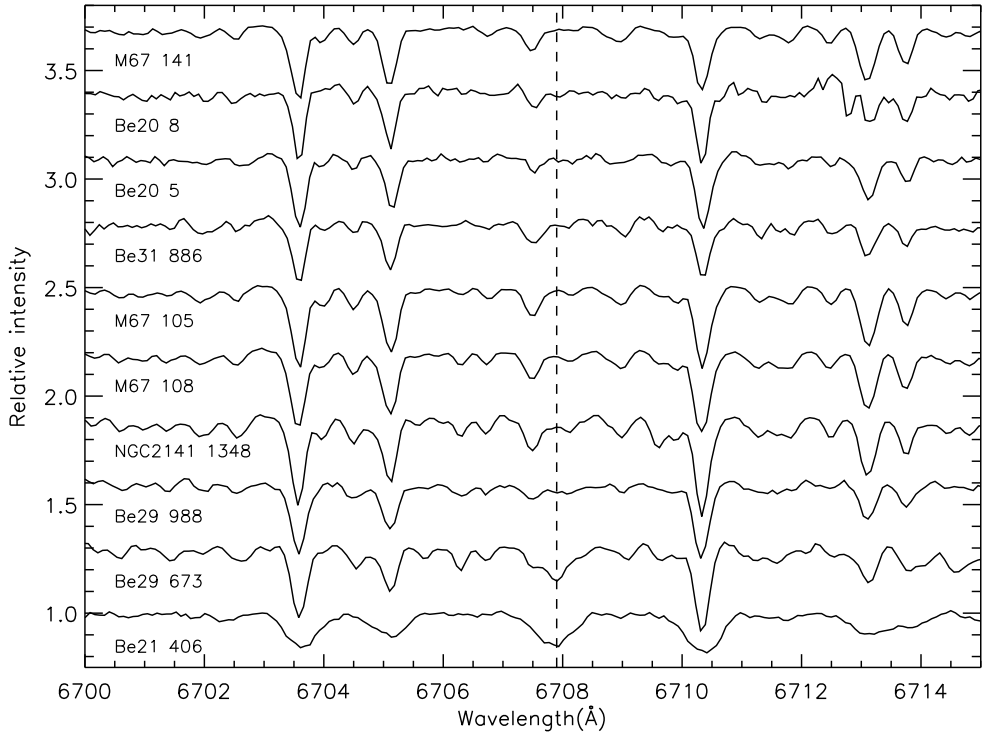}
\caption{Spectra centered on the Li line near 6708\AA. The spectra
are ordered by \teff~starting with the warmest star M67 141 and
continuing to the coolest star Be 29 673. 
Be 21 T406 is placed at the bottom and the position
of the Li feature is identified by the vertical line. Note the large
rotational broadening and the strength of the Li feature
in Be 21 T406.\label{fig:li}}
\end{figure}

During the abundance analysis, we noticed that Be 21 T406 exhibited 
an unusual spectrum. In Figure \ref{fig:li} we show 
the observed spectra for all stars centered on the Li feature near 6708\AA.
Be 21 T406 clearly exhibits a larger rotational broadening than the other
giants. (The additional line broadening presumably due to rotation,
compared to the other stars, 
is seen throughout the spectrum of Be 21 T406, not just near the Li line.) 
Be 21 T406 also displays an absorption feature whose wavelength
is coincident with Li 6708\AA. Due to the large rotational broadening, we
were unable to identify clean Fe lines from which we would have 
determined spectroscopic 
stellar parameters via the techniques described earlier in the paper. Given
that the photometric and spectroscopic parameters are in fair agreement
for the rest of the sample, we
therefore adopted the photometric parameters for this star and
synthesized the Li region to estimate the abundance. We assumed a 
limb darkening coefficient of 0.75 \citep{gray92} and found that a 
rotational broadening of 16 \kms was necessary to best fit the line profile. 
The derived Li abundance is log~$\epsilon$(Li)~=~0.2 using the same line list
employed by \citet{reddy02} but assuming a contribution solely from 
$^{7}$Li, i.e., neglecting $^{6}$Li. Since Li is easily 
destroyed in stars, we do not expect to observe Li in red giants whose large
convective envelopes cycle their atmospheres down to hot layers. 
With the possible exception of Be 29 673  
there is no absorption feature near 6708\AA~in
any of the stars. While it is quite unusual to be able to measure Li in 
such a cool star with a large rotational broadening, we would like to
compare the Li abundance with other giants to see if the Li detection
is significant. \citet{ramirez02} studied
the globular cluster M71 and measured Li (or upper limits) for several
cool giants. M71 has a metallicity comparable to the photometric estimate
for Be 21. At \teff$\simeq$4200K, \citeauthor{ramirez02} derived upper
limits of log~$\epsilon$(Li)~=~$-$0.7. Compared to giants in M71 at a similar
\teff, Be 21 T406 has a Li abundance enhanced by a factor of 10. 

One easy conclusion to draw is that the lithium enhancement is related to
the unusually high rotational velocity. However, there is contradictory
evidence from \citet{hill99}, who identified a giant Be 21 T33 with 
log~$\epsilon$(Li)~=~3.0, but no sign of rotational broadening. 
Be 21 T33, had a similar
(presumably low/normal) 
rotational velocity compared with the other giants they observed. 
One possibility considered by \citet{carney03} that can explain rapidly
rotating Li rich giants is that a planetary companion is engulfed as the
convective envelope expands while the star ascends the giant branch. 
In this scenario, \citet{carney03} suggested 
that the orbital angular momentum of the companion is converted into
rotational angular momentum in the primary accompanied by an increase
in the Li abundance. Self enrichment is a plausible alternate
explanation for a high Li abundance in evolved stars \citep{charbonnel00}. 
Since the rotational velocities appear
distinct between Be 21 T406 and Be 21 T33, the 
Li enhancement likely results from different mechanisms. 

A subordinate Li line exists near 6104\AA, however, we do not detect 
this line. Figure 3 in \citet{hill99} shows that the 6104\AA~Li line is weak even 
when log~$\epsilon$(Li)~=~3.0. Given the lower Li abundance and increased 
rotational velocity in Be 21 T406, the non-detection of the 6104\AA~Li line
is unsurprising. 

\section{DISCUSSION}

\subsection{Abundance trends versus age}

First we reinvestigate whether the open clusters exhibit any abundance trends
as a function of age. As mentioned earlier, we adopt the age
determinations by \citet{salaris04}. Note that the outer disk open clusters
are Be 20, NGC 2141, Be 29, and Be 31. M67 is not an outer disk open cluster
but instead is representative of the solar neighborhood. 
The absence of a strong correlation 
between age and metallicity has been well established for open clusters 
\citep{friel95,friel02,friel03,chen03,salaris04}. 
We emphasize that systematic differences 
between the various studies can introduce or
mask subtle abundance trends. 
Studies of a large number of clusters
analyzed in a consistent and homogeneous manner are essential for
identifying subtle abundance trends versus physical properties.
It is also important to note that the term ``metallicity'' is vague,
even though it is usually written as [Fe/H]. Low-resolution spectral
indices, or photometric colors, are affected by elements other than
iron, and which may not scale with iron. Further, many photometric
colors are affected by uncertainties in reddening and, of course, in the 
selection of stars by which the indices or colors are calibrated against stars
whose [Fe/H] values have been determined by high-resolution spectroscopy,
which is not vulnerable to reddening or variations in [X/Fe].

Figure 5 in \citet{friel02} demonstrates that there is no clear trend between 
[Fe/H] and age. If an age-metallicity relation really does exist among the 
open clusters, then systematic differences between the various studies 
must be considerable or the open clusters belong to different stellar 
populations as we have noted previously. 
\citet{bdp03} found a decrease in metallicity with increasing 
age. An age-metallicity relation also appears to exist among thick disk stars 
\citep*{bensby04}. The lack of an age-metallicity relation in the studies by
\citet{bdp93} and \citet{nordstrom04} may have been caused by the merger 
of stellar populations (thin disk, thick disk, and halo) with
quite distinct chemical enrichment histories as noted earlier. It is not clear 
that this applies to the open clusters whose ages are generally younger than 
the thick disk and halo. While the lack of an age-metallicity 
relation in open clusters may be
due to observational errors, it could also be due to an additional source of
chemical enrichment. This may be especially important because the open clusters
span a much wider range of Galactocentric distances than do the local thin and
thick disk stars, and hence the open clusters may probe into realms where other 
merger events may have occurred.

\begin{figure}
\epsscale{0.7}
\plotone{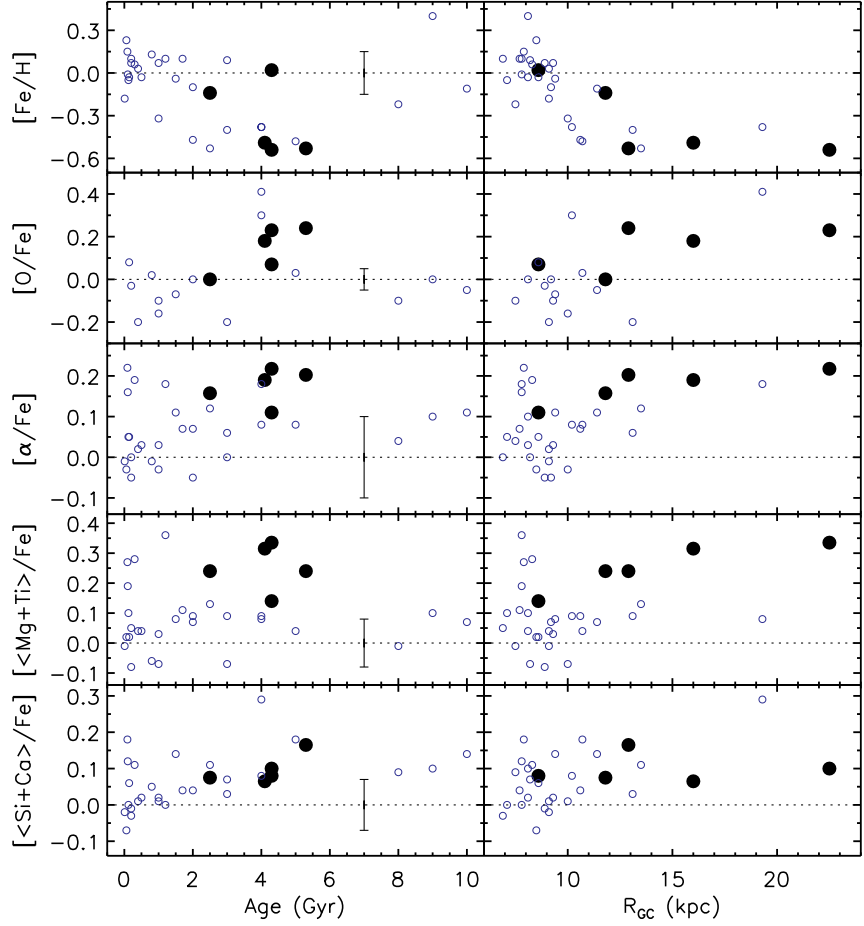}
\caption{Abundance ratios versus age (left) and versus Galactocentric 
distance (right). The filled black circles represent the cluster abundances 
derived in this study while the open blue circles  
represent various abundance determinations for clusters taken
from the literature. 
In the middle panel, $\alpha$ is the average of Mg, Si, Ca, and Ti. 
A representative error bar for the abundance
ratios derived in this study is shown.\label{fig:av_abund1}}
\end{figure}

\begin{figure}
\epsscale{0.7}
\plotone{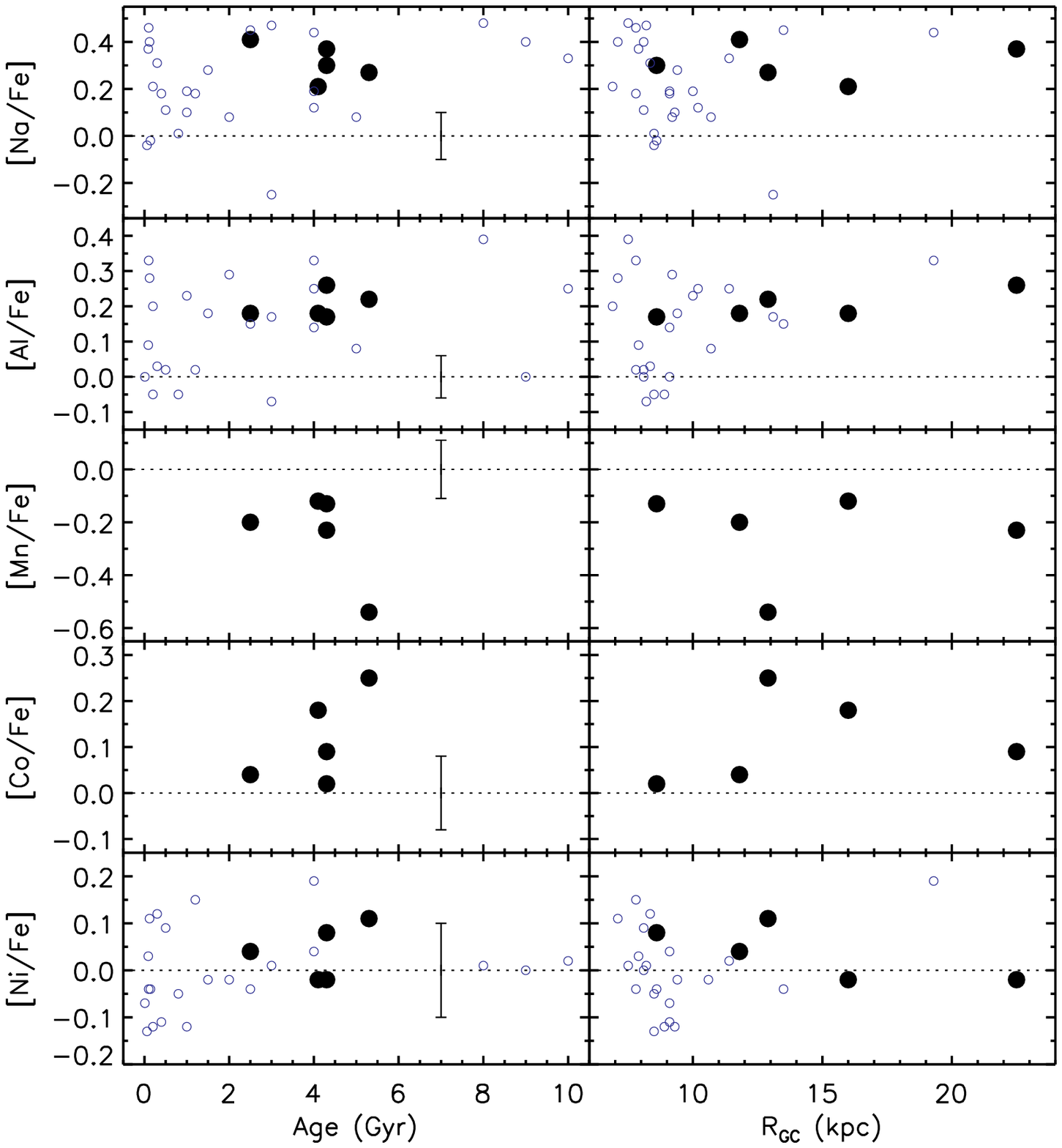}
\caption{Same as Figure \ref{fig:av_abund1} but for different abundance 
ratios.\label{fig:av_abund2}}
\end{figure}

\begin{figure}
\epsscale{0.7}
\plotone{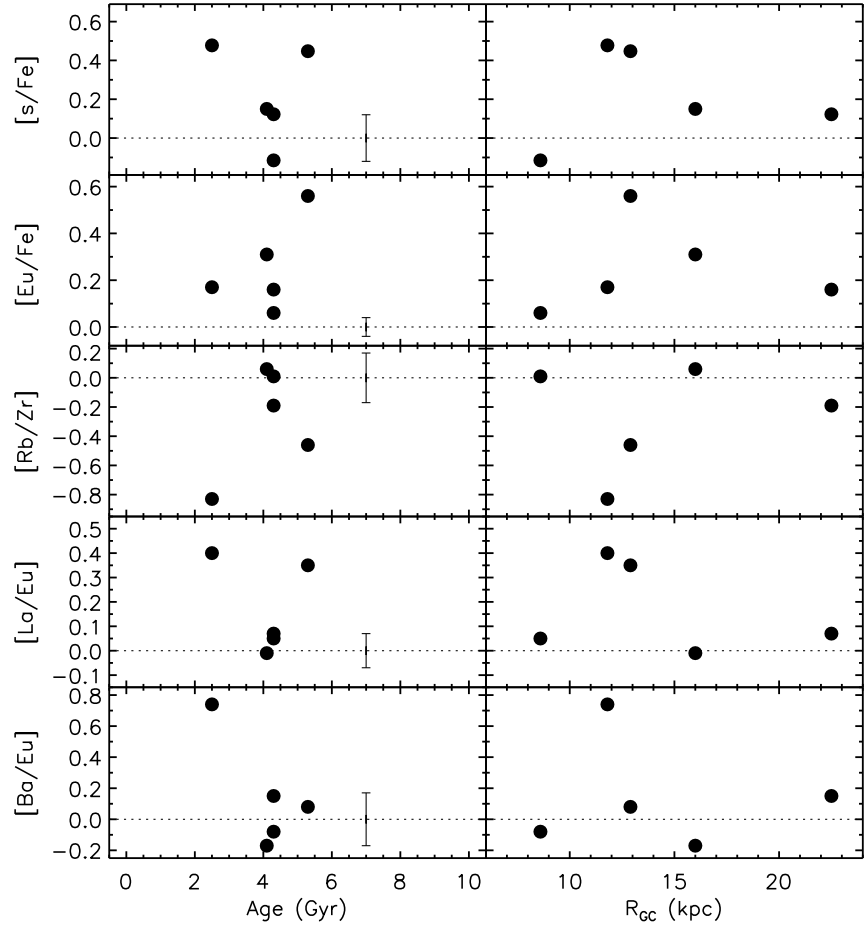}
\caption{Same as Figure \ref{fig:av_abund1} but for different abundance
ratios. In the upper panel, $s$ is the average of
Rb, Zr, Ba, and La.\label{fig:av_abund3}}
\end{figure}

In Figures \ref{fig:av_abund1}-\ref{fig:av_abund3} 
we plot abundance ratios versus age 
for each cluster including numerous open clusters whose values were
drawn from the literature (see \citealt{friel04}).\footnote{
The open cluster data were taken from 
\citet{bragaglia01},
\citet{brown96}, 
\citet{carraro04}, 
\citet*{cayrel85}, 
\citet{edvardsson95},
\citet{friel03}, 
Friel et al.\ (2005, AJ in press), 
\citet{gonzalez00}, 
\citet{gratton94}, 
\citet{hamdani00}, 
\citet{hill99}, 
\citet{king00}, 
\citet{luck94}, 
\citet{pasquini04}, 
\citet{paulson03}, 
\citet{peterson98}, 
\citet{schuler03},
\citet{schuler04}, 
\citet{smith87}, and 
\citet{wilden02}.}
(The results for our clusters reported earlier by \citealt{friel04} were
preliminary, and are superseded by the results reported here.) 
We confirm the apparent lack of a correlation between [Fe/H] and age. Nor can we 
identify any clear trends between [Na/Fe], [Al/Fe], [Ni/Fe], [O/Fe], and 
[$\alpha$/Fe] versus age ($\alpha$~=~$<$Mg+Si+Ca+Ti$>$). \citet{friel03}
also found no trends between these abundance ratios and
age. The three open clusters with ages between 8 and 10 Gyr do not have
enhanced [O/Fe] or [$\alpha$/Fe], while some younger clusters do manifest
enhancements. Age does not appear to be the key variable in this
sample. But Figure \ref{fig:av_abund1} shows that Galactocentric
distance is the vital parameter. 

In Figures 
\ref{fig:av_abund2}-\ref{fig:av_abund3} we plot [Mn/Fe], [Co/Fe], 
[$s$/Fe], and [Eu/Fe] versus age where $s$ is the average of
Rb, Zr, Ba, and La. For these elements, the data do not 
span a large range in age so more observations 
are required before a conclusive statement can be offered on the
presence of abundance trends with age. It would be astonishing if Mn, 
Co, Rb, Zr, Ba, La, or Eu exhibit a trend with 
age given the behavior of the other elements. 

\subsection{Abundance trends versus Galactocentric distance}

In Figures \ref{fig:av_abund1}-\ref{fig:av_abund3} we plot 
abundance ratios versus Galactocentric distance, $R_{\rm GC}$. 
As with the age-metallicity relation, prior studies of radial
abundance gradients using open clusters have focused primarily 
upon the iron abundance due to a lack of data for other elements. 
Linear functions are usually imposed to fit the radial abundance 
gradients of [Fe/H] and it has been shown that [Fe/H] decreases linearly with 
increasing Galactocentric distance in the vicinity of the solar 
neighborhood. A striking fact is that beyond $R_{\rm GC}~=~$10 to 12 kpc, 
the metallicity gradient appears to vanish. 
Rather than decreasing with increasing
distance, the iron abundance appears constant with [Fe/H] $\approx -0.5$.
This result is consistent with \citet{friel02}, who had only
two clusters at such large distances and whose metallicities are
similar to the basement value. 
\citet{carraro04} also studied Be~29, as well as Saurer~A, both of which lie 
at large Galactocentric distances. Their results for Be~29 are similar to 
ours, as we noted previously, and that for Saurer~A ([Fe/H] = $-0.38$, 
$R_{\rm GC} \approx 19$ kpc) the constancy of [Fe/H] at large $R_{\rm GC}$ 
values persists. The disappearance of a radial abundance gradient is a 
challenge to explain. 

The abundance ratios for the $\alpha$ elements in the open clusters depend 
on Galactocentric distance. We see that [$\alpha$/Fe] 
is enhanced with respect to solar in the outer disk. 
The local clusters ($6 \le R_{\rm GC} \le 10$)
have a lower [$\alpha$/Fe] than the clusters 
beyond $R_{\rm GC}~=~$10 to 12 kpc. Not
surprisingly, the behavior of O mimics the $\alpha$ elements. 
This difference in [$\alpha$/Fe] reveals that the history of star formation 
has been rather different in the intermediate age clusters of the outer 
disk than in the solar neighborhood. We may wish to inquire, however, 
whether all of these abundances reflect the initial composition within 
the clusters or if the present abundances have been altered by internal 
nucleosynthesis and mixing.

\begin{figure}
\epsscale{0.9}
\plotone{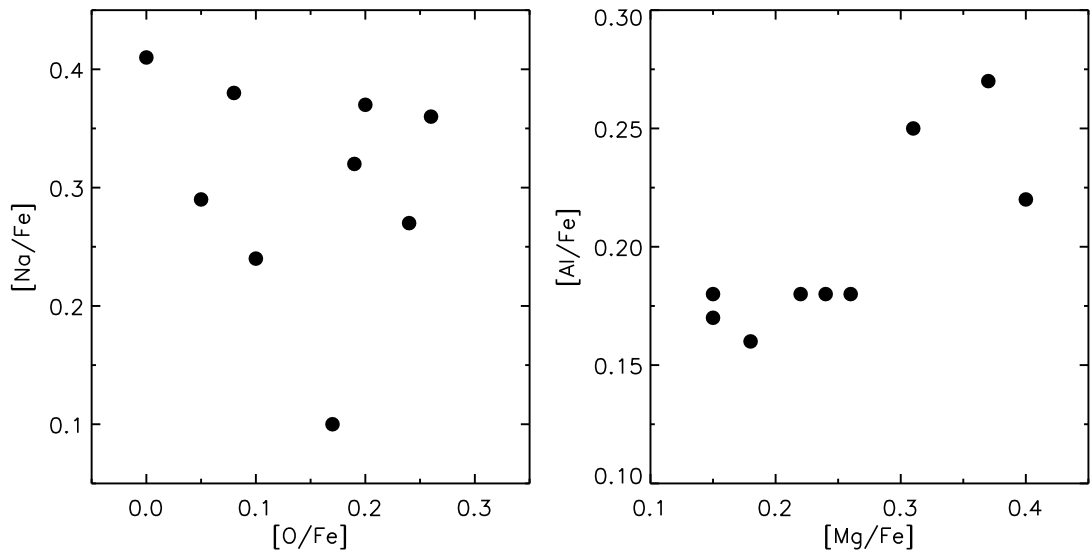}
\caption{[O/Fe] versus [Na/Fe] (left) and [Mg/Fe] versus [Al/Fe] (right) for
the open cluster stars. 
\label{fig:onamgal}}
\end{figure}

The open cluster giants are luminous, highly evolved stars. Their convective
envelopes may penetrate down to layers in which hydrogen-burning is taking 
place via the CNO cycle, the Ne-Na chain, and/or the Mg-Al chain. This leads
to the possibility for burning O into N, Ne into Na, and Mg into Al such that the
measured O, Na, Mg, and Al may not reflect their initial abundances (e.g., 
\citealt*{salaris02}). In Figure \ref{fig:onamgal} we plot [O/Fe] versus 
[Na/Fe] and [Mg/Fe] versus [Al/Fe]. If mixing has taken place within these stars,
we would expect O and Na to be anticorrelated as well as Mg and Al. 
Since we do not find such anticorrelations, we conclude that there are
no obvious signs of mixing within the observed stars. 

The abundances of Na and Al do not correlate with Galactocentric distance. 
For the open clusters, the ratios [Na/Fe] and [Al/Fe] are generally enhanced
with respect to the solar value though there is a large dispersion. 
\citet{carraro04} found a similar pattern from their analysis of two
distant open clusters. 

The abundances of Mn, Co, Ni, Rb, Zr, Ba, and La do not exhibit any 
obvious trends with Galactic location. 
The abundance of Co appears to track Eu while Mn is underabundant.
For the neutron-capture elements 
Rb, Zr, Ba, La, and Eu, the dispersion in abundance ratios [X/Fe] 
in the outer disk is considerably greater than for other
elements. While the data are sparse, we note that the neutron-capture 
elements are generally enhanced in the outer disk with the exception of Rb. 
Eu, in particular, appears to share the behavior of [$\alpha$/Fe] and show
enhanced levels relative to Fe in the outer disk. 

\subsection{Comparisons with different stellar populations}

A simple question that we can ask is at a given [Fe/H], how do the old open 
cluster abundance ratios [X/Fe] compare with different stellar populations: 
thin disk, thick disk, halo, bulge, and dwarf spheroidals. Before we make 
these comparisons, we first consider the chemical evolution of a 
closed system. Two ``global'' parameters, age and star formation rate, 
dictate the evolution, as we discussed in Section 1. Basically,
a sufficiently old system will have experienced the contribution 
from Type II supernovae, but not from Type Ia supernovae.
Such a system would have large ratios (i.e., higher than solar) of 
[$\alpha$/Fe], [O/Fe], and [Eu/Fe]. For a younger system, we expect to 
observe low (i.e., close to solar) ratios of [$\alpha$/Fe], [O/Fe], and 
[Eu/Fe] since sufficient time has passed to allow both Type II and Type Ia 
supernovae to evolve and eject their material. At an intermediate
age, Type Ia supernovae will have begun their contribution such that the ratios
[$\alpha$/Fe], [O/Fe], and [Eu/Fe] would lie below an older system
but above a younger system. Due to the vulnerability of O to proton-captures
in the hydrogen-burning shells of evolved giants and the magnitude of
the errors for [Eu/Fe], [$\alpha$/Fe] is perhaps the most meaningful measure
of the ``age'' of a system. 

Compared to the outer disk clusters, the oldest ($>$ 6 Gyr) open clusters 
with 7.5 $\le$ $R_{\rm GC}$ $\le$ 11.5 are underabundant 
in [$\alpha$/Fe]. The key aspect may be that the oldest open clusters are in the 
solar neighborhood. That is, they all lie 
within the apparently magic 10 to 12 kpc limit,
outside of which the radial metallicity gradient seems to disappear. The
low [$\alpha$/Fe] in the oldest open clusters just means the clusters formed 
well after Type Ia supernovae had begun to 
contribute and that star formation in the solar 
neighborhood had already commenced. The higher [$\alpha$/Fe] of the outer disk
clusters suggests that they formed shortly after a major burst of star 
formation. Their young ages indicate that this burst occurred recently,
within the last 4-6 Gyr. (Our abundances combined with existing color-magnitude
diagrams should enable more reliable ages to be determined for the
open clusters.)

The star formation rate must also be taken into account. Once star formation
commences, a slower star formation rate means that the transition from
Type II supernovae to Type Ia supernovae occurs at 
a lower metallicity. For the reasons given above, 
we again identify [$\alpha$/Fe] as
the preferred abundance ratio for tracking the star formation rate. 
The interpretation of the distinct [$\alpha$/Fe] versus [Fe/H] patterns 
observed in solar neighborhood thin disk and thick disk stars is that
the two stellar populations have experienced different star formation rates. 

While star formation does not
necessarily need to commence everywhere at once, it appears to have done so in
the halo \citep{rosenberg99,salaris02a,deangeli05}. Furthermore, the star 
formation rate must have been rapid given the near-constancy of [$\alpha$/Fe] 
\citep{carney96}. 

The star formation rate provides a primary cause for a Galactic radial abundance 
dependence. A slower star formation rate in the lower density outer disk would 
lead to a currently lower mean metallicity and a higher gas-to-star mass
ratio. Both of these are observed. Therefore, we would expect the outer disk
to show a transition from Type II supernovae to Type 
Ia supernovae at a lower [Fe/H] than in the solar
neighborhood if star formation began at roughly the same time. That we find
enhanced [$\alpha$/Fe] in the outer disk implies a different star formation
history compared to the solar neighborhood and the implications of this are 
discussed in more detail later. 

Are there stellar populations with similar
elemental abundance distributions like those we have found in the outer
disk? 

\subsubsection{Thin and thick disks}

In Figures \ref{fig:abund_gce1}-\ref{fig:abund_gce3}
we plot abundance ratios [X/Fe] versus [Fe/H] for the outer disk 
open clusters. Included in these figures are the abundance ratios for 
numerous open clusters taken from the literature (see \citealt{friel04}) 
as well as thin disk and
thick disk stars from various sources. Our comparison thin disk sample 
was a subset of stars from \citet{bdp93} and \citet{bdp03} which 
had probabilities $\ge$ 80\% of thin disk membership according to
the calculations by \citet{venn04}. The thin disk sample was
further restricted by considering only those stars with ages
in the range 2-6 Gyr to directly compare stars with similar ages to
the open clusters.
The comparison thick disk sample was a subset of stars drawn from 
\citet{bdp93}, \citet{prochaska00}, \citet*{bensby03,bensby04o}, and \citet{brewer05}
which had probabilities $\ge$ 80\% of thick disk membership again
based on the computations by \citet{venn04}. Thick disk stars
tend to be older than the thin disk (e.g., \citealt{bensby04} and
\citealt{brewer05}) and therefore no thick disk stars have comparable
ages to the outer disk open clusters. First we consider the products of 
Type II supernovae, namely O and $\alpha$. 

For the three most metal-poor outer disk
open clusters (Be 20, Be 29, and Be 31), the [O/Fe] values are 
consistent with the lower envelope defined by the thick disk. M67, the
solar neighborhood open cluster, and NGC 2141 have [O/Fe] ratios that
follow the thin disk. 

For the four outer disk open clusters we 
analyzed, [$\alpha$/Fe] exceeds the values found in thin disk stars 
at the clusters' metallicities. 
Since O and $\alpha$ elements (Mg, Si, Ca, and Ti) are primarily products of 
Type II supernovae, we would expect enhancements in [$\alpha$/Fe] compared to 
thin disk stars given that [O/Fe] is overabundant relative to the thin disk. 
The enhanced ratios of [$\alpha$/Fe] in the outer disk indicate intense star 
formation during which [Fe/H] was driven to the current observed levels.
The star formation proceeded sufficiently fast that Type Ia supernovae did not
have time to contribute to the ISM from which the present outer disk open
clusters formed. This implies a more recent beginning to star formation
(or the beginning of the major burst) in the outer disk than in the solar
neighborhood. 

Na appears enhanced with respect to the thin disk and thick disk 
at all values of [Fe/H]. Al appears enhanced compared to the thin disk 
with values similar to those of the thick disk. 
In their study of the open cluster IC 4651, \citet{pasquini04}
found Na was enhanced in giants by 0.2 dex relative to the dwarfs and proposed
that this was a result of dredge-up of Na 
freshly synthesized in the H-burning shell
via the Ne-Na chain. 
Abundance differences within an open cluster have also been reported 
by \citet{tautvaisiene00} for M67. Clump stars were noted to 
display enhancements in N and Na along with depletions of C compared with the
giants. In this case, \citet{tautvaisiene00} suggested that the additional 
mixing of material exposed to the CN cycle and the Ne-Na chain 
takes place after the core He flash at the tip of the first ascent red giant
branch. 
Recall that in Figure \ref{fig:onamgal}, we found no evidence for ``deep 
mixing'' within our sample. 

The abundance of Ni in the open clusters appears to follow the trends
defined by the thin and thick disks, that is, [Ni/Fe] is solar. Mn and Co
also tend to follow the thin and thick disks with slight underabundances
of [Mn/Fe] and overabundances of [Co/Fe]. 

Next we consider the $s$-process elements Rb, Zr, Ba, 
and La. For three of our open clusters, Be 20, Be 29,
and M67, the abundance ratio [$s$/Fe] (where $s$~=~$<$Rb+Zr+Ba+La$>$) 
is in accord with thin and thick disk stars at 
the same metallicity. The remaining two clusters,
Be 31 and NGC 2141, show large enhancements in [$s$/Fe]. 
Unfortunately, for the two clusters with large excesses,
the abundance measurements come from only one star. One explanation for 
overabundances of these $s$-process elements is mass transfer from a companion 
AGB star, now a white dwarf. Identifying variations in radial velocity 
may help support the mass transfer scenario. 
If other stars in Be 31 and NGC 2141 show $s$-process enhancements, then these
open clusters would be similar to the globular cluster M4 which shows
overabundances of Ba and La independent of evolutionary 
status \citep{ivans99}. The interpretation
is that M4's natal cloud was already enriched in $s$-process elements. The
globular cluster $\omega$ Cen also contains a population of stars that show
excess $s$-process abundances \citep{smith00}. This globular cluster is unique 
since it also shows a spread in Fe abundance. Low to intermediate mass AGB 
stars are believed to be responsible for the $s$-process
excess and a hiatus or decrease in the star formation rate may have allowed AGB 
ejecta to mix into the ISM before star formation restarted. 
It would be extremely interesting to study more stars in these two open 
clusters to see if other cluster members show enhancements in Zr, Ba, and La.

For four of the open clusters, the abundance ratios [Eu/Fe] match 
the thin disk. For the remaining cluster, Be 31, the Eu abundance 
slightly exceeds the thin disk and is comparable to the thick disk.
In Be 31, Eu is enhanced relative to Fe such 
that [Eu/Fe] $\simeq$ [$<$Rb+Zr+Ba+La$>$/Fe]
$\simeq$ 0.5. In this cluster we could only measure the abundances for a
single star. Since Eu is synthesized primarily by the $r$-process, we cannot
invoke pollution from an AGB star to explain the excess. Instead, an excess
in Eu can be explained if the star formed from primordial gas enriched by
Type II supernovae and/or without a contribution from Type Ia supernovae. 
In this scenario, the Eu excess should be accompanied
by O, Mg, Si, Ca, and Ti. It is difficult to understand the origin of the Eu
enhancement in this outer disk open cluster given that we do not see large 
values of [$\alpha$/Fe] relative to the other outer disk open clusters 
clusters. That is, while [$\alpha$/Fe] exceeds the solar value, it is similar 
to other outer disk open clusters
which do not show such extreme enhancements of [Eu/Fe]. Observations of 
more cluster members are required to confirm and understand the origin of the
Eu excess in Be 31. Interestingly, the abundance ratio [Co/Fe] appears to 
follow [Eu/Fe] for all clusters as anticipated if Co is synthesized in 
massive stars.

We conclude that neither the local thin nor thick 
disk populations provide exact templates 
to match the abundances seen in the outer disk open clusters. 
Specifically, measurements of O, Na, Al, and $\alpha$ clearly suggest a 
different nucleosynthetic history for the outer disk when compared with
the thin and thick disks. In addition to the chemical composition differences,
the thick disk may be eliminated as a possible source of the open clusters
based on age considerations: the thick disk is much older than the clusters
\citep{bensby04,brewer05}. Note that the composition of the nearby cluster 
M67 is representative of the thin disk and differs from the outer disk open 
clusters. 

\subsubsection{Halo}

The metallicity distribution function for the halo peaks around [Fe/H]~=~$-1.6$ 
with a metal-rich tail extending as high as [Fe/H]~=~$-0.5$ \citep{laird88,ryan91}. 
The outer disk open clusters are not as metal-poor as the bulk of the halo
nor are they as old as the halo. Based on the Fe abundances and 
ages, it is unlikely that the outer disk open clusters 
can be associated with the halo. Kinematic considerations would also suggest
no plausible connection between the halo and outer disk open clusters. 
Nonetheless, studies of halo stars and clusters offer us tools to explore
the nucleosynthetic history of the outer disk. 

Often applied to halo stars, abundance ratios between different 
neutron-capture elements can provide a detailed insight into their 
nucleosynthesis. We examine more carefully abundance 
ratios between the neutron-capture elements to gain further insight
into the chemical enrichment history of the outer disk open clusters. 

\citet{sneden94} identified
a metal-poor star, CS22892-052, with large excesses of the neutron-capture 
elements. The abundance pattern of these heavy elements represents a 
scaled solar ``pure
$r$-process'' distribution. Though rare, additional metal-poor stars have 
since been identified in which the neutron-capture 
elements follow a scaled solar $r$-process pattern. 
The open clusters have abundance ratios [La/Eu] and [Ba/Eu] that vary from
cluster to cluster. At one extreme, NGC 2141 has [La/Eu]~=~0.4 and [Ba/Eu]~=~0.74
and at the other end Be 20 has [La/Eu]~=~$-$0.01 and [Ba/Eu]~=~$-$0.17. 
All clusters have [La/Eu] and [Ba/Eu] ratios considerably 
greater than the pure $r$-process value. (Assuming the \citet{burris00} 
assignments for the $s$-process and $r$-process 
fractions, the pure $r$-process values are [La/Eu]~=~$-$0.6 and [Ba/Eu]~=~$-$0.82
while the pure $s$-process values are [La/Eu]~=~1.4 and [Ba/Eu]~=~1.45.)
This demonstrates that the gas from which the outer disk open 
clusters formed had been polluted by Type II supernovae as well as AGB stars.
This is not surprising given that the open clusters have $s$-process enhancements
and are considerably more metal-rich than CS22892-052. Although some halo stars
as metal-rich as [Fe/H]~=~$-1.0$ show a pure $r$-process distribution for heavy elements,
other halo stars as metal-poor as [Fe/H]~=~$-$2.0 show evidence for $s$-process 
production \citep{simmerer04}. Note that for all clusters, the abundances
of heavy $s$ elements (Ba + La) exceeds the light $s$ element (Zr) abundances. 

While the solar Rb abundance is attributed in equal parts to the $s$-process and 
$r$-process, Rb offers a potential 
diagnostic of the neutron density at the site of the $s$-process, and therefore
the mass of the AGB star(s) responsible for their 
synthesis \citep*{busso99}. The abundance 
of Rb relative to other light $s$-process 
elements such as Sr, Y, and Zr is controlled by the neutron density 
at the site of the $s$-process.
This sensitivity arises from the 10.7 year half-life of $^{85}$Kr. At low
neutron densities, $^{85}$Kr decays to $^{85}$Rb. At higher neutron densities,
$^{85}$Kr will capture neutrons and form $^{86}$Kr and then $^{87}$Kr which
decays to $^{87}$Rb (effectively stable since the half-life is
$5 \times 10^{10}$ years). The key feature is that the neutron-capture 
cross-sections differ by a factor of 10 between the two Rb isotopes 
($\sigma_{87}$~=~$\sigma_{85}/10$). While
stellar Rb isotope ratios cannot be measured, at low neutron densities the 
ratio Rb/Zr can be up to 10 times lower than at high neutron densities. 
The neutron density at the site of the $s$-process 
is controlled by the mass of the AGB star 
(see \citealt{lambert95} for more details on the use of Rb in measuring
the neutron density of the $s$-process in stars). 
Inspection of Figure 14 in \citet{smith00} suggests that the $s$-process
elements in the three most metal-poor outer disk open clusters (Be 20, Be 29, 
and Be 31) are likely to have been produced in 1.5 -- 3M$_\odot$ AGB stars. We
note that Be 20 has a higher ratio [Rb/Zr] than 
Be 29 and Be 31, though we could only measure
Rb in one of the two cluster stars. This higher [Rb/Zr] ratio could be 
due to the $s$-process elements being synthesized in higher mass AGB stars.
If the $s$-process elements observed in 
Be 20 were produced in AGB stars more massive than 3M$_\odot$, we
would also expect high ratios of $^{25}$Mg/$^{24}$Mg 
(and $^{26}$Mg/$^{24}$Mg) due to the
$^{22}$Ne($\alpha,n$)$^{25}$Mg reaction providing the neutron source
rather than the dominant neutron source $^{13}$C($\alpha,n$)$^{16}$O 
in low mass stars. This could be verified by measurements of Mg isotope
ratios in Be 20 though additional observations with 
high resolution and high S/N would be required. 

\subsubsection{The Galactic Bulge}

While detailed abundance measurements of stars in the Galactic bulge 
are rare, we may compare the limited bulge data with the equally limited data
for the outer disk open clusters. Based on location, kinematics, and age, 
it is unlikely that the outer disk open clusters can be associated with the 
bulge. In addition to the large physical separation, there is a considerable 
difference in age, with the bulge being much older, at least 10 
Gyr according to \citet{zoccali03}. Nevertheless, we 
conduct a comparison of the chemical compositions starting with a more careful 
investigation of the behavior of the $\alpha$ elements. 
As \citet{mcwilliam94} point out, the observed abundances of 
O, Mg, Si, S, Ca, and Ti in thin disk and halo stars exhibit a common pattern.
These elements show an increasing abundance relative to
Fe as [Fe/H] decreases. From a nucleosynthetic perspective, Mg, Si, S, Ca, and Ti
are not predicted to follow O exactly. \citet{woosley95} presented
extensive nucleosynthesis calculations for massive stars characterizing the
yields either by the mass ejected or by the production factor (the 
mass fraction of a nuclide in the ejecta normalized to the standard 
solar mix). Inspection of the \citet{woosley95}
production factors for metal-poor supernovae show variations of up to a factor 
of three between O, Mg, Si, Ca, and Ti. Ti is a well known case for which 
observations show an $\alpha$-like behavior but based on theoretical yields,
Ti is expected to scale with Fe. 

Observations of K giants in the Galactic bulge have shown that the abundances
of [Mg/Fe] and [Ti/Fe] are enhanced but that [Si/Fe] and [Ca/Fe]
are not as strongly enhanced and act in a similar manner as thin disk stars
\citep{mcwilliam94}. It is difficult to interpret such abundance patterns. 
Mg and Ti offer evidence that the star formation was rapid and intense with
the nucleosynthesis controlled by massive stars. Si and Ca present a conflicting
case in which a longer duration of star formation allowed Type Ia supernovae to 
play a prominent role. \citeauthor{mcwilliam94} showed that modifications
to the adopted stellar parameters cannot result in 
all $\alpha$ elements giving the same [$\alpha$/Fe] ratio. 
Clearly nature is more complex than our understanding. Slight enhancements
for [Eu/Fe] in the bulge giants offer further support for rapid star
formation in the bulge.

\begin{figure}
\epsscale{0.9}
\plotone{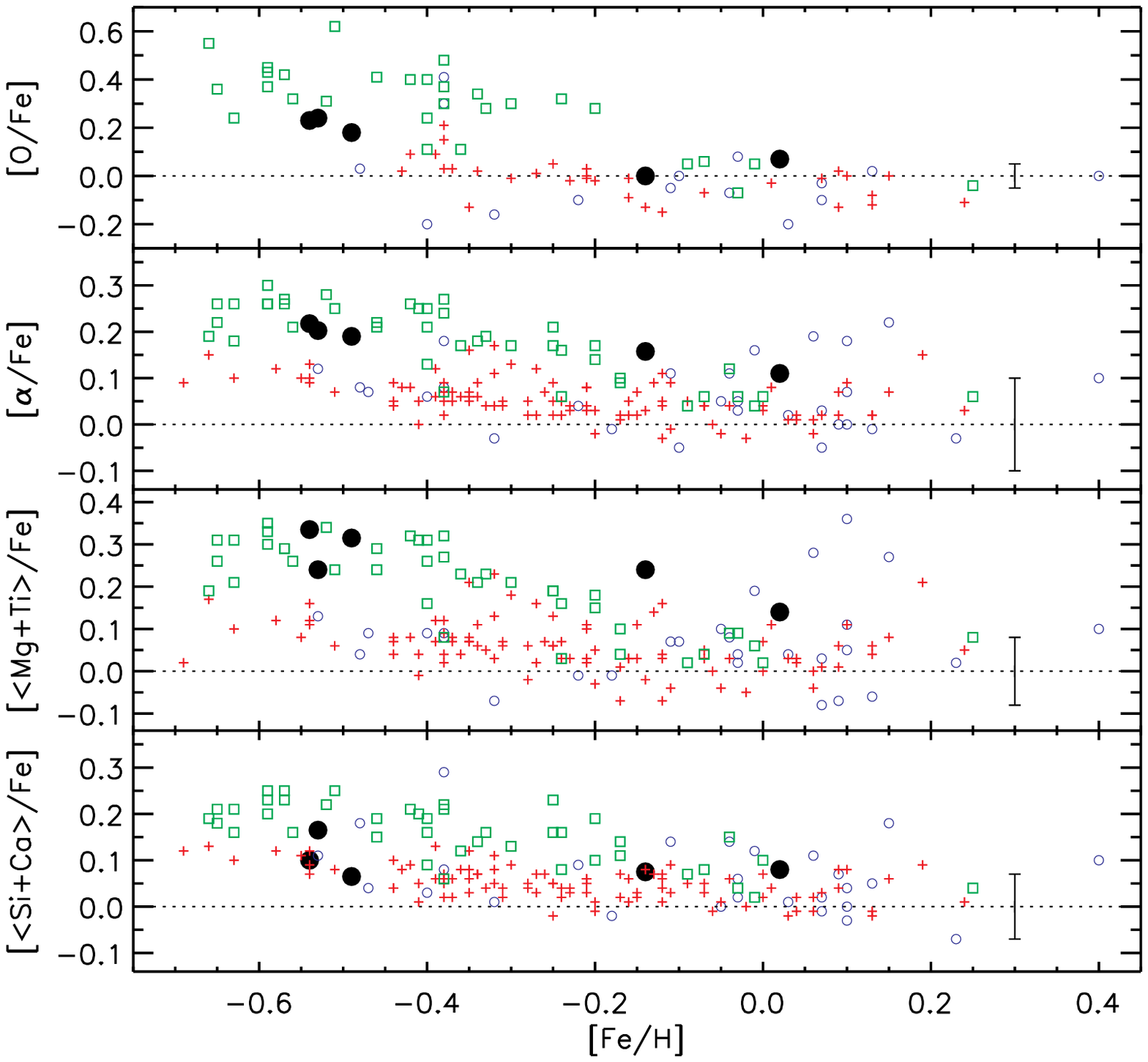}
\caption{Abundance ratios [X/Fe] versus [Fe/H]. The filled black
circles represent the cluster abundances derived in this study while 
the open blue circles represent various abundance determinations for 
open clusters taken
from the literature. The red plus signs are thin disk stars 
(taken from \citealt{bdp93} and \citealt{bdp03})
and the green squares are thick disk stars (taken from 
\citealt{bensby03,bensby04o}, \citealt{brewer05}, \citealt{bdp93}, and 
\citealt{prochaska00}).
(See text for further details on field star selection.) 
A representative error bar for our measured abundance
ratios is shown.\label{fig:abund_gce1}}
\end{figure}

\begin{figure}
\epsscale{0.9}
\plotone{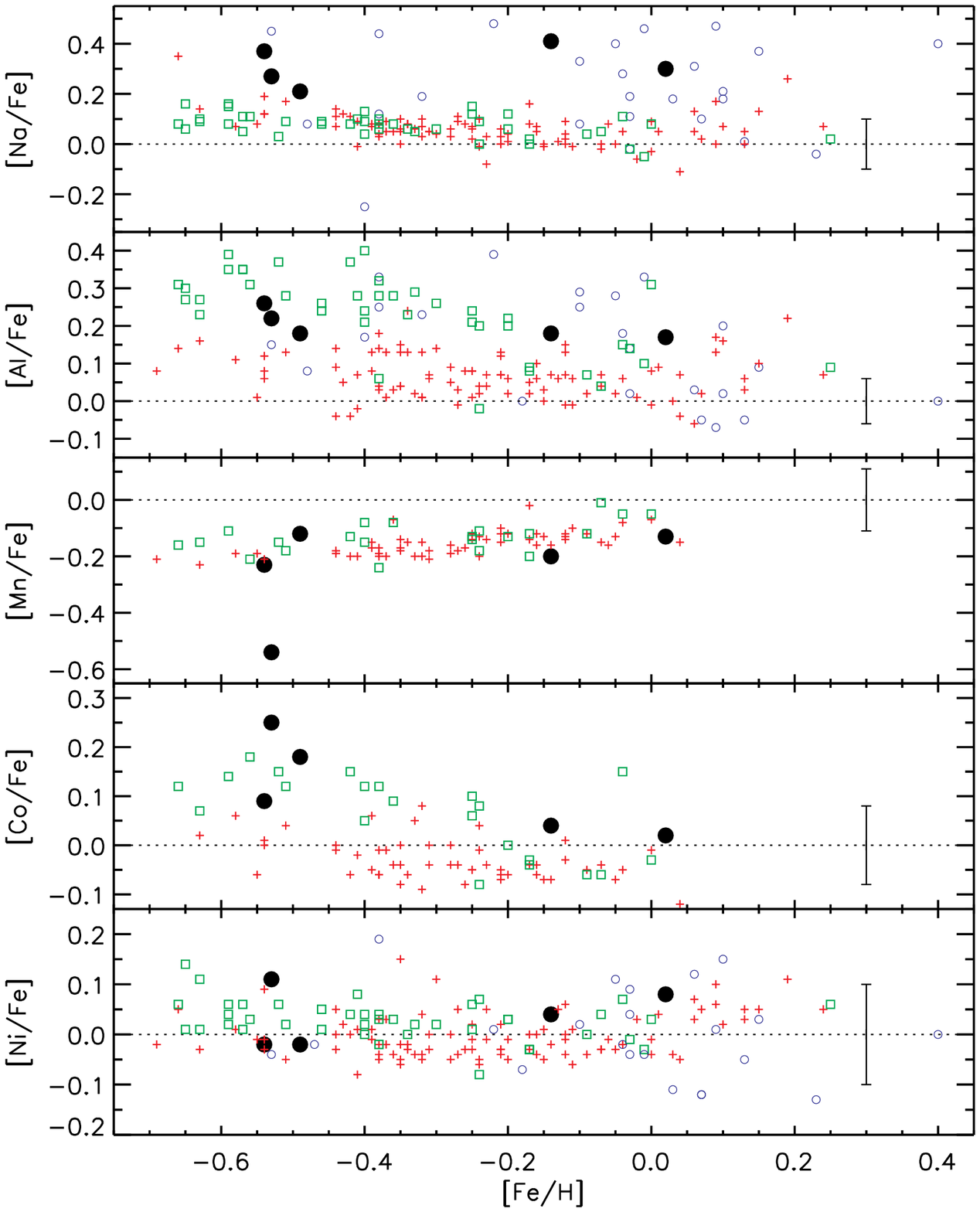}
\caption{Same as Figure \ref{fig:abund_gce1} but for different abundance
ratios.\label{fig:abund_gce2}}
\end{figure}

\begin{figure}
\epsscale{0.9}
\plotone{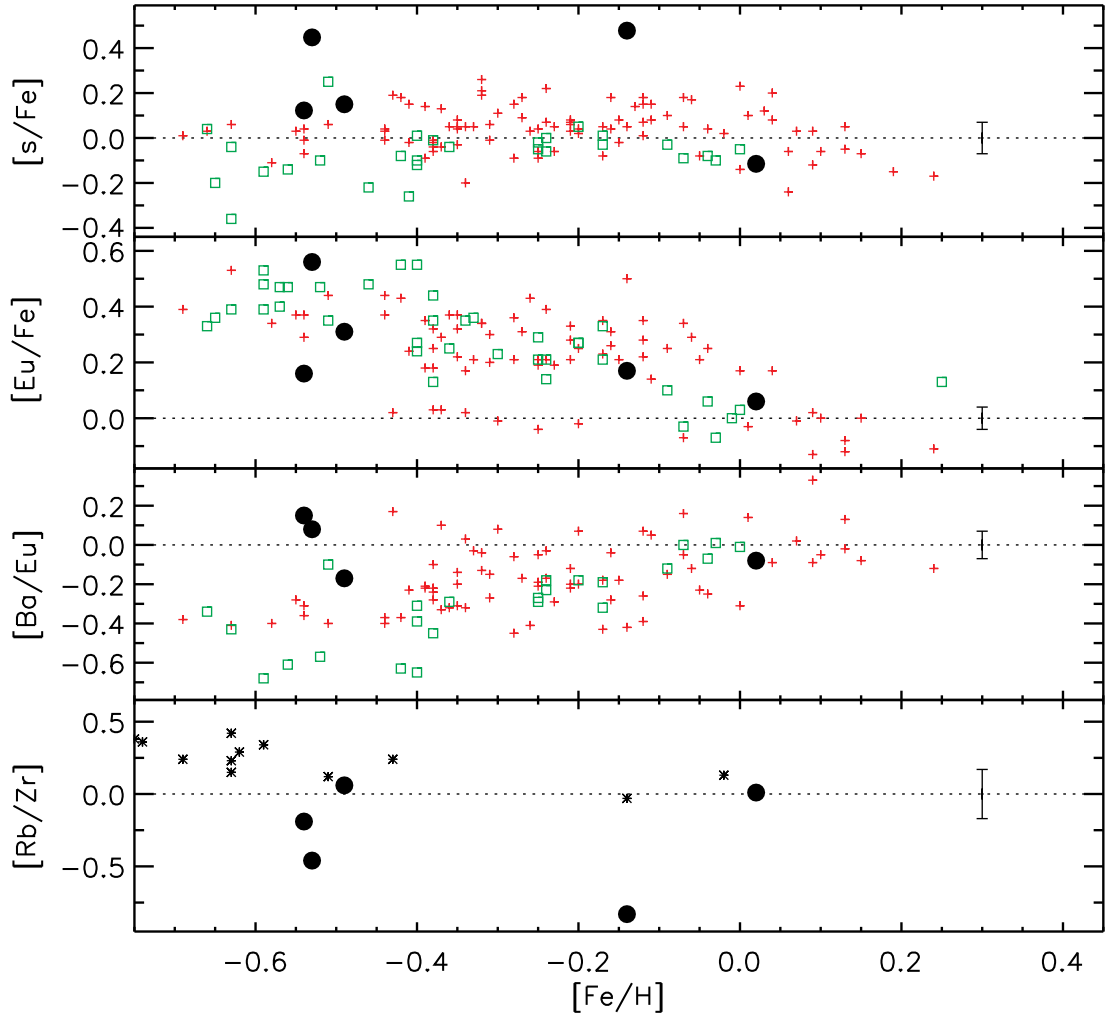}
\caption{Same as Figure \ref{fig:abund_gce1} but for different abundance
ratios. In the upper panel, $s$ is the average of Rb, Sr, Y, Zr, Ba, and La
where available. In the bottom panel, the asterisks represent data taken
from \citet{tomkin99}.\label{fig:abund_gce3}}
\end{figure}

Throughout this paper we have grouped Mg, Si, Ca, and Ti together, referring
to them as the ``$\alpha$'' elements. Oxygen should behave identically to
these $\alpha$ elements assuming no depletion from deep-mixing. 
Now we split the $\alpha$ elements into two separate
groups, Mg+Ti and Si+Ca. In Figures \ref{fig:av_abund1} and 
\ref{fig:abund_gce1}, we find that the enhancement in [$\alpha$/Fe] is driven
by enhanced [Mg/Fe] and [Ti/Fe]. We see that [Si/Fe] and [Ca/Fe]
are almost present in their solar proportions. Excluding the nearby cluster 
M67, we find a mean difference for the outer disk open clusters of 
[$<$Mg+Ti$>$/Fe]$-$[$<$Si+Ca$>$/Fe]~=~0.2. Inspection of Table \ref{tab:err}
demonstrates that no adjustment to the stellar parameters will force
Mg, Si, Ca, and Ti to have the same [$\alpha$/Fe] value. When we
re-examine Figure \ref{fig:av_abund1}, we see that [$<$Si+Ca$>$/Fe] is essentially
constant as a function of Galactocentric position, but [$<$Mg+Ti$>$/Fe]
follows [O/Fe] where the abundance ratios increase with increasing
Galactocentric radius. We see in Figure \ref{fig:abund_gce1} that [$<$Si+Ca$>$/Fe]
and [O/Fe] mimic the field stars at the same metallicity whereas [$<$Mg+Ti$>$/Fe]
is enhanced with respect to the field stars. 
It is difficult to understand the origin of the difference between Mg+Ti 
and Si+Ca. While the observed $\alpha$ element abundances could potentially 
be used to probe the mass of the polluting supernovae, we do not see any yields in 
\citet{woosley95} with low Si and Ca along with high Ti and Mg. 
At this stage, all we are able to say is that the bulge and 
outer disk clusters seem to have experienced a 
similar but not shared nucleosynthetic history
with regard to the synthesis of the $\alpha$ elements and that our
present understanding of the stellar nucleosynthesis of the $\alpha$ elements 
is incomplete. 

The $\alpha$ element abundance patterns are not unique to the \citet{mcwilliam94} 
study of the bulge and this study of the outer disk open clusters. 
We also note that the globular cluster NGC 6553 has an inner-disk/bulge-like 
space velocity and appears to share the bulge's unusual pattern of $\alpha$ 
element abundances \citep{cohen99}. \citet*{pompeia03} studied a sample
of nearby dwarfs with kinematics signifying an inner disk or bulge origin. 
In their sample, Si and Ca followed the bulge trend and were underabundant
with respect to Mg. That the Ti abundances did not follow the bulge's pattern
may suggest that the sample is not representative of the bulge. 

Additional elemental abundance ratios point to a similar, but not shared, 
nucleosynthetic history between the bulge and the outer disk open clusters. 
\citet{mcwilliam94} found large enhancements of [Al/Fe] in bulge giants. While
the enhancement is not as extreme, we also observe excess [Al/Fe] abundances 
in the outer disk. \citet{mcwilliam94} also found Sc and Co to be overabundant in
bulge stars. We tried to determine abundances for Sc but no useful lines
could be identified in our sample. For Co, the most distant open clusters show slight
enhancements in [Co/Fe] signifying yet another possible connection between
the outer disk and the bulge populations. 

\subsubsection{Dwarf spheroidals}

In the standard picture of hierarchical structure formation, 
galaxies form via the accretion of smaller structures. Our Galaxy is
believed to have formed via such a process \citep{searle78} with the
Sagittarius dwarf spheroidal galaxy being an 
example of an on-going merger event \citep*{ibata95}.
The nearby dwarf spheroidal galaxies are potential surviving candidates 
from which the Galaxy may have formed and that the outer disk may have been
accreting. However, observations reveal that
at a given [Fe/H], the nearby dwarf spheroidal galaxies exhibit very 
different [$\alpha$/Fe] ratios to those measured in the halo 
(see \citealt{venn04} and references therein). Generally the dwarf
spheroidals have lower [$\alpha$/Fe] at a given [Fe/H]. 
Such observations preclude the current dwarf spheroidal galaxies as being
the building blocks from which the Galaxy formed. Based on enhanced
[$\alpha$/Fe] ratios in the open clusters in this study, we can 
extend this conclusion to the outer disk. That is, accretion of dwarf 
galaxies with compositions similar to the current dwarf spheroidals
is unlikely to have built up the outer disk. However, the Mn
abundances in one cluster offer evidence for a possible connection between the 
outer disk and dwarf spheroidals. 

Our measurements of Mn in the outer disk open clusters 
are very well matched to the thin and thick disks with [Mn/Fe]
decreasing with decreasing [Fe/H]. \citet*{mcwilliam03} has measured
Mn abundances in the Galactic bulge and showed that [Mn/Fe] mimics 
the behavior seen in the solar neighborhood while stars in the 
Sagittarius dwarf are more deficient in [Mn/Fe] at a given [Fe/H]. 
\citet{mcwilliam03} suggested that the yields of Mn from Type Ia and Type II
supernovae are metallicity dependent, a view also shared by \citet{shetrone03}.
Given that four of our clusters show [Mn/Fe] that match the thin disk but have
excess [$\alpha$/Fe], we agree that the synthesis of Mn probably takes place
in both Type Ia and Type II supernovae with metallicity dependent yields
rather than by Type Ia supernovae as suggested by \citet{gratton89}.
\citet{mcwilliam03} propose a search for low [Mn/Fe] that would reveal the 
signature of accretion of low-mass systems into the Galaxy. One cluster, 
Be 31, appears to have an unusually low Mn abundance with a value similar to 
stars of comparable metallicity in the Sagittarius dwarf. However, 
this cluster has enhanced [$\alpha$/Fe] which is
not seen in dwarf spheroidals at the same [Fe/H], for example, \citet{bonifacio04}
find sub-solar [$\alpha$/Fe] in a metal-rich population ($-0.8 \le {\rm [Fe/H]}
\le 0.0$) in the Sagittarius dwarf. Further complicating the
nucleosynthetic history of this cluster are the enhanced ratios of [Zr/Fe], 
[Ba/Fe], [La/Fe], and [Eu/Fe]. Observations of additional members of Be 31
are required to determine if the low [Mn/Fe] is peculiar to the individual 
star that was studied or if Mn is deficient in all cluster members. 
Overall, Mn offers little evidence that the outer disk is comprised by 
dwarf spheroidals. 

\subsubsection{Cepheids, OB stars, and Planetary nebulae}

Cepheids, OB stars, and planetary nebulae (and H II regions) have also been used 
to search for metallicity gradients in the outer disk. (Cepheids will be discussed 
in considerable detail in a future paper in this series.)
We offer some brief comparisons with the open clusters though we reiterate 
that possibly different techniques and tools are applied to the various analyses.
None of these other objects extend to the same 
large Galactocentric distances that the open clusters encompass.
Furthermore, radial abundance gradients are believed to have evolved with time 
making the different age ranges spanned by the diverse classes of objects 
an important issue to consider. 

We note first that Cepheids are high mass stars 
with very short lifetimes relative to the
open cluster giants and therefore reflect the present abundances in the ISM. 
Like the open clusters, 
Cepheids in the range 10 $\le$ $R_{\rm GC}$ (kpc) $\le$ 15 also show evidence
for a differing behavior for [Fe/H] compared with the solar neighborhood and
Galactic center \citep{andrievsky02c,luck03,andrievsky04}. 
For the inner disk, the Cepheids show a
steep radial abundance gradient for [Fe/H]. In
the solar neighborhood, the abundance is roughly constant with a mean value of 
[Fe/H] $\approx$ 0.0. For $R_{\rm GC} \ge$ 10 kpc, the abundance is
constant and the mean value for Cepheids is 
[Fe/H] $\approx -0.3$ which is greater than the basement value seen in 
the open clusters. In particular, the 
Cepheids display a radial abundance discontinuity in 
[Fe/H] around $R_{\rm GC}~=~$10 kpc.
This radial abundance discontinuity was first noted by 
\citet*{twarog97} in a study of open clusters based on photometric metallicities. 
At the risk of overinterpreting the data, our results are consistent 
with a radial abundance discontinuity albeit one 
that may occur closer to $R_{\rm GC}~=~$12 kpc. However, the elements 
Na, Al, $<$Si+Ca$>$, Mn, and Ni do not appear to change with $R_{\rm GC}$. 
The solar neighborhood Cepheids tend to have solar ratios for [O/Fe] and 
[$\alpha$/Fe] as anticipated for a population younger and slightly more 
metal-rich than the open clusters. The Cepheids do not
exhibit the peculiar abundance pattern for the $\alpha$ elements seen in the bulge
and open clusters though they do show enhanced ratios for $s$-process elements and 
Eu. It would be extremely useful to measure abundance ratios in more distant
Cepheids to compare with the open clusters. 

Hot, young OB stars not only suggest a radial abundance gradient in the disk, 
they also offer evidence for a decreasing gradient with time when compared
with other classes of objects \citep{daflon04}. Further, abundances in OB stars
are in general agreement with the composition discontinuity displayed
by Cepheids. Unfortunately the data do not extend to sufficiently large Galactocentric
distances to explore whether OB stars show an abundance plateau in the outer disk. 
Planetary nebulae also provide evidence for flatter radial abundance gradients,
or a possibly constant abundance, in the outer disk compared with the
solar neighborhood \citep*{costa04}. In general, Cepheids, OB stars, and planetary 
nebulae offer additional evidence for different abundance trends in the outer disk
relative to the solar neighborhood. 

\subsection{Comparison with model predictions}

Having compared the abundance patterns in the outer disk with existing stellar
templates (thin disk, thick disk, halo, bulge, and dwarf spheroidals), we find 
that similarities are present for particular abundance ratios. However, 
none of the stellar populations provides a precise match to the abundances 
measured in the outer disk open clusters. Age comparisons and kinematic
considerations make it even more difficult to associate the open clusters
with any known stellar population. Therefore, we shift our 
attention to theoretical predictions. 

As discussed by \citet{carraro04}, 
\citet*{chiappini01} have presented a chemical evolution model (their model C), that 
reproduces both the linear decrease in [Fe/H] in the solar neighborhood
as well as a basement (or slight increase) for the outer disk. 
In their model, the Galaxy forms as a 
result of two separate gas infall episodes. The first gas infall 
episode leads to the rapid collapse of primordial gas into the halo and bulge
components. For the halo, star formation does not cease if the gas density
falls below a critical value. 
The second gas infall episode forms the thin disk via the accretion of 
primordial gas along with trace amounts of halo gas. Star formation only
proceeds if the gas density exceeds a 
threshold value. For the innermost regions
of the disk, the gas falls in at a faster rate than in the outermost
regions. Star formation occurs more rapidly, leading to a more complete 
conversion of gas into stars. 
This accounts for the linear decrease in [Fe/H] with increasing
Galactocentric distance. For the outer disk, the halo and disk gas densities
are comparable such that gas left-over from the halo is incorporated into 
the star formation that produces the stellar content of the outer 
disk. For the inner regions, including the solar
neighborhood, the disk gas always dominates the halo gas. At a certain
point in the outer disk, [Fe/H] stops decreasing and begins to 
reach a minimum value (or even increase again) due to the non-negligible 
contribution from the pre-enriched halo gas. 
The predictions from the \citet{chiappini01} model provide a very good match 
to the observed iron abundances.

Setting aside the subtle differences between [$<$Mg+Ti$>$/Fe] and 
[$<$Si+Ca$>$/Fe] discussed earlier in the paper, the
global enhancement in [$\alpha$/Fe] in the outer disk is consistent with 
predictions from the \citet{chiappini01} model. If the outer parts of 
the disk formed from halo gas, then we would also expect halo-like abundances 
for the $\alpha$ elements. Further, the model 
would predict enhanced [Eu/Fe] in the outer disk due to the dominance of halo
gas which is also consistent with the observations. However, 
halo stars have low ratios of [Na/Fe] and [Al/Fe] with respect to the solar value
(e.g., \citealt{mcwilliam95,ryan96,cayrel04}). If the model
correctly describes the formation of the outer disk, then 
it is unusual that [$\alpha$/Fe] is in accord with the halo
but that Na and Al differ from the halo. 
Unfortunately, the \citet{chiappini01} do not offer 
predictions for Na and Al in the outer disk.

We are also concerned with conflicting kinematics between the halo and the disk. 
For example, consideration 
of the angular momentum may preclude the possibility that the outer disk 
is comprised by left-over halo gas. Is it physically realistic
that the halo (a population with negligible angular momentum)
can contribute to the outer disk (a very high angular momentum ensemble)?
Based on angular momentum arguments, 
\citet*{carney90} and \citet{wyse92} have shown 
that the gas left over from the halo most likely resides in the bulge. 
The good agreement between our data and the \citet{chiappini01} predictions
might suggest that the outer disk and halo may have a similar but not shared
star formation history. But we have dismissed such an idea earlier in the
paper citing the different [Fe/H] ranges, ages, and kinematics. Therefore, 
specific details regarding the chemo-dynamical evolution of the outer
disk may require refinement and need to be incorporated into the chemical
evolution model. We look forward to detailed predictions of a larger 
ensemble of elemental abundance ratios in the outer disk. 

While additional chemical evolution models offer a fair match to observed
abundance ratios in the solar neighborhood (e.g., \citealt*{hou00,alc01}), they 
predict a continually declining iron abundance with increasing Galactocentric
radius, in conflict with our observations. The discontinuity in [Fe/H] displayed by 
the Cepheids has been interpreted in several ways. \citet{twarog97} offer a
qualitative argument that the limiting value of [Fe/H] (due in part to the 
logarithmic nature of [Fe/H] and the nucleosynthetic yield i.e., stellar yields,
initial mass function, gas dynamics etc.) for the outer disk is simply lower than 
for the inner disk. \citet{twarog97} also consider that ``$R_{\rm GC}~=~$10 kpc 
is a reflection of the original boundary of the newly formed'' thick disk. 
\citet{andrievsky02c}, \citet{luck03}, and \citet{andrievsky04} 
suggest that the radial abundance
discontinuity is a result of inefficient mixing processes due to a small
radial component of the gas velocity near the co-rotation resonance as
described by \citet*{lepine01}. However, 
it is not obvious how the abundance ratios in the outer disk -- 
high [$\alpha$/Fe], peculiar $\alpha$ element pattern, excess Na, Al, and
$s$-process abundances -- fit into this scenario. 
An additional concern is that the lack of a radial metallicity gradient
apparently persists over a very large range of distances, well beyond the
co-rotation radius. 

\subsection{Speculations: remnants of a merger event(s)}

Despite the relative youth of the outer disk open clusters, the abundance ratios 
[$\alpha$/Fe] and [Eu/Fe] point to a recent beginning of star formation, 
contrary to what is 
seen in the solar neighborhood. The outer disk star formation was probably not 
slow and steady since we do not see strong signs of Type Ia supernovae 
contributions, though we do see the effects of contamination from low mass AGB stars. 

\citet{chiappini01} argued that the outer disk formed from a second reservoir
of gas that had experienced a different star formation history. They suggested
the halo as a likely source, but we note that 
angular momentum considerations probably mean that the halo is 
an inappropriate choice. However, the general idea has merit. Could a merger
with an object on a prograde, low-inclination orbit provide the source of gas?
Similarly, could interactions between such an object and the Galaxy
have triggered an intense burst of star
formation in an otherwise unpolluted gaseous outer disk? If so, the ages of the
oldest outer disk open clusters are probably a chronometer of when such an event
took place. \citet{janes94} suggested that indeed the old open clusters may
have formed during bursts of star formation triggered by mergers or interactions
with external systems. \citet*{scott95} noted that two old clusters,
NGC 6791 and Be 17, have unusually eccentric orbits. 

An excess of stars has recently been identified in the direction of the 
Galactic anticenter (e.g., \citealt{newberg02}, \citealt{ibata03}, 
\citealt{rocha03}, and \citealt{yanny03}). \citet{frinchaboy04} refer 
to this stellar overdensity as the Galactic
anticenter stellar structure (GASS) due to its ambiguous shape, orientation,
size, and origin. The GASS has also been referred to as the Monoceros ring 
and might be related to Canis Major galaxy \citep{martin04,martin05}. 
Five globular clusters (Pal 1, NGC 2808, NGC 5286, 
NGC 2298, and BH 176) are located near the GASS and lie in a 
peculiar string-like configuration, not found elsewhere among the globular 
cluster system and resembling a tidal stream. Pal 1 and BH 176 are unusual
globular clusters because they are both young and metal-rich and therefore
possibly ``transitional'' clusters between young globular clusters
and massive old open clusters. Four distant Galactic open clusters
(AM 2, Tombaugh 2, Be 29, and Sau A) are located near the GASS and appear to
extend the string-like configuration defined by the four globular clusters
associated with the GASS.  
\citet{frinchaboy04} searched for more clusters extrapolating from the alignment
of the 9 clusters. An additional 7 clusters were found, further extending the 
string-like sequence. 

\citet{frinchaboy04} suggested that the GASS system and associated clusters 
are possibly the result of tidal accretion of a dwarf satellite galaxy. 
The center of the GASS would presumably be the nucleus of the accreted
dwarf galaxy. 
\citet{bellazzini04} suggested that the accreted Canis Major 
galaxy is centered at $(l,b)~=~(244\arcdeg,-8\arcdeg)$.
\citet{crane03} also argued that the GASS is a satellite galaxy currently in the 
process of being tidally disrupted. Evidence for this scenario
include: (1) the velocity dispersion for M giants associated with the 
GASS is smaller than for stars in the thin disk; (2)
the GASS shows a trend between Galactocentric radial velocity 
($V_{\rm GSR}$) and Galactic longitude ($l$) indicating a non-circular orbit; 
(3) a wide metallicity spread $-1.6 \le $[Fe/H]$ \le -0.4$ has been found,
assuming all the above-mentioned clusters belong(ed) to the galaxy. 

We caution that it is difficult to assign GASS membership based solely
upon radial velocities. Members of the GASS on a low-inclination prograde 
orbit will behave very similarly to the outer Galactic disk since they're both 
responding to the same gravitational potential. Nonetheless, \citet{frinchaboy04}
suggest that Be 29 is a GASS member while Be 31 is not. 
In Table \ref{tab:clustervelocities}, we include
$V_{\rm GSR}$ for the outer disk open clusters. Combined with Galactic
longitude, a comparison can be made
with Figure 2 in \citet{frinchaboy04} to see how well the outer disk open clusters
match the $l$-$V_{\rm GSR}$ distribution defined by the GASS. Be 21 and
NGC 2141 fall close to the $l$-$V_{\rm GSR}$ relation while Be 20 does not. 
When we compare
[Mg/Fe], [Si/Fe], [Ca/Fe], and [Ti/Fe] for Be 29 and Be 31, they are essentially
identical. It is unusual that open clusters with comparable ages but 
presumably very different origins would both have similar and peculiar
abundance ratios. 

NGC 2298 is one of the globular clusters possibly 
associated with the GASS to which \citet{salaris98} 
assign an age of 11.7 $\pm$ 1.1 Gyr. \citet*{mcwilliam92} found $\alpha$ 
enhancements in NGC 2298 similar to those we 
observe in the outer disk open clusters. It is 
very difficult to understand how the globular cluster NGC 2298 and the open
cluster Be 29 with wildly different ages and Fe abundances but presumably a 
common origin in a small galaxy can have similar [$\alpha$/Fe] ratios. The only 
contrived possibility is through an episodic merger picture where infall of 
metal-poor gas triggers a brief period of star formation. But it seems more
likely that episodic mergers would happen to our Galaxy rather than 
accretion of a smaller satellite. 

\citet{penarrubia04} also warn that position and radial velocities alone are 
insufficient to assign membership. \citet{martin04} suggested that the
globular clusters NGC 1851, 1904, 2298, and 2808 may be associated with Canis
Major. However, \citet{penarrubia04} considered proper-motions for these 
clusters and showed that they are kinematically distinct from Canis Major. 

That the open clusters appear to have experienced a star 
formation and nucleosynthetic history unlike any known Galactic 
population is consistent with the outer disk being formed, in part, by
a merger event. (The similarity to the Galactic bulge discussed in 
Section 7.3.3, recall, is only similar, not shared.) 
If the merger event was caused by the GASS, then the similar and unusual
compositions would suggest that all outer disk open clusters are associated 
with the merger. This would imply that the radial velocity criterion for
assigning GASS membership is inadequate. An 
origin from a merger event is appealing since it
would help explain the lack of an age-metallicity relation in the outer 
disk open clusters. Since age-metallicity
relations have now been shown to exist for the thin and thick disk, the lack of
an age-metallicity relation for the open clusters suggests that they 
represent a collection of different stellar populations. 
There are two ways in which the open clusters can be attributed to 
a possible merger event. Either they were created by 
the merger event out of Galactic material or they were part of the victim 
galaxy. If the outer disk open clusters were 
originally members of the victim, then star formation must have commenced 
recently and proceeded sufficiently rapidly 
to reach [Fe/H]~=~$-$0.5 without much contamination from Type Ia supernovae.
Moreover, for a given [Fe/H] the GASS would 
have had [$\alpha$/Fe] ratios unlike those observed in the current
dwarf spheroidals. Such a situation seems somewhat improbable given that
recent observations of candidate members of Canis Major show solar or 
sub-solar ratios of [$\alpha$/Fe] \citep{sbordone05}. A more plausible scenario 
may be that the outer disk open clusters were created from star formation 
triggered by the merger event. 
Another possibility is that the outer disk grew in spurts through
continual or episodic infall of gas and/or dwarf galaxies with the 
GASS/Canis Major representing one end of the mass spectrum.
A test of this would be the
identification and abundance analysis of additional open clusters at large
Galactocentric distances with an emphasis on clusters both younger and
older than those of this study.

If a merger is the preferred explanation for the formation of the outer 
disk, such a scenario must account for the basement in [Fe/H] that persists 
over 10 kpc, the enhanced [$\alpha$/Fe] indicative of rapid star formation
for these young clusters, and abundance ratios that do not match any known
stellar population. 
Ultimately, more data are required to characterize and understand the 
evolution of the outer disk.

\section{CONCLUSIONS AND SUMMARY}

In this paper we have measured radial velocities for a number of stars
in old, distant open clusters using high resolution spectra. Candidates
were selected based on radial velocities and location in color-magnitude 
diagrams. Using spectrum synthesis of the
highest quality data, we conducted a detailed abundance analysis 
to derive the chemical compositions of Fe, 
O, Na, Mg, Al, Si, Ca, Ti, Mn, Co, Ni, Rb, Zr, Ba, La, and Eu
in stars representing four old open clusters and the nearby cluster M67. 
These species represent various
$\alpha$, Fe-peak, and neutron-capture elements whose synthesis occurs in
via different nucleosynthetic processes in different stellar sites. 
The spectra revealed that Be 21 T406 is a rapidly rotating Li-rich star. 
This is the second 
Li-rich star found in this sparse cluster and may be a result of the ingestion of 
a planetary companion during the ascent of the red giant branch.

We first searched for abundance trends versus age. No strong correlations
were found between any abundance and age, a result previously noted.
The lack of an age-metallicity relation may be due to the open clusters being
members of different stellar populations with different origins and
chemical enrichment histories. 
Next we searched for abundance trends versus Galactocentric radius. We found 
that certain abundance ratios showed a dependence upon Galactic location. In 
the outer disk, beyond the apparently ``magic'' radius of $R_{\rm GC}~=~$10 to 
12 kpc, the iron abundance reached a constant 
level of [Fe/H] $\approx -0.5$. This plateau 
deviates from the linear decrease with increasing Galactocentric radius 
defined by clusters located in the solar neighborhood. We also found that [O/Fe], 
[$\alpha$/Fe], and [Eu/Fe] are enhanced relative to solar in the outer disk.
The younger Cepheids offer evidence for a plateau in [Fe/H] in the outer disk
as well. Our open cluster
results are compatible with the radial abundance discontinuity displayed
by Cepheids though the magnitude and location of the discontinuity differ
between Cepheids and open clusters, no doubt due to the chemical evolution
that has occurred since the clusters and the Cepheids formed, a time span
of several billion years. 

We compared the abundance ratios in the outer disk open clusters to well studied
Galactic populations. The abundance ratios [O/Fe] and [$\alpha$/Fe] are 
enhanced and comparable to the thick disk. Despite some composition
similarities, the open clusters are young and probably not members of an 
older population such as the thick disk. 
The high abundances of $\alpha$ elements indicate rapid star formation such that 
Type Ia supernovae did not have sufficient time to evolve 
and contribute to the chemical evolution. 
Na and Al are overabundant compared to thin disk stars at the same [Fe/H]. 
Ni and Mn are in good agreement with the thin disk while Co may be 
slightly enhanced. Be 31 and NGC 2141
show $s$-process enhancements and Be 31 also shows large values of [Eu/Fe]. 

Consideration of the iron abundances, ages, and kinematics suggest no connection
between the outer disk and the halo. Relative abundance 
ratios of the neutron-capture elements show that the
outer disk open clusters have varying ratios of [La/Eu] and [Ba/Eu]. That
is, the ratio of $s$-process to $r$-process material differs from cluster to
cluster. Further, none of the clusters exhibit a scaled solar pure $r$-process 
distribution, revealing that AGB stars have contributed to the nucleosynthetic
history of the clusters. For the three most
metal-poor open clusters (Be 20, Be 29, and Be 31), 
consideration of the abundance ratio [Rb/Zr] shows 
that the $s$-process material was likely synthesized
in 1-3M$_\odot$ AGB stars. For Be 20, a slightly higher ratio of [Rb/Zr] suggests
that a 5M$_\odot$ AGB star may have played a role in the chemical evolution 
of the proto-cluster gas. 

While there are considerable differences in kinematics, ages, and formation
histories between the bulge and the outer disk, both display 
a puzzling pattern within the individual $\alpha$ elements.
Mg and Ti show enhanced ratios with respect to Fe whereas Si and Ca are in 
their solar proportions. The origin of the discrepancy is not known. 
Enhancements in Al, Co, and Eu are also common to the bulge giants and 
outer disk open clusters. These abundance ratios indicate that the
bulge and outer disk may have a common but not shared nucleosynthetic history.
Based on low [$\alpha$/Fe] in the current dwarf spheroidals, there does not
appear to be any connection between the outer disk and dwarf galaxies,
a conclusion that has also been reached for the halo. 
One cluster, Be 31, has a low [Mn/Fe] ratio, similar to stars of comparable 
metallicity in the 
Sagittarius dwarf spheroidal galaxy. In short, while similarities may 
be found for some abundance ratios, the outer disk open cluster abundance 
patterns do not perfectly match any existing stellar template. 
They appear to signify an unusual chemical evolutionary process or processes. 

We note that we could only observe one or two 
stars per cluster. It is necessary to confirm
our results by studying additional stars in these clusters, particularly those
that show unusual abundance ratios. Future studies of a range of 
elements in large numbers of open clusters analyzed in a uniform and 
homogeneous manner are required to advance our knowledge of the evolution
of the Galactic disk.  

A chemical evolution model by \citet{chiappini01} accurately predicts the
plateau of [Fe/H]~=~$-$0.5 beyond $R_{\rm GC}~=~$12 kpc as well as enhanced 
[$\alpha$/Fe] and [Eu/Fe] in the outer disk. According to the model, the
outer disk forms from gas left-over from the halo. However, such a notion
does not seem 
plausible based on angular momentum arguments. The basic idea
of the outer disk forming from material with a different nucleosynthetic
history or histories appears to a requirement. 

Some evidence suggests that a significant merger 
event is currently taking place in the outer
disk. 
That the outer disk open cluster abundance
patterns do not match any Galactic population allows the possibility that the 
open clusters may be associated with one or more merger 
events. If the open clusters were originally members
of a single victim galaxy, then the [$\alpha$/Fe] is unlike any current dwarf 
galaxy. Perhaps the outer disk open clusters were instead created from the merger 
event. If so, then the current ages of the open clusters would be a measure of when 
the event took place. The star formation resulting from the merger would then
have had to have been sufficiently rapid to drive the metallicity to
[Fe/H]~=~$-$0.5 without allowing Type Ia supernovae to contribute, though AGB 
stars must have contaminated the proto-cluster gas. A merger event would mean
that the open clusters represent different stellar populations and therefore
should not present a clear age-metallicity relation. 
An alternative explanation is that the outer Galactic disk has literally grown
steadily, if episodically, due to infall of metal-poor gas and small
galaxies with the Canis Major galaxy being only the latest addition. If such
infall triggered star formation for a relatively brief time, we could, perhaps,
explain the enhanced [$\alpha$/Fe] ratios, as well as the lack of a metallicity
gradient. Of great interest
is the identification and abundance analysis of additional open clusters
residing in the outer disk to test the merger hypothesis. Can younger and older
outer disk open clusters be identified and if so, what are their compositions?

\acknowledgments

We warmly thank the referee, Eileen Friel, for many helpful suggestions and 
comments. We are extremely grateful to the National Science Foundation for
their financial support through grants grants AST 96-19381, AST 99-88156, 
and AST 03-05431 to the University of North Carolina. 

\clearpage

\begin{deluxetable}{lrrrrrrrrrr}
\tabletypesize{\footnotesize}
\tablecaption{Observed Clusters \label{tab:clusters}}
\tablehead{
\colhead{\sc Cluster}  & \colhead{R.\ A.\tablenotemark{a}} &
\colhead{\sc Dec.\tablenotemark{a}} & \colhead{$\ell$}    & \colhead{$b$} &
\colhead{[Fe/H]}  &  \colhead{$\delta V$} & \colhead{MAI\tablenotemark{b}} &
\colhead{d\tablenotemark{c}} & \colhead{d\tablenotemark{d}} &
\colhead{\sc Ref} }

\startdata
Berkeley 20 & 05:32:36 & +00:11:24 & 203.51 & $-17.28$ & $-0.61$ & 2.1 & 4.1 & 8.1 & 8.6 & 1 \\
Berkeley 21 & 05:51:45 & +21:48:20 & 186.84 & $-2.51$ & $-0.97$ & 1.6 & 2.2 & 5.8 & 6.1 & 2 \\
NGC 2141 & 06:03:00 & +10:29:40 & 198.08 & $-5.78$ & $-0.26$ & 1.6 & 2.5 & 4.3 & 3.9 & 3 \\
Berkeley 29 & 06:53:02 & +16:56:20 & 197.95 & +7.98 & $-0.18$ & 2.1 & 4.3 & 8.6 & 14.8 & 4 \\
Berkeley 31 & 06:57:38 & +08:17:19 & 206.25 & +5.12 & $-0.40$ & 2.3 & 5.3 & 3.7 & \nodata & 5 \\
M67 & 08:51:18 & +11:50:00 & 215.66 & +31.91 & $+0.02$ & 2.3 & 4.3 & 0.8 & 0.8 & 6 \\
\enddata
\tablenotetext{a}{J2000.0}
\tablenotetext{b}{Taken from \citet{salaris04}, in Gyrs.}
\tablenotetext{c}{Distance in kpc from \citet{janes94}}
\tablenotetext{d}{Distance estimate obtained using red clump stars (this paper)}
\tablerefs{(1) \citet{macminn94}; (2) \citet{tosi98}; (3) 
\citet{rosvick95};
(4) \citet{kaluzny94}; (5) \citet{guetter93}; (6) \citet{montgomery93}
}
\end{deluxetable}

\begin{deluxetable}{lrrrrrrr}
\tabletypesize{\footnotesize}
\tablecaption{Photometric Data \label{tab:photometry}}
\tablewidth{0pt}
\tablehead{
\colhead{\sc Star} & \colhead{\sc R.\ A.\tablenotemark{a}} & \colhead{\sc Dec\tablenotemark{a}} & 
  \colhead{$V$} &   \colhead{$B-V$} & \colhead{$V-I_{C}$} & \colhead{$K$\tablenotemark{b}} & 
  \colhead{$J-K$\tablenotemark{b}} 
}

\startdata%
 \noalign{\vskip +0.5ex}
 \multicolumn{8}{c}{\bf Berkeley\ 20} \cr
 \noalign{\vskip  .8ex}%
 \hline
 \noalign{\vskip -2ex}\\
 5    & 05:32:37.8 & $+$00:11:09 & 14.80 & \nodata & 1.49 & 11.35 & 0.91 \\
 8    & 05:32:38.8 & $+$00:11:21 & 15.15 & \nodata & 1.43 & 11.85 & 0.88 \\
 22   & 05:32:36.8 & $+$00:11:50 & 16.90 & \nodata & 1.18 & 14.31: & 0.50: \\
 28   & 05:32:38.1 & $+$00:11:17 & 17.11 & \nodata & 1.16 & 14.21: & 0.73: \\
\cutinhead{\bf Berkeley\ 21}
 39   & 05:51:38.4 & $+$21:47:21 & 15.67 & 1.77 & 2.11 & 10.80 & 1.10 \\
 51   & 05:51:42.0 & $+$21:48:04 & 15.69 & 1.70 & 2.06 & 10.96 & 1.10 \\
 67   & 05:51:44.8 & $+$21:48:52 & 14.99 & 1.95 & 2.29 & 9.69 & 1.26 \\
 88   & 05:51:49.4 & $+$21:47:00 & 15.82 & 1.73 & 2.09 & 11.00 & 1.09 \\
\cutinhead{\bf NGC\ 2141} 
 1007  & 06:02:50.8 & $+$10:30:28 & 13.27 & \nodata & 1.85 & 8.98 & 1.08 \\
 1286  & 06:02:56.3 & $+$10:29:05 & 14.81 & \nodata & 1.49 & 11.31 & 0.81 \\
 2066  & 06:02:58.4 & $+$10:26:38 & 14.18 & \nodata & 1.62 & 10.33 & 0.93 \\
 514  & 06:03:00.0 & $+$10:32:24 & 14.09 & \nodata & 1.61 & 10.33 & 0.93 \\
 1997\tablenotemark{c} & 06:03:00.3 & $+$10:28:45 & 14.96 & \nodata & 1.53 & 11.63: & 0.80: \\
 1333\tablenotemark{c}  & 06:03:00.3 & $+$10:28:44 & 15.09 & \nodata & 1.47 & 11.63: & 0.80: \\
 1348  & 06:03:01.6 & $+$10:28:34 & 13.25 & \nodata & 1.94 & 8.86 & 1.07 \\ 
 1821  & 06:03:07.6 & $+$10:26:48 & 14.13 & \nodata & 1.61 & 10.32 & 0.93 \\
\cutinhead{\bf Berkeley\ 29}
 412   & 06:53:01.6 & $+$16:56:22 & 16.64 & 0.98 & 1.07 & 14.27 & 0.56: \\
 1032  & 06:53:03.4 & $+$16:55:09 & 16.56 & 0.97 & 1.05 & 14.17: & 0.61: \\
 988   & 06:53:03.9 & $+$16:55:16 & 14.59 & 1.56 & 1.59 & 10.90 & 0.98 \\
 673   & 06:53:04.1 & $+$16:55:56 & 14.38 & 1.67 & 1.81 & 10.27 & 1.07 \\
 556   & 06:53:04.3 & $+$16:56:03 & 16.60 & 0.98 & 1.07 & 14.14: & 0.60: \\
 241   & 06:53:07.1 & $+$16:57:13 & 14.48 & 1.61 & 1.69 & 10.60 & 1.02 \\
 801   & 06:53:08.1 & $+$16:55:41 & 16.58 & 0.96 & 1.06 & 14.23: & 0.60: \\
\cutinhead{\bf Berkeley\ 31\tablenotemark{d}}
 1065 & 06:57:34.7 & $+$08:15:42 & 16.28 & 0.93 & 1.10 & 14.30: & 0.56: \\
 886  & 06:57:37.4 & $+$08:15:59 & 14.67 & 1.33 & 1.30 & 11.68 & 0.78 \\
 720  & 06:57:39.6 & $+$08:15:24 & 16.32 & 1.08 & 1.19 & 13.65 & 0.77: \\
 728  & 06:57:39.6 & $+$08:16:22 & 16.52 & 0.98 & 1.14 & 14.01: & 0.62: \\
 666  & 06:57:41.0 & $+$08:17:20 & 13.84 & 1.35 & 1.33 & 11.66:\tablenotemark{e} & 0.78: \\
\cutinhead{\bf M\ 67}
 105  & 08:51:17.1 & $+$11:48:16 &   10.30  & 1.26 & 1.23 & 7.39 & 0.76 \\
 108  & 08:51:17.5 & $+$11:45:22 &  9.72 & 1.37 & 1.36 &  6.49 & 0.83 \\
 141  & 08:51:22.8 & $+$11:48:03 &   10.48  & 1.11 & 1.08 & 7.94 & 0.62 \\
\enddata
\tablenotetext{a}{J2000.0}
\tablenotetext{b}{Values with errors greater than 0.05 mag are indicated with a colon.} 
\tablenotetext{c}{These two stars are blended in the 2MASS results.}
\tablenotetext{d}{$V$ and $B-V$ data are from \citet{guetter93}, and the 
$V-I_{C}$ values
are from \citet{phelps94}.} 
\tablenotetext{e}{The stellar image is elongated.}
\end{deluxetable}

\begin{deluxetable}{lrrrlr}
\tabletypesize{\footnotesize}
\tablecaption{Radial Velocity Standards \label{tab:vradstandards}}
\tablehead{
\colhead{\sc Star} & \colhead{R.\ A.\tablenotemark{a}} &
\colhead{\sc Dec\tablenotemark{a}} & \colhead{$V$}  &
\colhead{\sc Sp.\ Type}  &  \colhead{V$_{\rm rad}$\tablenotemark{b}} 
}

\startdata
HD 26162 & 04:09:10.0 & +19:36:33 & 5.5 & K1 III & $+23.9 \pm 0.6$ \\
BD+6 648 & 04:13:11.9 & +06:35:51 & 9.1 & K0 III & $-141.4 \pm 1.0$ \\
HD 80170 & 09:16:57.1 & $-39$:24:05 & 5.3 & K5 III-IV & $0.0 \pm 0.2$ \\
HD 90861 & 10:29:53.7 & +28:34:52 & 7.2 & K2 III & $+36.3 \pm 0.4$ \\
\enddata
\tablenotetext{a}{J2000.0}
\tablenotetext{b}{Radial velocities are taken the {\em Astronomical
Almanac 1999}, except for BD+6~648, whose value was taken from
\citet{carney86}.}
\end{deluxetable}

\begin{deluxetable}{lllllll}
\tabletypesize{\footnotesize}
\tablecaption{Spectroscopic Observations \label{tab:spectra}}
\tablewidth{0pt}
\tablehead{
\colhead{\sc Star\tablenotemark{a}} &  
  \colhead{$<$V$>$} &  \colhead{\sc Exp.\ Time} & \colhead{HJD$-2,450,000$} & \colhead{S/N} & 
  \colhead{V$_{\rm rad}$ (\kms)\tablenotemark{b}} & \colhead{No.\ Ap.} }

\startdata%
 \noalign{\vskip +0.5ex}
 \multicolumn{7}{c}{\bf Berkeley\ 20} \cr
 \noalign{\vskip  .8ex}%
 \hline
 \noalign{\vskip -2ex}\\
 5*   & 14.80 & 1x\phn600\,s & 0762.7637 & \phn10 & $+77.4 \pm 0.6$ & 10 \\
 5*   & & 4x60\,m & 0831.6946 & \phn74 & $+77.1 \pm 1.0$ & 30 \\
 8*   &  15.15 & 1x\phn900\,s & 0762.7803 & \phn18 & $+80.1 \pm 0.5$ & 10 \\
 8*   & & 4x60\,m & 0832.6566 & \phn56 &  $+77.6 \pm 0.6$ & 23 \\
 22*  & 16.90 & 1x2400\,s    & 0762.8159 & \phn12 & $+78.9 \pm 1.2$ & 8 \\
 28*  & 17.11 & 1x2400\,s    & 0762.8515 & \phn11 & $+80.6 \pm 1.1$ & 9 \\
\cutinhead{\bf Berkeley\ 21}
 39*  & 15.67 & 1x\phn900\,s & 1184.6872 & \phn18 & $-0.2 \pm 0.6$ & 27 \\
 51*  & 15.69 & 1x\phn900\,s & 1184.6709 & \phn\phn9 & $-1.0 \pm 1.0$ & 27 \\
 67*  & 14.99 & 1x\phn600\,s & 1184.7200 & \phn13 & $+2.9 \pm 1.7$ & 27 \\
 67*  & & 4x60\,m & 1185.8721 & 101 & $+2.8 \pm 0.8$ & 37 \\
 88*  & 15.82 & 1x\phn900\,s & 1184.7030 & \phn17 & $-4.1 \pm 0.6$ & 27 \\
\cutinhead{\bf NGC\ 2141} 
 1007* & 13.27 & 1x\phn150\,s & 1186.8603 & \phn25 &  $+24.4 \pm 0.6$ & 36 \\
 1286* & 14.81 & 1x\phn600\,s & 1186.8466 & \phn20 &  $+23.0 \pm 1.0$ & 32 \\
 2066* & 14.18 & 1x\phn400\,s & 1186.8124 & \phn23 &  $+24.8 \pm 0.5$ & 35 \\
 514* & 14.09 & 1x\phn300\,s & 1186.8692 & \phn20 &  $+23.3 \pm 0.9$ & 34 \\
 1997 & 14.96 & 1x\phn900\,s & 1186.8029 & \phn23 & $-7.9 \pm 2.4$ & 21 \\
 1333* & 15.09 & 1x\phn700\,s & 1186.8258 & \phn20 &  $+23.5 \pm 1.3$ & 30 \\
 1348* & 13.25 & 1x\phn200\,s & 1186.8506 & \phn25 &  $+24.5 \pm 0.8$ &  37 \\
 1348* & & 4x40\,m & 1205.6854 & 130 &  $+24.7 \pm 0.5$ & 26 \\
 1821* & 14.13 & 1x\phn300\,s & 1186.8774 & \phn19 & $+24.8 \pm 1.0$ & 34 \\
\cutinhead{\bf Berkeley\ 29}
 412*  & 16.64 &    3x2400\,s & 0809.9655 & \phn14 & $+24.4 \pm 1.4$ & 24 \\
 1032* & 16.56 &    1x3600\,s & 0811.8746 & \phn\phn7 & $+25.7 \pm 1.6$ & 4 \\
 988*  & 14.59 & 1x\phn600\,s & 0811.9174 & \phn\phn9 & $+24.6 \pm 0.6$ & 5 \\
 988*  & & 5x60\,m & 1185.9525\tablenotemark{c} & 107 & $+24.4 \pm 0.9$ & 34 \\
 673*  & 14.38 & 1x\phn600\,s & 0811.9777 & \phn12 & $+24.6 \pm 0.5$ & 5 \\
 673*  & & 4x60\,m & 1183.8724 & 115 & $+24.6 \pm 0.5$ & 43 \\
 556*  & 16.60 &    1x3600\,s & 0811.9645 & \phn10 & $+24.7 \pm 2.8$ & 3 \\
 241*  & 14.48 &    1x1800\,s & 0811.9055 & \phn21 & $+23.9 \pm 0.2$ & 5 \\
 801*  & 16.58 &    2x3000\,s & 0810.8871 & \phn\phn8 & $+24.9 \pm 1.5$ & 11 \\
\cutinhead{\bf Berkeley\ 31}
 1065 & 16.32 &    1x2400\,s & 1204.7858 & \phn12 &  $+66.2 \pm 0.7$ & 27 \\
 886* & 14.77 & 1x\phn600\,s & 1205.7676 & \phn10 &  $+56.4 \pm 1.0$ & 29 \\
 886* & & 4x60\,m & 1206.7021 & \phn60 &  $+56.6 \pm 0.3$ & 23 \\
 720  & 16.42 &    1x2400\,s & 1204.7479 & \phn12 &  $+48.4 \pm 0.7$ & 27 \\
 728* & 16.61 &    1x3600\,s & 1205.7524 & \phn13 &  $+55.0 \pm 1.2$ & 25 \\
 666  & 13.95 & 1x\phn400\,s & 1205.7817 & \phn13 &  $+74.0 \pm 0.4$ & 22 \\
\cutinhead{\bf M\ 67}
 105* & 10.30  & 1x30\,m & 1187.0231 &  224 &  $+33.4 \pm 0.8$ & 44 \\
 108* &  9.72 & 1x20\,m & 1187.0433 &  197 &  $+33.5 \pm 0.7$ & 44 \\
 141* & 10.48  & 1x30\,m & 1186.9982 &  209 &  $+33.0 \pm 0.8$ & 44 \\
\enddata
\tablenotetext{a}{An asterisk designates stars we believe to be cluster members.}
\tablenotetext{b}{Radial velocity error refers to the standard deviation.}
\tablenotetext{c}{The spectra were taken over a period of three nights. The HJD
refers to the mid-exposure of all five observations.}
\end{deluxetable}

\begin{deluxetable}{cccccccc}
\tabletypesize{\footnotesize}
\tablecaption{Mean Radial Velocities of the Distant Open Clusters \label{tab:clustervelocities}}
\tablehead{
\colhead{Cluster}  &  \colhead{$<V_{\rm rad}>$} &
\colhead{$\sigma$(mean)} & \colhead{$\sigma$} & 
\colhead{$V_{\rm GSR}$\tablenotemark{a} } & 
\colhead{$l$} & \colhead{$b$} &
\colhead{No.\ Stars} 
}
\startdata
Berkeley 20 & +78.9 & 0.7 & 1.4 & +61.7 & 203.51 & $-17.28$ & 4 \\
Berkeley 21 & $-0.6$ & 1.4 & 2.9 & $-13.0$ & 186.84 & $-2.51$ & 4 \\
NGC 2141 & +24.1 & 0.3 & 0.7 & +8.9 & 198.08 & $-5.78$ & 8 \\
Berkeley 29 & +24.7 & 0.2 & 0.5 & +11.5 & 197.95 & +7.98 & 7\\
Berkeley 31 & +55.7 & 0.7 & 1.0 & +40.5 & 206.25 & +5.12 & 2\\
\enddata
\tablenotetext{a}{Following \citet{frinchaboy04}, $V_{\rm GSR}$ is the 
Galactocentric radial velocity assuming a solar apex 
$(\alpha,\delta)$~=~(18$^{\rm h}$,30$\arcdeg$) at 20 \kms and  
220 \kms rotational velocity for the local 
standard of rest.}
\end{deluxetable}

\begin{deluxetable}{lrrrrrrrr}
\tabletypesize{\scriptsize}
\tablecaption{Red Clump Data \label{tab:redclump}}
\tablehead{
\colhead{\sc Cluster}  & \colhead{\sc No.\ Stars} & \colhead{$<K>$} &
\colhead{$<J-K>$} & \colhead{E($J-K$)}    & \colhead{E($B-V$)} &
\colhead{($m-M$)$_{0}$}  &  \colhead{d (kpc)} & \colhead{$R_{\rm GC}$}}

\startdata
Berkeley 20 & 3 & 13.18 & 0.70 & 0.21: & 0.38: & 14.66: & 8.6: & 16.0\\
Berkeley 21 & 10 & 12.62 & 0.93 & 0.50 & 0.91 & 13.92 & 6.1 & 14.1 \\
NGC 2141 &  10 & 11.51 & 0.79 & 0.24 & 0.44 & 12.97 & 3.9 & 11.8 \\
Berkeley 29 & 5 & 14.20 & 0.59 & 0.02 & 0.04 & 15.85 & 14.8 & 22.5 \\
Berkeley 31\tablenotemark{a} & \nodata & \nodata & \nodata & \nodata & 0.13 & 13.6  & 5.3  & 12.9 \\
M67 & 3 & 7.96 & 0.615 & 0.015 & 0.03 & 9.56 & 0.8 & 8.6 \\
\enddata
\tablenotetext{a}{These values are derived from the work of \citet{guetter93}}
\end{deluxetable}

\begin{deluxetable}{lrrrrrrrrr}
\tabletypesize{\footnotesize}
\tablecaption{Atmospheric Parameters \label{tab:parameters}}
\tablehead{
\colhead{\sc Star} & \colhead{[Fe/H]\tablenotemark{a}} & 
  \colhead{$T_{\rm eff}$\tablenotemark{b}} & 
  \colhead{log $g$\tablenotemark{b}} & 
  \colhead{$T_{\rm eff}$\tablenotemark{c}} &   
  \colhead{log $g$\tablenotemark{c}} & 
  \colhead{$\xi_{\rm t}$\tablenotemark{c}} &
  \colhead{[Fe/H]\tablenotemark{c}} &
  \colhead{$T_{\rm eff}$\tablenotemark{d}} &   
  \colhead{log $g$\tablenotemark{d}}
}

\startdata%
 \noalign{\vskip +0.5ex}
 \multicolumn{9}{c}{\bf Berkeley\ 20} \cr
 \noalign{\vskip  .8ex}%
 \hline
 \noalign{\vskip -2ex}\\
 5  & $-0.61$  & 4610 & 1.6 & 4500 & 1.8 & 1.58 & $-0.53$ & 4620 & 1.6 \\
 8  & $-0.61$  & 4750 & 1.8 & 4590 & 2.2 & 1.42 & $-0.45$ & 4760 & 1.8 \\
\cutinhead{\bf Berkeley\ 21}
 67 & $-0.97$  & 4310 & 1.1 & \nodata & \nodata & \nodata & [$-0.54$] & 4180 & 1.1 \\
\cutinhead{\bf NGC\ 2141} 
 1348 & $-0.26$ & 4100 & 1.2 & 4100 & 1.2 & 1.33 & $-0.14$ & 4090 & 1.2 \\ 
\cutinhead{\bf Berkeley\ 29}
 988 & $-0.18$  & 3900 & 0.9 & 3980 & 1.0 & 1.31 & $-0.56$ & 4090 & 0.9 \\
 673 & $-0.18$  & 3750 & 0.6 & 3830 & 0.6 & 1.30 & $-0.52$ & 3890 & 0.6 \\
\cutinhead{\bf Berkeley\ 31\tablenotemark{e}}
 886 & $-0.41$ & 4390 & 2.2 & 4490 & 1.9 & 1.22 & $-0.53$ & 4400 & 2.2 \\
\cutinhead{\bf M\ 67}
 105 & +0.05  & 4390 & 2.1 & 4390 & 2.1 & 1.18 & +0.06 & 4340 & 2.1 \\
 108 & +0.05  & 4180 & 1.7 & 4200 & 1.6 & 1.38 & $-0.01$ & 4140 & 1.7 \\
 141 & +0.05  & 4700 & 2.3 & 4700 & 2.3 & 1.34 & 0.00 & 4640 & 2.3 \\
\enddata
\tablenotetext{a}{From \citet{janes94}.}
\tablenotetext{b}{Estimates obtained using the reddening and distance
estimates from Table~\ref{tab:redclump} and the photometry from
Table~\ref{tab:photometry}.}
\tablenotetext{c}{Quantities derived using the spectroscopic methods
described in the text.}
\tablenotetext{d}{Re-derived values from photometry using the spectroscopic
metallicities.}
\tablenotetext{e}{The photometric temperatures and gravities
were derived using reddening and distance determinations
from \citet{guetter93}.}

\end{deluxetable}

\begin{deluxetable}{lccrclccrclccr} 
\tabletypesize{\tiny}
\rotate
\tablecolumns{14} 
\tablewidth{0pc} 
\tablecaption{Line list\label{tab:line}}
\tablehead{ 
\colhead{Wavelength(\AA)} &
\colhead{Species} &
\colhead{LEP(eV)} &
\colhead{log $gf$} &
\colhead{} &
\colhead{Wavelength(\AA)} &
\colhead{Species} &
\colhead{LEP(eV)} &
\colhead{log $gf$} &
\colhead{} &
\colhead{Wavelength(\AA)} &
\colhead{Species} &
\colhead{LEP(eV)} &
\colhead{log $gf$} 
}
\startdata
6300.30 & O I & 0.00 & $-$9.717 & & 5837.70 & Fe I & 4.29 & $-$2.340 & & 6750.15 & Fe I & 2.42 & $-$2.621 \\
6363.78 & O I & 0.02 & $-$10.185 & & 5853.16 & Fe I & 1.49 & $-$5.280 & & 6756.56 & Fe I & 4.29 & $-$2.750 \\
5688.19 & Na I & 2.11 & $-$0.420 & & 5855.09 & Fe I & 4.60 & $-$1.547 & & 6786.86 & Fe I & 4.19 & $-$1.850 \\
6154.23 & Na I & 2.10 & $-$1.530 & & 5856.10 & Fe I & 4.29 & $-$1.640 & & 6810.26 & Fe I & 4.60 & $-$1.000 \\
6160.75 & Na I & 2.10 & $-$1.230 & & 5858.79 & Fe I & 4.22 & $-$2.260 & & 6971.93 & Fe I & 3.02 & $-$3.390 \\
6318.72 & Mg I & 5.11 & $-$1.970 & & 5909.97 & Fe I & 3.21 & $-$2.640 & & 7112.17 & Fe I & 2.99 & $-$3.044 \\
6319.24 & Mg I & 5.11 & $-$2.220 & & 5916.25 & Fe I & 2.45 & $-$2.994 & & 7189.15 & Fe I & 3.07 & $-$2.796 \\
6965.41 & Mg I & 5.75 & $-$1.510 & & 5927.80 & Fe I & 4.65 & $-$1.090 & & 7223.66 & Fe I & 3.01 & $-$2.269 \\
7387.69 & Mg I & 5.75 & $-$0.870 & & 5933.80 & Fe I & 4.64 & $-$2.230 & & 7401.69 & Fe I & 4.18 & $-$1.660 \\
6696.02 & Al I & 3.14 & $-$1.340 & & 5956.69 & Fe I & 0.86 & $-$4.608 & & 7710.36 & Fe I & 4.22 & $-$1.129 \\
6698.67 & Al I & 3.14 & $-$1.640 & & 5969.58 & Fe I & 4.28 & $-$2.730 & & 7723.20 & Fe I & 2.28 & $-$3.617 \\
7835.31 & Al I & 4.02 & $-$0.470 & & 6012.21 & Fe I & 2.22 & $-$4.070 & & 7941.09 & Fe I & 3.27 & $-$2.331 \\
7836.13 & Al I & 4.02 & $-$0.310 & & 6019.36 & Fe I & 3.57 & $-$3.360 & & 4993.36 & Fe II & 2.80 & $-$3.480 \\
5690.43 & Si I & 4.93 & $-$1.751 & & 6027.05 & Fe I & 4.07 & $-$1.106 & & 5234.62 & Fe II & 3.22 & $-$2.150 \\
5793.07 & Si I & 4.93 & $-$1.843 & & 6054.08 & Fe I & 4.37 & $-$2.310 & & 5325.55 & Fe II & 3.22 & $-$3.220 \\
6125.02 & Si I & 5.61 & $-$1.506 & & 6105.13 & Fe I & 4.55 & $-$2.050 & & 5414.07 & Fe II & 3.22 & $-$3.750 \\
6145.01 & Si I & 5.62 & $-$1.362 & & 6120.24 & Fe I & 0.91 & $-$5.970 & & 5425.26 & Fe II & 3.20 & $-$3.370 \\
6155.13 & Si I & 5.62 & $-$0.786 & & 6145.42 & Fe I & 3.37 & $-$3.600 & & 5991.38 & Fe II & 3.15 & $-$3.557 \\
6166.44 & Ca I & 2.52 & $-$1.142 & & 6151.62 & Fe I & 2.17 & $-$3.299 & & 6084.11 & Fe II & 3.20 & $-$3.808 \\
6169.04 & Ca I & 2.52 & $-$0.797 & & 6157.73 & Fe I & 4.08 & $-$1.320 & & 6149.26 & Fe II & 3.89 & $-$2.724 \\
6169.56 & Ca I & 2.53 & $-$0.478 & & 6159.38 & Fe I & 4.61 & $-$1.970 & & 6247.56 & Fe II & 3.89 & $-$2.329 \\
6455.60 & Ca I & 2.52 & $-$1.290 & & 6165.36 & Fe I & 4.14 & $-$1.490 & & 6369.46 & Fe II & 2.89 & $-$4.250 \\
6064.63 & Ti I & 1.05 & $-$1.888 & & 6173.34 & Fe I & 2.22 & $-$2.880 & & 6416.92 & Fe II & 3.89 & $-$2.740 \\
6091.17 & Ti I & 2.27 & $-$0.367 & & 6180.20 & Fe I & 2.73 & $-$2.637 & & 6432.68 & Fe II & 2.89 & $-$3.708 \\
6312.24 & Ti I & 1.46 & $-$1.496 & & 6200.31 & Fe I & 2.61 & $-$2.437 & & 6456.38 & Fe II & 3.90 & $-$2.075 \\
6336.10 & Ti I & 1.44 & $-$1.687 & & 6219.28 & Fe I & 2.20 & $-$2.433 & & 7224.49 & Fe II & 3.89 & $-$3.243 \\
6013.53 & Mn I & 3.07 & $-$0.251 & & 6229.23 & Fe I & 2.84 & $-$2.846 & & 7711.72 & Fe II & 3.90 & $-$2.543 \\
6016.67 & Mn I & 3.08 & $-$0.100 & & 6232.64 & Fe I & 3.65 & $-$1.283 & & 6189.00 & Co I & 1.71 & $-$2.450 \\
6021.80 & Mn I & 3.08 & 0.034 & & 6246.32 & Fe I & 3.60 & $-$0.894 & & 6455.03 & Co I & 3.63 & $-$0.250 \\
4802.88 & Fe I & 3.69 & $-$1.531 & & 6265.13 & Fe I & 2.17 & $-$2.550 & & 6632.45 & Co I & 2.28 & $-$2.000 \\
5044.21 & Fe I & 2.85 & $-$2.034 & & 6270.22 & Fe I & 2.86 & $-$2.500 & & 5846.99 & Ni I & 1.68 & $-$3.210 \\
5054.64 & Fe I & 3.64 & $-$1.938 & & 6271.28 & Fe I & 3.33 & $-$2.728 & & 6086.28 & Ni I & 4.26 & $-$0.515 \\
5242.49 & Fe I & 3.63 & $-$0.980 & & 6297.79 & Fe I & 2.22 & $-$2.740 & & 6175.37 & Ni I & 4.09 & $-$0.535 \\
5321.11 & Fe I & 4.43 & $-$1.106 & & 6301.50 & Fe I & 3.65 & $-$0.766 & & 6177.24 & Ni I & 1.83 & $-$3.510 \\
5322.04 & Fe I & 2.28 & $-$2.840 & & 6322.69 & Fe I & 2.59 & $-$2.426 & & 6204.60 & Ni I & 4.09 & $-$1.140 \\
5326.14 & Fe I & 3.57 & $-$2.130 & & 6336.82 & Fe I & 3.68 & $-$0.916 & & 6635.12 & Ni I & 4.42 & $-$0.828 \\
5365.40 & Fe I & 3.57 & $-$1.040 & & 6353.84 & Fe I & 0.91 & $-$6.477 & & 6772.32 & Ni I & 3.66 & $-$0.987 \\
5367.48 & Fe I & 4.41 & 0.430 & & 6355.03 & Fe I & 2.84 & $-$2.403 & & 7800.29 & Rb I & 0.00 & 0.130 \\
5379.57 & Fe I & 3.69 & $-$1.530 & & 6411.65 & Fe I & 3.65 & $-$0.734 & & 6127.44 & Zr I & 0.15 & $-$1.060 \\
5491.84 & Fe I & 4.18 & $-$2.250 & & 6469.19 & Fe I & 4.84 & $-$0.770 & & 6134.55 & Zr I & 0.00 & $-$1.280 \\
5618.63 & Fe I & 4.21 & $-$1.292 & & 6574.23 & Fe I & 0.99 & $-$5.004 & & 6143.20 & Zr I & 0.07 & $-$1.100 \\
5701.55 & Fe I & 2.56 & $-$2.216 & & 6575.02 & Fe I & 2.59 & $-$2.727 & & 5853.64 & Ba II & 0.60 & $-$1.010 \\
5705.47 & Fe I & 4.30 & $-$1.420 & & 6581.21 & Fe I & 1.50 & $-$4.705 & & 5805.77 & La II & 0.13 & $-$1.560 \\
5741.85 & Fe I & 4.25 & $-$1.689 & & 6609.11 & Fe I & 2.56 & $-$2.692 & & 6390.48 & La II & 0.32 & $-$1.410 \\
5775.08 & Fe I & 4.22 & $-$1.310 & & 6648.08 & Fe I & 1.01 & $-$5.918 & & 6645.11 & Eu II & 1.38 & 0.204 \\
5778.45 & Fe I & 2.59 & $-$3.480 & & 6699.16 & Fe I & 4.59 & $-$2.170 & & \nodata & \nodata & \nodata & \nodata \\
5811.92 & Fe I & 4.14 & $-$2.430 & & 6739.52 & Fe I & 1.56 & $-$4.820 & & \nodata & \nodata & \nodata & \nodata \\
\enddata

\end{deluxetable}

\begin{deluxetable}{lcccr} 
\tabletypesize{\footnotesize}
\tablecolumns{5} 
\tablewidth{0pc} 
\tablecaption{Measured solar abundances\label{tab:solar}}
\tablehead{ 
\colhead{log~$\epsilon$(X)} &
\colhead{\citet{grevesse98}} &
\colhead{This study} &
\colhead{$\sigma$} &
\colhead{N}
}
\startdata
O & 8.83 & 8.86 & \nodata & 1 \\
Na & 6.33 & 6.28 & 0.05 & 3 \\
Mg & 7.58 & 7.49 & 0.01 & 2 \\
Al & 6.47 & 6.27 & 0.06 & 4 \\
Si\tablenotemark{a} & 7.55 & 7.55 & 0.00 & 5 \\
Ca & 6.36 & 6.43 & 0.10 & 4 \\
Ti & 5.02 & 4.92 & 0.03 & 4 \\
Mn & 5.39 & 5.42 & 0.09 & 3 \\
Fe I & 7.50 & 7.54 & 0.09 & 57 \\
Fe II & 7.50 & 7.54 & 0.09 & 15 \\
Co & 4.92 & 4.91 & 0.08 & 3 \\
Ni & 6.25 & 6.30 & 0.13 & 7 \\
Rb & 2.60 & 2.58 & \nodata & 1 \\
Zr & 2.60 & 2.80 & 0.09 & 3 \\
Ba & 2.13 & 2.50 & \nodata & 1 \\
La\tablenotemark{b} & 1.17 & 1.13 & \nodata & \nodata \\
Eu\tablenotemark{b} & 0.51 & 0.52 & \nodata & \nodata \\
\enddata

\tablenotetext{a}{For Si we used an inverted solar analysis
to derive the $gf$ values.}

\tablenotetext{b}{For La we adopted 1.13 \citep{la} and for Eu we
adopted 0.52 \citep{eu}}

\end{deluxetable}

\begin{deluxetable}{lrrrcr} 
\tabletypesize{\footnotesize}
\tablecolumns{6} 
\tablewidth{0pc} 
\tablecaption{Arcturus abundances\label{tab:arc}}
\tablehead{ 
\colhead{Species} &
\colhead{\citet{peterson93}} &
\colhead{\citet{carraro04}} &
\colhead{This study} &
\colhead{$\sigma$} &
\colhead{N}
}
\startdata
{\rm [O/Fe]} & 0.40 & 0.39 & 0.30 & 0.02 & 2 \\
{\rm [Na/Fe]} & 0.30 & 0.24 & 0.15 & 0.08 & 3 \\
{\rm [Mg/Fe]} & 0.40 & 0.46 & 0.45 & 0.14 & 4 \\
{\rm [Al/Fe]} & 0.40 & 0.27 & 0.28 & 0.08 & 4 \\
{\rm [Si/Fe]} & 0.40 & 0.25 & 0.35 & 0.06 & 5 \\
{\rm [Ca/Fe]} & 0.30 & 0.32 & 0.22 & 0.09 & 4 \\
{\rm [Ti/Fe]} & 0.30 & 0.22 & 0.26 & 0.03 & 4 \\
{\rm [Mn/Fe]} & \nodata & \nodata & $-$0.25 & 0.06 & 3 \\
{\rm [Fe I/H]} & $-$0.50 & $-$0.51 & $-$0.56 & 0.13 & 50 \\
{\rm [Fe II/H]} & $-$0.50 & $-$0.51 & $-$0.59 & 0.12 & 12 \\
{\rm [Co/Fe]} & \nodata & \nodata & 0.12 & 0.08 & 3 \\
{\rm [Ni/Fe]} & 0.00 & 0.16 & $-$0.02 & 0.06 & 7 \\
{\rm [Rb/Fe]} & \nodata & \nodata & 0.08 & \nodata & 1 \\
{\rm [Zr/Fe]} & \nodata & \nodata & $-$0.27 & 0.08 & 3 \\
{\rm [Ba/Fe]} & \nodata & \nodata & $-$0.09 & \nodata & 1 \\
{\rm [La/Fe]} & \nodata & \nodata & 0.03 & 0.09 & 2 \\
{\rm [Eu/Fe]} & \nodata & \nodata & 0.29 & \nodata & 1 \\
\enddata
\end{deluxetable}

\begin{deluxetable}{lrrrrrrrrrrr} 
\tabletypesize{\footnotesize}
\tablecolumns{12} 
\tablewidth{0pc} 
\tablecaption{Mean stellar abundances\label{tab:abund}}
\tablehead{ 
\colhead{Species} &
\colhead{Abundance} &
\colhead{$\sigma$} &
\colhead{N} &
\colhead{} &
\colhead{Abundance} &
\colhead{$\sigma$} &
\colhead{N} &
\colhead{} &
\colhead{Abundance} &
\colhead{$\sigma$} &
\colhead{N}
}
\startdata
 \noalign{\vskip +0.5ex}
 \hline
 \noalign{\vskip +0.5ex}
\colhead{} & \multicolumn{3}{c}{\bf Be 20 5} & & \multicolumn{3}{c}{\bf Be 20 8} & & \multicolumn{3}{c}{\bf Be 29 673} \\
 \noalign{\vskip  .8ex} \hline
 \noalign{\vskip -2ex}\\
{\rm [O/Fe]} & 0.19 & 0.05 & 2 & & 0.17 & \nodata & 1 & & 0.20 & 0.04 & 2 \\
{\rm [Na/Fe]} & 0.32 & 0.08 & 3 & & 0.10 & 0.07 & 2 & & 0.37 & 0.19 & 3 \\
{\rm [Mg/Fe]} & 0.26 & 0.19 & 4 & & 0.22 & 0.09 & 4 & & 0.31 & 0.05 & 4 \\
{\rm [Al/Fe]} & 0.18 & 0.13 & 4 & & 0.18 & 0.06 & 4 & & 0.25 & 0.06 & 4 \\
{\rm [Si/Fe]} & 0.08 & 0.05 & 5 & & 0.03 & 0.06 & 4 & & 0.20 & 0.06 & 4 \\
{\rm [Ca/Fe]} & 0.08 & 0.04 & 4 & & 0.06 & 0.05 & 4 & & 0.04 & 0.04 & 4 \\
{\rm [Ti/Fe]} & 0.45 & 0.11 & 4 & & 0.33 & 0.09 & 4 & & 0.37 & 0.06 & 4 \\
{\rm [Mn/Fe]} & $-$0.07 & 0.04 & 2 & & $-$0.16 & 0.05 & 2 & & $-$0.27 & 0.19 & 2 \\
{\rm [Fe I/H]} & $-$0.49 & 0.21 & 37 & & $-$0.40 & 0.17 & 43 & & $-$0.52 & 0.15 & 24 \\
{\rm [Fe II/H]} & $-$0.57 & 0.09 & 5 & & $-$0.49 & 0.17 & 5 & & $-$0.52 & 0.21 & 6 \\
{\rm [Co/Fe]} & 0.20 & 0.08 & 3 & & 0.15 & 0.04 & 4 & & 0.08 & \nodata & 1 \\
{\rm [Ni/Fe]} & 0.00 & 0.11 & 7 & & $-$0.04 & 0.13 & 7 & & $-$0.04 & 0.10 & 7 \\
{\rm [Rb/Fe]} & 0.11 & \nodata & 1 & & \nodata & \nodata & \nodata & & $-$0.16 & \nodata & 1 \\
{\rm [Zr/Fe]} & 0.08 & 0.06 & 3 & & 0.02 & 0.10 & 2 & & 0.08 & 0.15 & 2 \\
{\rm [Ba/Fe]} & 0.10 & \nodata & 1 & & 0.17 & \nodata & 1 & & 0.32 & \nodata & 1 \\
{\rm [La/Fe]} & 0.28 & 0.02 & 2 & & 0.32 & 0.04 & 2 & & 0.16 & 0.07 & 2 \\
{\rm [Eu/Fe]} & 0.33 & \nodata & 1 & & 0.29 & \nodata & 1 & & 0.11 & \nodata & 1 \\
 \noalign{\vskip +0.5ex}
 \hline 
 \noalign{\vskip +0.5ex}
\colhead{} & \multicolumn{3}{c}{\bf Be 29 988} & & \multicolumn{3}{c}{\bf Be 31 886} & & \multicolumn{3}{c}{\bf NGC 2141 1348} \\
 \noalign{\vskip  .8ex} \hline
 \noalign{\vskip -2ex}\\
{\rm [O/Fe]} & 0.26 & 0.05 & 2 & & 0.24 & 0.08 & 2 & & 0.00 & 0.06 & 2 \\
{\rm [Na/Fe]} & 0.36 & 0.27 & 3 & & 0.27 & 0.10 & 3 & & 0.41 & 0.04 & 3 \\
{\rm [Mg/Fe]} & 0.37 & 0.13 & 4 & & 0.40 & 0.10 & 4 & & 0.24 & 0.14 & 4 \\
{\rm [Al/Fe]} & 0.27 & 0.09 & 4 & & 0.22 & 0.13 & 4 & & 0.18 & 0.07 & 4 \\
{\rm [Si/Fe]} & 0.16 & 0.10 & 3 & & 0.20 & 0.14 & 5 & & 0.05 & 0.19 & 4 \\
{\rm [Ca/Fe]} & 0.00 & 0.04 & 4 & & 0.13 & 0.05 & 4 & & 0.10 & 0.04 & 4 \\
{\rm [Ti/Fe]} & 0.29 & 0.10 & 4 & & 0.08 & 0.09 & 4 & & 0.24 & 0.11 & 4 \\
{\rm [Mn/Fe]} & $-$0.19 & 0.06 & 3 & & $-$0.54 & 0.12 & 3 & & $-$0.20 & 0.12 & 3 \\
{\rm [Fe I/H]} & $-$0.56 & 0.19 & 25 & & $-$0.57 & 0.23 & 35 & & $-$0.18 & 0.15 & 23 \\
{\rm [Fe II/H]} & $-$0.55 & 0.17 & 8 & & $-$0.49 & 0.18 & 7 & & $-$0.10 & 0.27 & 4 \\
{\rm [Co/Fe]} & 0.09 & 0.03 & 2 & & 0.25 & 0.08 & 2 & & 0.04 & 0.04 & 2 \\
{\rm [Ni/Fe]} & 0.00 & 0.09 & 7 & & 0.11 & 0.12 & 7 & & 0.04 & 0.11 & 7 \\
{\rm [Rb/Fe]} & $-$0.07 & \nodata & 1 & & $-$0.11 & \nodata & 1 & & $-$0.20 & \nodata & 1 \\
{\rm [Zr/Fe]} & 0.06 & 0.13 & 3 & & 0.35 & 0.07 & 3 & & 0.63 & 0.06 & 2 \\
{\rm [Ba/Fe]} & 0.29 & \nodata & 1 & & 0.64 & \nodata & 1 & & 0.91 & \nodata & 1 \\
{\rm [La/Fe]} & 0.30 & 0.02 & 2 & & 0.91 & 0.07 & 2 & & 0.57 & 0.07 & 2 \\
{\rm [Eu/Fe]} & 0.20 & \nodata & 1 & & 0.56 & \nodata & 1 & & 0.17 & \nodata & 1 \\
 \noalign{\vskip +0.5ex}
 \hline
 \noalign{\vskip +0.5ex}
\colhead{} & \multicolumn{3}{c}{\bf M 67 105} & & \multicolumn{3}{c}{\bf M 67 108} & & \multicolumn{3}{c}{\bf M 67 141} \\
 \noalign{\vskip  .8ex} \hline
 \noalign{\vskip -2ex}\\
{\rm [O/Fe]} & 0.05 & 0.07 & 2 & & 0.08 & 0.04 & 2 & & 0.10 & 0.06 & 2 \\
{\rm [Na/Fe]} & 0.29 & 0.10 & 3 & & 0.38 & 0.08 & 3 & & 0.24 & 0.10 & 3 \\
{\rm [Mg/Fe]} & 0.15 & 0.06 & 4 & & 0.15 & 0.13 & 4 & & 0.18 & 0.04 & 4 \\
{\rm [Al/Fe]} & 0.17 & 0.05 & 4 & & 0.18 & 0.04 & 4 & & 0.16 & 0.06 & 4 \\
{\rm [Si/Fe]} & 0.08 & 0.11 & 5 & & 0.09 & 0.13 & 5 & & 0.11 & 0.08 & 5 \\
{\rm [Ca/Fe]} & 0.08 & 0.05 & 4 & & 0.03 & 0.03 & 4 & & 0.09 & 0.03 & 4 \\
{\rm [Ti/Fe]} & 0.17 & 0.04 & 4 & & 0.13 & 0.03 & 4 & & 0.05 & 0.05 & 4 \\
{\rm [Mn/Fe]} & $-$0.08 & 0.04 & 3 & & $-$0.12 & 0.12 & 3 & & $-$0.20 & 0.03 & 3 \\
{\rm [Fe I/H]} & 0.03 & 0.15 & 30 & & $-$0.05 & 0.15 & 27 & & $-$0.01 & 0.12 & 37 \\
{\rm [Fe II/H]} & 0.09 & 0.12 & 6 & & 0.03 & 0.12 & 5 & & 0.01 & 0.09 & 8 \\
{\rm [Co/Fe]} & 0.02 & 0.11 & 3 & & 0.03 & 0.14 & 3 & & 0.01 & 0.09 & 3 \\
{\rm [Ni/Fe]} & 0.11 & 0.11 & 7 & & 0.07 & 0.06 & 7 & & 0.06 & 0.09 & 7 \\
{\rm [Rb/Fe]} & $-$0.31 & \nodata & 1 & & $-$0.28 & \nodata & 1 & & $-$0.22 & \nodata & 1 \\
{\rm [Zr/Fe]} & $-$0.32 & 0.03 & 3 & & $-$0.27 & 0.05 & 3 & & $-$0.26 & 0.04 & 3 \\
{\rm [Ba/Fe]} & $-$0.05 & \nodata & 1 & & $-$0.04 & \nodata & 1 & & 0.02 & \nodata & 1 \\
{\rm [La/Fe]} & 0.11 & 0.02 & 2 & & 0.09 & 0.07 & 2 & & 0.13 & 0.04 & 2 \\
{\rm [Eu/Fe]} & 0.04 & \nodata & 1 & & 0.09 & \nodata & 1 & & 0.05 & \nodata & 1 \\
\enddata

\end{deluxetable}

\begin{deluxetable}{lrrr} 
\tabletypesize{\footnotesize}
\tablecolumns{4} 
\tablewidth{0pc} 
\tablecaption{Abundance dependences on model parameters for
Be 20 5\label{tab:err}}
\tablehead{ 
\colhead{Species} &
\colhead{\teff~+~100K} &
\colhead{log $g$~+~0.2} &
\colhead{$\xi_{\rm t}$~+~0.2} 
}
\startdata
{\rm [O/Fe]} & 0.02 & 0.00 & 0.05 \\
{\rm [Na/Fe]} & 0.11 & $-$0.07 & 0.02 \\
{\rm [Mg/Fe]} & 0.06 & $-$0.07 & 0.01 \\
{\rm [Al/Fe]} & 0.10 & $-$0.06 & 0.02 \\
{\rm [Si/Fe]} & $-$0.02 & $-$0.02 & 0.03 \\
{\rm [Ca/Fe]} & 0.13 & $-$0.06 & $-$0.04 \\
{\rm [Ti/Fe]} & 0.18 & $-$0.06 & $-$0.03 \\
{\rm [Mn/Fe]} & 0.11 & $-$0.05 & $-$0.03 \\
{\rm [Fe I/H]}  & 0.07 & 0.02 & $-$0.09 \\
{\rm [Fe II/H]} & $-$0.11 & 0.11 & $-$0.03 \\
{\rm [Co/Fe]} & 0.11 & $-$0.02 & 0.05 \\
{\rm [Ni/Fe]} & 0.05 & $-$0.02 & 0.01 \\
{\rm [Rb/Fe]} & 0.14 & $-$0.07 & 0.06 \\
{\rm [Zr/Fe]} & 0.24 & $-$0.05 & 0.03 \\
{\rm [Ba/Fe]} & 0.05 & 0.07 & $-$0.20 \\
{\rm [La/Fe]} & 0.04 & 0.02 & 0.03 \\
{\rm [Eu/Fe]} & 0.01 & 0.03 & 0.03 \\
\enddata
\end{deluxetable}

\begin{deluxetable}{lrrrr} 
\tabletypesize{\footnotesize}
\tablecolumns{5} 
\tablewidth{0pc} 
\tablecaption{Mean abundances in M67\label{tab:m67}}
\tablehead{ 
\colhead{Species} &
\colhead{\citet{tautvaisiene00}} &
\colhead{This study} &
\colhead{$\sigma$} &
\colhead{N}
}
\startdata
{\rm [O/Fe]} & 0.02 & 0.07 & 0.05 & 6 \\
{\rm [Na/Fe]} & 0.19 & 0.30 & 0.10 & 9 \\
{\rm [Mg/Fe]} & 0.09 & 0.16 & 0.08 & 12 \\
{\rm [Al/Fe]} & 0.14 & 0.17 & 0.05 & 12 \\
{\rm [Si/Fe]} & 0.10 & 0.09 & 0.11 & 15 \\
{\rm [Ca/Fe]} & 0.04 & 0.07 & 0.06 & 12 \\
{\rm [Ti/Fe]} & 0.04 & 0.12 & 0.07 & 12 \\
{\rm [Fe/H]} & $-$0.03 & 0.02 & 0.14 & 113 \\
{\rm [Ni/Fe]} & 0.04 & 0.08 & 0.10 & 21 \\
{\rm [Zr/Fe]} & $-$0.17 & $-$0.28 & 0.04 & 9 \\
{\rm [Ba/Fe]} & 0.04 & $-$0.02 & 0.05 & 3 \\
{\rm [La/Fe]} & 0.13 & 0.11 & 0.06 & 6 \\
{\rm [Eu/Fe]} & 0.07 & 0.06 & 0.02 & 3 \\
\enddata
\end{deluxetable}

\clearpage

\clearpage

\clearpage

\clearpage

\clearpage

\end{document}